\documentclass[longauth]{aa} 
\usepackage{txfonts}
\usepackage{xcolor}
\usepackage[]{graphicx}
\usepackage{caption}
\usepackage{subcaption}
\usepackage{longtable}
\usepackage{tabularx}
\usepackage{lscape}
\usepackage[breaklinks=true]{hyperref} 
\usepackage{supertabular}
\usepackage{array}
\usepackage{textgreek}
\usepackage{float}
\usepackage{placeins}
\usepackage{comment}

\graphicspath{
    {.} % document root dir
    {paper_figures/}
}

\usepackage{thmtools}
\usepackage{multirow}
\usepackage{tikz}

\usepackage{algorithm}
\usepackage[noend]{algpseudocode}
\makeatletter
\renewcommand{\ALG@name}{Algorithm}
\makeatother
\usetikzlibrary{matrix,plotmarks,shapes,arrows,decorations.pathmorphing}

\DeclareMathOperator\erf{erf}
\DeclareMathOperator{\Tr}{Tr}
\DeclareMathOperator{\diag}{diag}

\DeclareMathOperator{\BIC}{BIC}

\newcommand{\argmin}{\operatornamewithlimits{argmin}}

\declaretheoremstyle{nospacetheorem}
\declaretheorem[style=nospacetheorem,name=A]{hypothesis}

\newcommand{\lae}{\ensuremath{\mathrm{Ly}\alpha} emitter}
\newcommand{\laes}{\ensuremath{\mathrm{Ly}\alpha} emitters}

\newcommand{\vmpc}{\ensuremath{\mathrm{cMpc^{3}}}}

\newcommand{\msun}{\ifmmode M_{\odot} \else M$_{\odot}$\fi}

\newcommand{\kms}{\ensuremath{\mathrm{km\,s^{-1}}}}
\newcommand{\degree}{\ensuremath{^\circ}}

\newcommand{\ergsluma}[1]{\ensuremath{\mathrm{10^{#1} erg\,s^{-1}}}}
\newcommand{\ergslumb}[2]{\ensuremath{\mathrm{{#1} \times 10^{#2} \, erg\,s^{-1}}}}

\newcommand{\ergs}{\ensuremath{\mathrm{erg\,s^{-1}\,cm^{-2}\,\AA^{-1}}}}
\newcommand{\ergsa}[1]{\ensuremath{\mathrm{10^{#1}\,erg\,s^{-1}\,cm^{-2}\,\AA^{-1}}}}
\newcommand{\ergsb}[2]{\ensuremath{\mathrm{{#1} \times 10^{#2}\,erg\,s^{-1}\,cm^{-2}\,\AA^{-1}}}}

\newcommand{\ergsline}{\ensuremath{\mathrm{erg\,s^{-1}\,cm^{-2}}}}
\newcommand{\ergslineb}[2]{\ensuremath{\mathrm{#1 \times 10^{#2}\,erg\,s^{-1}\,cm^{-2}}}}

\newcommand{\erglinesurf}[2]{\ensuremath{\mathrm{#1 \times 10^{#2}\,erg\,s^{-1}\,cm^{-2}\,arcsec^{-2}}}}

\newcommand{\ergsurfb}{\ensuremath{\mathrm{erg\,s^{-1}\,cm^{-2}\,arcsec^{-2}}}}

\newcommand{\lya}{Ly\textalpha}
\newcommand{\ha}{H\textalpha}
\newcommand{\hb}{H\textbeta}

\newcommand{\oii}{[O\,{\sc ii}]}
\newcommand{\oiid}{[O\,{\sc ii}]\textlambda\textlambda3726,3729}
\newcommand{\oiiid}{[O\,{\sc iii}]\textlambda4959,5007}
\newcommand{\oiiia}{[O\,{\sc iii}]\textlambda4959}
\newcommand{\oiiib}{[O\,{\sc iii}]\textlambda5007}

\newcommand{\ciii}{C\,{\sc iii}]}
\newcommand{\ciiid}{C\,{\sc iii}]\textlambda\textlambda1907,1909}

\newcommand{\civ}{C\,{\sc iv}}
\newcommand{\civd}{C\,{\sc iv}\,\textlambda\textlambda1548,1550}

\newcommand{\mgii}{Mg\,{\sc ii}}
\newcommand{\mgiid}{Mg\,{\sc ii}\,\textlambda\textlambda2797,2803}
\newcommand{\heii}{He\,{\sc ii}}

\newcommand{\mosaic}{\textsf{MOSAIC}}
\newcommand{\udft}{\textsf{UDF-10}}
\newcommand{\mxdf}{\textsf{MXDF}}

\newcommand{\drt}{\textsf{DR2}}
\newcommand{\dro}{\textsf{DR1}}

\newcommand{\origin}{\textsf{ORIGIN}}
\newcommand{\odhin}{\textsf{ODHIN}}
\newcommand{\nbext}{\textsf{NBEXT}}
\newcommand{\pymarz}{\textsf{pyMarZ}}
\newcommand{\mpdaf}{\textsf{MPDAF}}
\newcommand{\pfit}{\textsf{pyPlatefit}}
\newcommand{\sex}{\textsf{SExtractor}}
\newcommand{\se}{\textsf{SourceInspector}}
\newcommand{\psfrec}{\textsf{muse-psfr}}
\newcommand{\prospector}{\textsf{Prospector}}
\newcommand{\magphys}{\textsf{Magphys}}
\newcommand{\amused}{\textsf{AMUSED}}

\newcommand{\hubble}{\textsf{\textit{Hubble}}}
\newcommand{\hudf}{\textsf{HUDF}}
\newcommand{\uvudf}{\textsf{UVUDF}}
\newcommand{\tdhst}{\textsf{3D-HST}}
\newcommand{\candels}{\textsf{CANDELS v2}}
\newcommand{\astrod}{\textsf{ASTRODEEP}}

\begin{document}

\title{The MUSE \textit{\hubble}\ Ultra Deep Field surveys: Data release II 
\thanks{Based on observations made with ESO telescopes at the La Silla Paranal Observatory under the programs 094.A-0289(B), 095.A-0010(A), 096.A-0045(A), 096.A-0045(B) and 1101.A-0127}
}

\author{
Roland Bacon\inst{1} \and Jarle Brinchmann\inst{2} \and Simon Conseil\inst{3} \and Michael Maseda\inst{4} \and Themiya Nanayakkara\inst{5} \and Martin Wendt\inst{6,7} \and Raphael Bacher\inst{1,8} \and David Mary\inst{9} \and Peter M. Weilbacher\inst{6} \and Davor Krajnovi\'c \inst{6} \and Leindert Boogaard\inst{16} \and Nicolas Bouch\'e\inst{1} \and Thierry Contini\inst{14} \and Benoit Epinat\inst{3,18} \and Anna Feltre\inst{11} \and Yucheng Guo\inst{1} \and Christian Herenz\inst{15} \and Wolfram Kollatschny\inst{17} \and Haruka Kusakabe\inst{12} \and Floriane Leclercq\inst{10} \and L\'eo Michel-Dansac\inst{1} \and Roser Pello\inst{3} \and Johan Richard\inst{1} \and Martin Roth\inst{6} \and Gregory Salvignol\inst{1} \and Joop Schaye\inst{13} \and Matthias Steinmetz\inst{6} \and Laurence Tresse\inst{3} \and Tanya Urrutia\inst{6} \and Anne Verhamme\inst{12} \and Eloise Vitte\inst{12} \and Lutz Wisotzki\inst{6} \and Sebastiaan L. Zoutendijk\inst{13}
}
   
 \titlerunning{The MUSE HUDF surveys: Data release II}

\institute{
   Univ Lyon, Univ Lyon1, Ens de Lyon, CNRS, Centre de Recherche Astrophysique de Lyon UMR5574, F-69230, Saint-Genis-Laval, France
    \and Instituto de Astrof{\'\i}sica e Ci{\^e}ncias do Espaço, Universidade do Porto, CAUP, Rua das Estrelas, PT4150-762 Porto, Portugal  
    \and Aix Marseille Universit\'e, CNRS, CNES, LAM (Laboratoire d'Astrophysique de Marseille) UMR 7326, 13388, Marseille, France 
    \and Department of Astronomy, University of Wisconsin, 475 N. Charter Street, Madison, WI 53706, USA
    \and Centre for Astrophysics and Supercomputing, Swinburne University of Technology, Melbourne, VIC 3122, Australia
    \and Leibniz-Institut f{\"u}r Astrophysik Potsdam (AIP), An der Sternwarte 16, 14482 Potsdam, Germany
    \and Institut f{\"u}r Physik und Astronomie, Universit{\"a}t Potsdam,Karl-Liebknecht-Str. 24/25, D-14476 Golm, Germany
    \and Univ. Grenoble Alpes, CNRS, Grenoble INP, GIPSA-lab, 11 rue des Math\'ematiques, Grenoble Campus BP46, F-38402, Saint Martin D'heres Cedex, France
    \and Laboratoire Lagrange, CNRS, Universit\'e C\^ote d'Azur, Observatoire de la C\^ote d'Azur, CS 34229, 06304, Nice, France 
    \and Department of Astronomy, University of Texas at Austin, 2515 Speedway, Austin, TX 78712, USA
    \and INAF – Osservatorio di Astrofisica e Scienza dello Spazio di Bologna, Via P. Gobetti 93/3, 40129 Bologna, Italy
    \and Observatoire de Gen\`eve, Universit\'e de Gen\`eve, 51 Ch. des Maillettes, CH-1290 Versoix, Switzerland 
    \and Leiden Observatory, Leiden University, P.O. Box 9513, 2300 RA Leiden, The Netherlands 
    \and IRAP, Institut de Recherche en Astrophysique et Plan\'etologie, CNRS,  Universit\'e de Toulouse, 14, avenue Edouard Belin, F-31400 Toulouse, France
    \and European Southern Observatory, Av. Alonso de C\'ordova 3107, 763 0355 Vitacura, Santiago, Chile  
    \and Max Planck Institute for Astronomy, K\"onigstuhl 17, 69117, Heidelberg, Germany 
    \and Institut  f{\"u}r Astrophysik, Universit{\"a}t G{\"o}ttingen, Friedrich-Hund-Platz 1, D-37077 G{\"o}ttingen, Germany
    \and Canada-France-Hawaii Telescope, CNRS, 96743 Kamuela, Hawaii, USA
}

\date{2022-10-15 Accepted}

\abstract {
We present the second data release of the MUSE \textit{Hubble} Ultra-Deep Field surveys, which includes the deepest spectroscopic survey ever performed. The MUSE data, with their 3D content, amazing depth, wide spectral range, and excellent spatial and medium spectral resolution, are rich in information. Their location in the \textit{Hubble} ultra-deep field area, which benefits from an exquisite collection of ancillary panchromatic information, is a major asset. This update of the first release incorporates a new 141-hour adaptive-optics-assisted MUSE eXtremely Deep Field (\mxdf; 1 arcmin diameter field of view) in addition to the reprocessed 10-hour mosaic ($\rm 3 \times 3 \, arcmin^2$) and the single 31-hour deep field ($\rm 1 \times 1 \, arcmin^2$). All three data sets were processed and analyzed homogeneously using advanced data reduction and analysis methods. The $3\sigma$ point-source flux limit of an unresolved emission line reaches $\rm 3.1 \times 10^{-19}$ and $\rm 6.3 \times 10^{-20} \, \ergsline$ at 10- and 141-hour depths, respectively. 
We have securely identified and measured the redshift of 2221 sources, an increase of 41\% compared to the first release. With the exception of eight stars, the collected sample consists of 25 nearby galaxies ($z < 0.25$), 677 \oii\ emitters ($z=0.25-1.5$), 201 galaxies in the MUSE  redshift desert range ($z=1.5-2.8$), and 1308 \laes\ ($z=2.8-6.7$). 
This represents an order of magnitude more redshifts than the collection of all spectroscopic redshifts obtained before MUSE in the \textit{Hubble} ultra-deep field area (i.e., 2221 versus 292). At high redshift ($z > 3$), the difference is even more striking, with a factor of 65 increase (1308 versus 20).
We compared the measured redshifts against three published photometric redshift catalogs and find the photo-z accuracy  to be lower than the constraints provided by photo-z fitting codes.  
Eighty percent of the galaxies in our final catalog have an HST counterpart. These galaxies are on average faint, with a median AB F775W magnitude of 25.7 and 28.7 for the \oii\ and \lya\ emitters, respectively. Fits of their spectral energy distribution show that these galaxies tend to be low-mass star-forming galaxies, with a median stellar mass of $\rm 6.2 \times 10^8 \, M_\odot$ and a median star-formation rate of $\rm 0.4 \, M_\odot yr^{-1}$.
We measured the completeness of our catalog with respect to HST and found that, in the deepest 141-hour  area, 50\% completeness is achieved for an AB magnitude of 27.6 and 28.7 (F775W) at $z=0.8-1.6$ and $z=3.2-4.5$, respectively. 
Twenty percent of our catalog, or 424 galaxies, have no HST counterpart. The vast majority of these new sources are high equivalent-width $z>2.8$ \laes\ that are detected by MUSE thanks to their bright and asymmetric broad \lya\ line.
We release advanced data products, specific software, and a web interface to select and download data sets.
}


\maketitle

\section{Introduction}
\label{sec:intro}

In 2017 we published the first MUSE spectroscopic survey data release \citep[][hereafter DR1]{Bacon2017, Inami2017} in the \textit{Hubble} ultra-deep field \citep{Beckwith2006} area. These two papers were published together with eight other papers, providing a first glimpse into the richness of the data and their scientific impact. In that series of papers we investigated the photometric redshift properties of the sample \citep{Brinchmann2017}, the properties of \ciii\ emitters \citep{Maseda2017}, the spatially resolved stellar kinematics of galaxies at z=0.2--0.8 \citep{Guerou2017}, the faint end of the \lya\ luminosity function \citep[LF;][]{Drake2017}, the properties of Fe\,{\sc ii}$^*$ emission \citep{Finley2017}, the extended \lya\ halos \citep{Leclercq2017}, the evolution of the galaxy merger fraction \citep{Ventou2017}, and the equivalent-width properties of \laes\ \citep{Hashimoto2017}.

These initial studies did not exhaust the scientific content of the data set and were quickly followed by many others: the study of \mgii\ emission and absorption in star-forming galaxies \citep{Feltre2018}, the recovery of systemic redshifts from \lya\ line profiles \citep{Verhamme2018}, the covering fraction of \lya\ emission \citep{Wisotzki2018},  the low-mass end of the star formation sequence \citep{Boogaard2018}, the spatially resolved properties of \lya\ halos  \citep{Leclercq2020}, the evolution of the \lae\ fraction \citep{Kusakabe2020}, the molecular gas properties of high redshift galaxies \citep{Boogaard2019, Boogaard2020, Boogaard2021, Inami2020}, the study of high equivalent-width \laes\  \citep{Maseda2018, Maseda2020}, the study of \heii\ emission line properties \citep{Nanayakkara2019}, the rest-UV properties of \laes\ \citep{Feltre2020}, and the angular momentum of low-mass star-forming galaxies \citep{Bouche2021}.

Shortly after the completion of the MUSE  \textit{Hubble} Ultra-Deep Field (hereafter HUDF) survey, MUSE was successfully coupled to the ground-layer module of the VLT adaptive optics facility \citep[AOF;][]{Leibundgut2017, Paufique2018}. The new system offers improved spatial resolution as well as faster survey times by allowing a wider range of atmospheric conditions (i.e., seeing and airmass) for operations. 

The DR1 was based on two data sets: a $\rm 3 \times 3 \, arcmin^2$ mosaic of nine MUSE fields at a 10-hour depth (hereafter \mosaic) and a single $\rm 1 \times 1 \, arcmin^2$  31-hour depth field (hereafter \udft). The improved depth of \udft\ proved to be essential, especially for the study of the faint end of the \lae\ population and the study of diffuse emission. We then decided to push forward in depth and start a new adaptive-optics-assisted GTO survey in the same area with the goal of reaching a depth of over 100 hours.  This new survey, called the MUSE eXtremly Deep Field (\mxdf), was completed in January 2019. With an achieved depth of 141 hours, it is a key addition to the existing \mosaic\ and \udft\ data sets and is the deepest spectroscopic survey ever performed.

In addition to these new observations, we significantly improved the data-reduction and data-analysis system with respect to DR1 (Sect.~\ref{sec:dr1}). The \mosaic\ and \udft\ data sets were therefore reprocessed with the same tools and methodology used for the \mxdf\ processing. This second data release (DR2), which incorporates these three data sets, aims to provide a comprehensive and homogeneous deep spectroscopic survey in the HUDF iconic field.

The paper is organized as follows: the \mxdf\ observations are reported in Sect.~\ref{sec:obs} and the improved data reduction in Sect.~\ref{sec:dataredmain} and Appendix~\ref{sec:datared}. Section~\ref{sec:dataprop} presents the data properties. The processes of source detection and classification are described in Sect.~\ref{sec:dr2process}, and the resulting catalogs and sample properties are presented in Sect.~\ref{sec:results}. A summary and conclusions are given in Sect.~\ref{sec:conclusions}. Finally, the released advanced data products and software are detailed in Appendix~\ref{sec:products}.

Data reduction and catalog building for deep integral field unit (IFU) data is not yet common. Therefore, we aim to provide sufficient details on the methodology used in all our data processing and data analysis steps, to enable future users to perform similar deep fields. Readers not interested in these technical details can skip Sects.~\ref{sec:dataredmain}, \ref{sec:dataprop}, and \ref{sec:dr2process}.

\section{Observations}
\label{sec:obs}
In this section we report only recent observations of \mxdf\ (GTO Large Program 1101.A-0127, PI R. Bacon). For previous observations related to the \udft\ and \mosaic\ fields, we refer to \cite{Bacon2017}.
The observing campaign started in August 2018 and lasted until January 2019 for a total of 6 runs made during new moon periods. 

\begin{figure}[htbp]
\begin{center}
\includegraphics[width=1\columnwidth]{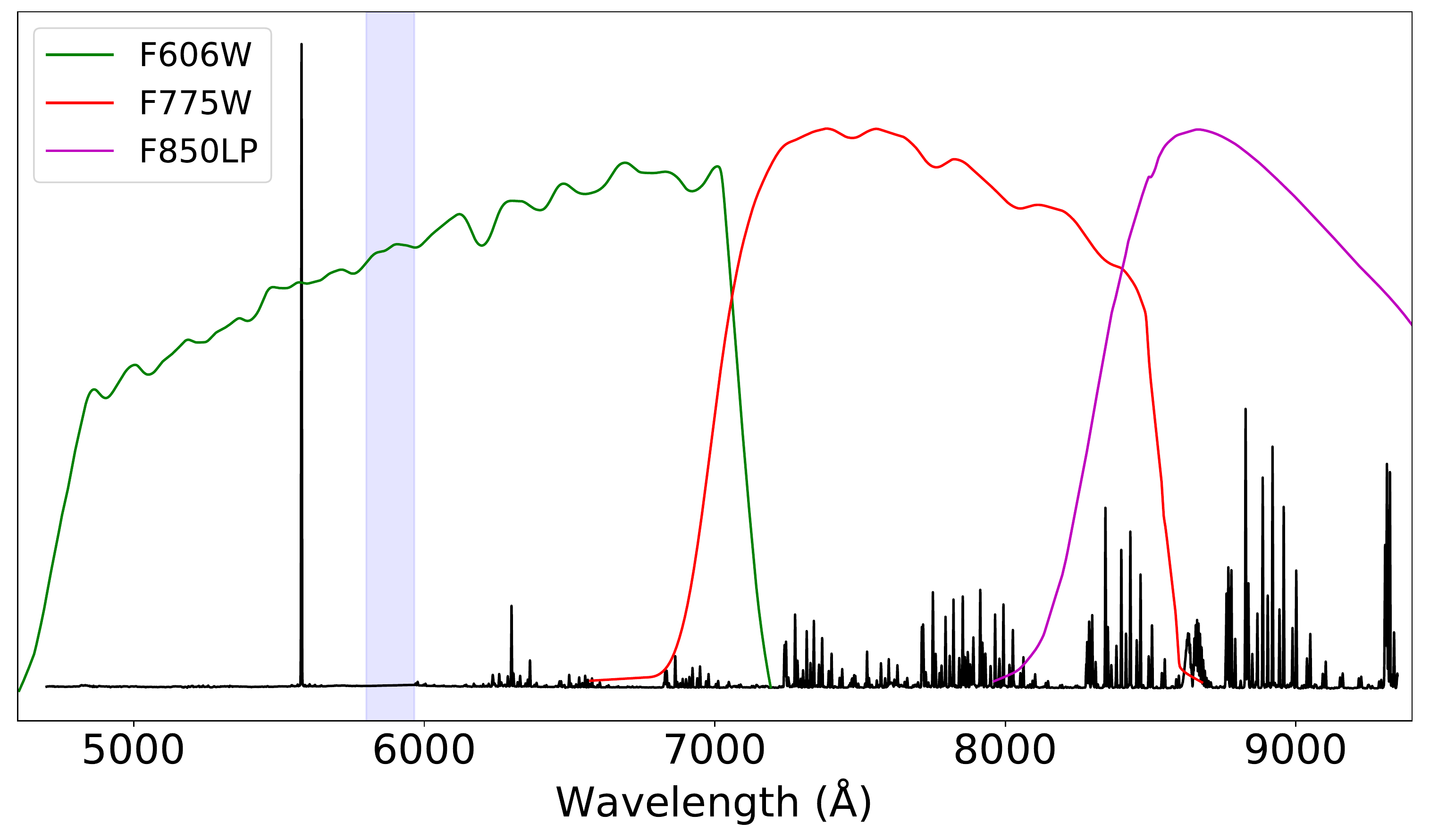}
\caption{MUSE spectral range. The typical sky spectrum (shown in black) is observed in one 25 min observing block. The blue shaded wavelength region shows the location of the sodium notch filter used in MUSE GLAO mode. The response curves of the HST ACS filters F606W, F775W, and F850LP are also indicated.
}
\label{fig:specrange}
\end{center}
\end{figure}

All observations were made with the VLT's AOF and GALACSI, its dedicated ground-layer adaptive optics (GLAO) system (\citealt{Kolb2017, Paufique2018}). With respect to non-AOF observations, the only change in the MUSE instrumental configuration is the notch filter that blocks bright light due to the four sodium laser guide stars in the 5800-5966~\AA\ wavelength range (Fig.~\ref{fig:specrange}). The AOF ran smoothly and achieved robust performance during all runs.

\begin{figure}[htbp]
\begin{center}
\includegraphics[width=1\columnwidth]{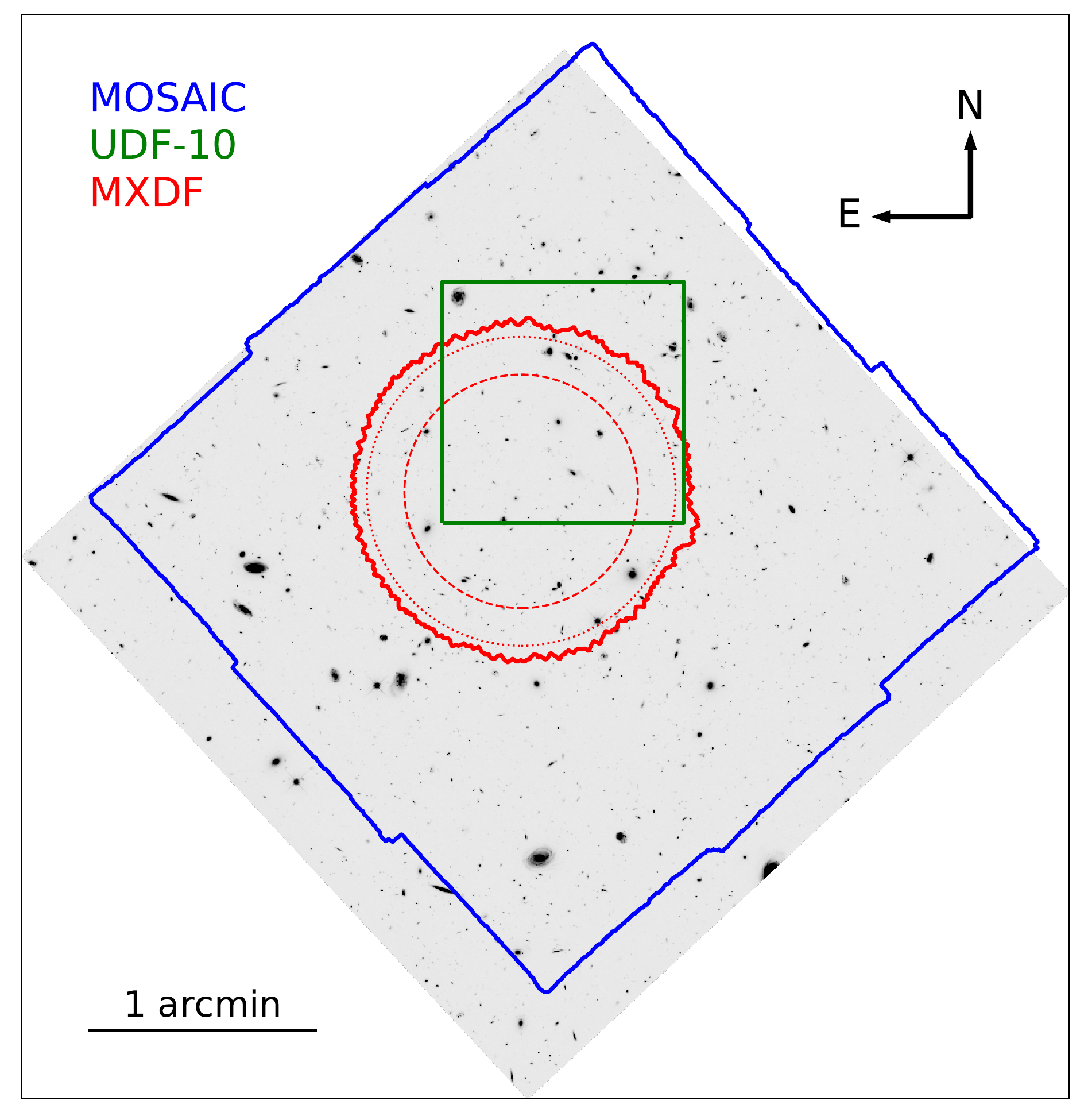}
\caption{Location of the three deep fields used in this paper: \mxdf\ (141-hour depth), \mosaic\ (10-hour depth), and \udft\ (31-hour depth) overlaid on the HST F775W UDF image. The dotted and dashed red circles show the \mxdf\ 10- and 100-hour exposure time isocontours, respectively.
}
\label{fig:fields}
\end{center}
\end{figure}

The location of the \mxdf\ field (Fig.~\ref{fig:fields}) was chosen (i) to be in the HUDF extremely deep field region with WFC3 deep imaging \citep{Illingworth2013} (ii) to match the AOF tip/tilt natural star requirement (i.e., V $>$ 18.5 in a 3.5 arcmin field of view and outside the MUSE field of view), and (iii) to have a usable slow-guiding star in the outer circle to compensate for the derotator's wobble and possible misalignment of the Nasmyth platform with respect to the telescope's focal plane.  

As shown in Table~\ref{tab:obs}, we observed the field over a wide range of atmospheric conditions, including poor seeing (1.2 arcsec) and up to high air mass (1.4). The ground layer fraction, a key parameter for GLAO observations -- the larger the better -- averaged 65\%, and rarely was less than 50\%. We observed some correlation between seeing and ground layer fraction: on many occasions, poor seeing is related to increased ground layer turbulence. This behavior of the Paranal atmospheric turbulence is fortunate because it makes the GLAO mode very effective. 

It is instructive to compare the observing conditions of the \udft\ and \mosaic\ campaigns with the \mxdf\ campaign (Table~\ref{tab:obs}). The first campaign, conducted before the AOF was commissioned, was limited to the natural seeing observing mode of MUSE. To maximize the final spatial resolution, we therefore decided to observe the UDF field only under good seeing conditions (seeing less than 0.8 arcsec) and at low air mass (less than 1.2). The consequence is that it took 2.5 years and 8 runs to accumulate the required telescope time. In contrast, the \mxdf\ campaign took only 6 months and 6 consecutive runs to reach 155 hours of integration time. As we will see in Sect.~\ref{sec:fsf} (Fig.~\ref{fig:iq_comp}), thanks to the performance of the AOF/GALACSI GLAO, the spatial resolution achieved by \mxdf\ is better than that of \udft\ and \mosaic, despite its relaxed atmospheric observing conditions.

\begin{table*}[!htbp]
\caption{Summary of \mxdf, \udft, and \mosaic\ observations.}   
\label{tab:obs}
\centering
\begin{tabular}{cccrrrrrrrrrrrr}
\hline 
Field & Runs\tablefootmark{a} & \multicolumn{1}{c}{Integ\tablefootmark{b}} & \multicolumn{3}{c}{Airmass\tablefootmark{c}} & \multicolumn{3}{c}{Seeing\tablefootmark{c,d} (\arcsec)} & \multicolumn{3}{c}{GL frac\tablefootmark{c,e}} \\ 
 & & & q50 & q10 & q90 & q50 & q10 & q90 & q50 & q10 & q90  \\
\hline 
\mxdf         & 08/18-01/19 (6) & 155 & 1.08 & 1.01 & 1.37 & 0.76 & 0.54 & 1.11 & 0.65 & 0.46 & 0.83 \\ 
\hline
\begin{tabular}{l} \udft\ \& \\ \mosaic \end{tabular} & 09/14-02/16 (8) & 116 & 1.05 & 1.00 & 1.19 & 0.75 & 0.60 & 0.86 \\ 
\hline 
\end{tabular} 
\tablefoot{
\tablefoottext{a}{Starting and end dates. The number of conducted runs is indicated in parentheses.}
\tablefoottext{b}{Total open shutter time in hours.}
\tablefoottext{c}{50\% (q50, median), 10\% (q10) and 90\% (q90) percentile.}
\tablefoottext{d}{Seeing is measured in the pointing direction with the AO telemetry for the \mxdf\ observations and with the telescope guiding camera for the \mosaic\ and \udft\ non-AO observations.}
\tablefoottext{e}{Fraction of 1 km ground-layer turbulence.}
}
\end{table*}

The \mxdf\ observing strategy also differs from the 90-degree rotation plus small offset dithering patterns commonly used in MUSE observations. Each observing block is similar to the scheme used in the first campaign: that is, a set of four 25 min exposures with successive 90\degree\ instrument rotations. But in order to further reduce the systematics and to break the horizontal and vertical patterns introduced by the instrument field splitter and slicer geometry, we systematically rotated the field of view by a few degrees between each observing block. Consequently, the final combined field of view is approximately circular (Fig.~\ref{fig:fields}) with a radius of 41\arcsec\ and 31\arcsec\ for respectively 10+ and 100+ hours of depth. The field center celestial coordinates are 53.16467 deg., -27.78537 deg. (J2000 FK5).
After the rejection of a few bad exposures due to satellite track contamination, poor final spatial resolution, or cloud absorption, the achieved final maximum depth was 141 hours. 

\section{Data reduction}
\label{sec:dataredmain}

\begin{figure*}[htbp]
\begin{center}
\includegraphics[width=1\textwidth]{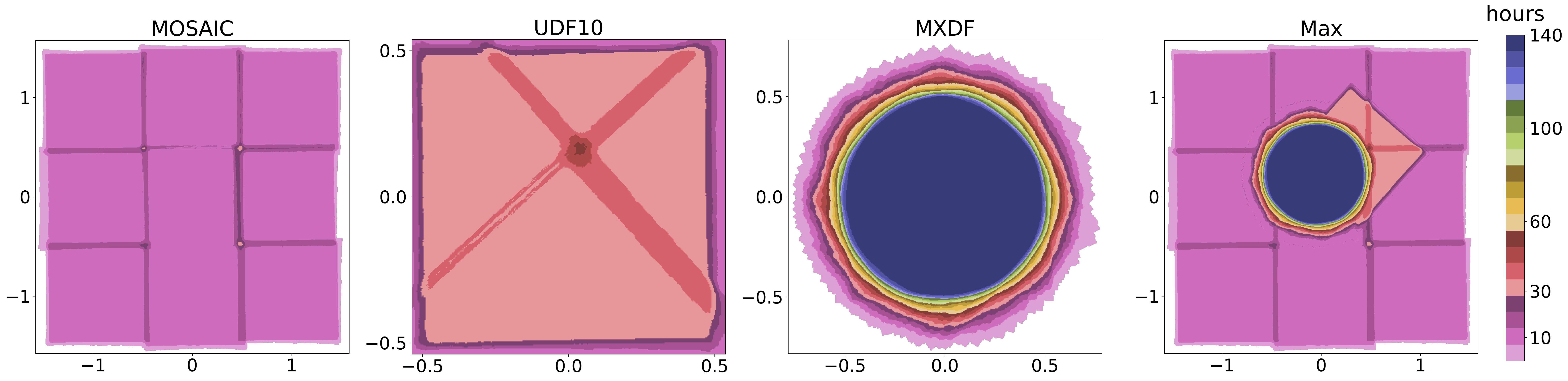}
\caption{Achieved depth in hours. The \mosaic, \udft, and \mxdf\ exposure maps are shown in the first three panels, starting from the left. The \textsf{Max} exposure map, computed as the maximum depth for each spaxel, is shown in the right panel.
\udft\ and \mxdf\ are to the north (top) and the east (left), respectively, while the \mosaic\ and \textsf{Max} exposure maps are rotated by 42\degree (see Fig.~\ref{fig:fields}). Axis labels are in arcmin.
}
\label{fig:depth}
\end{center}
\end{figure*}

After DR1, described in \cite{Bacon2017}, we continued to work on the data reduction process, which led to several major improvements. The overall process is similar to DR1, with important changes  in the self-calibration algorithm, in the sky-subtraction with the Zurich Atmospheric Purge (ZAP) software (\citealt{Soto2016}, Appendix~\ref{sec:zap}), and the use of a ``{superflat}.'' This process was first applied to the \udft\ and \mosaic\ (278 exposures) and then to the \mxdf\ (373 exposures). The resulting depth and color images for the three data sets are shown in Fig.~\ref{fig:depth} and Fig.~\ref{fig:rgb}, respectively.
A detailed description of this improved data reduction process is given in Appendix~\ref{sec:datared}.

\begin{figure*}[htbp]
\begin{center}
\includegraphics[width=0.9\textwidth]{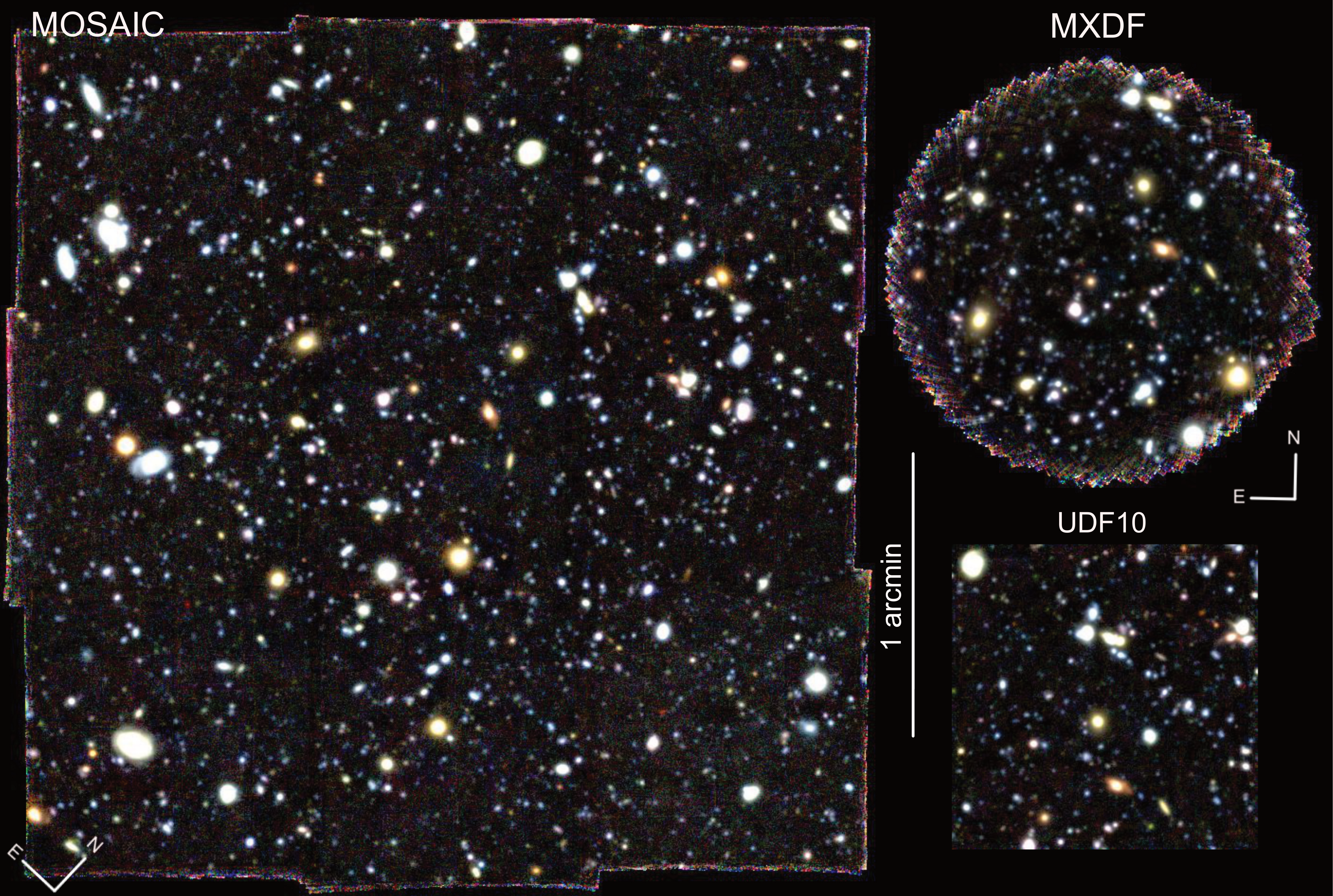}
\caption{Reconstructed pseudo-color images of the \mosaic, \mxdf, and \udft\ data sets. 
}
\label{fig:rgb}
\end{center}
\end{figure*}

\section{Data properties}
\label{sec:dataprop}

\subsection{Astrometry and photometry}
\label{sect:astro}
 
As in DR1, the datacube's world coordinate system has been matched to the \hubble\ ACS astrometry (Appendix~\ref{sec:datared}). By construction, the astrometry of the datacube should therefore be consistent with the \hudf\ published photometric catalogs: \uvudf\ \citep{Rafelski2015}, \tdhst\ \citep{Skelton2014}, and \candels\ \citep{Whitaker2019}. We note that \cite{Dunlop2017} and \cite{Franco2018} have shown that the \textit{Hubble} Space Telescope (HST) astrometry is offset from the \textsf{Gaia} DR2 catalogs \citep{Gaia2016, Gaia2018}. The offset, as measured by \cite{Whitaker2019}, amounts to $\Delta \text{Ra} = +0.094 \pm 0.042$ arcsec and $\Delta \text{Dec} = -0.26 \pm 0.10$ arcsec. In this paper, we have ignored this offset in order to maintain the same astrometric reference as the HST catalogs.

We measured the astrometric accuracy achieved by comparing source centroids in the HST ACS F775W high resolution image with the corresponding values measured in MUSE reconstructed image. We proceeded as follows: For each field, we derived the equivalent F775W broadband image by computing at each spaxel the transmission-weighted average flux of the corresponding F775W ACS filter  (Fig.~\ref{fig:specrange}). This reconstructed image is then convolved by the HST ACS point spread function (PSF). We approximated this PSF using a Moffat function\footnote{The circular 2D Moffat profile \citep{Moffat1969} has two parameters: the FWHM in arcsec and the $\rm \beta$ shape parameter, $\rm \beta = 1$ and $\rm \beta \gg 1$ for respectively a Lorentzian or a Gaussian distribution.}
 with a 0.085 arcsec full width at half maximum (FWHM) and $\beta = 1.60$. The HST ACS F775W image is convolved with the MUSE PSF value at the F775W reference wavelength (7750~\AA), that is, 0.45 arcsec FWHM and $\beta = 1.89$,  0.60 arcsec FWHM and $\beta = 2.80,$ and 0.63 arcsec FWHM and $\beta = 2.80$ for the \mxdf, \udft,\ and \mosaic\ data sets, respectively.
Source detection is performed on both convolved images with \sex\ \citep{Bertin1996}. We tune  \sex's detection parameters to obtain similar segmentation maps between HST and MUSE convolved images. The two catalogs are crossmatched and the mean offset and its standard deviation are derived using $3 \sigma$ sigma-clip statistics on the matched catalog. The astrometric differences between the MUSE datacubes and HST ACS F775W image are given in the upper part of Table~\ref{tab:astro_photo}. If we exclude the faint sources with $\text{AB} > 27$, the mean astrometric relative error is about 0.1 arcsec, which is half of a spaxel or about 1/6 of the spatial resolution. We note that astrometric errors are larger for \mxdf\ than for the \mosaic\ data set. Such an increase with depth is expected given the increase of source confusion.

We used the same crossmatched catalogs to evaluate the achieved accuracy in source AB magnitude for MUSE data sets with respect to HST photometry. The results, reported  in the lower part of Table~\ref{tab:astro_photo}, are based on \sex\ computed automatic aperture magnitude\footnote{\sex’s automatic aperture photometry routine is derived from Kron’s algorithm \citep{Kron1980}; see https://sextractor.readthedocs.io/en/latest/Photom.html for more detailed information.}. We note that the magnitude difference is below 0.1 for \mxdf\ sources with $\text{AB} < 27$.  As expected, the magnitude differences decrease with the source brightness and increase with depth.

\begin{table}
\caption{Estimated astrometric and photometric accuracy for the three data sets when compared to the HST ACS F775W image. Mean and standard deviation are given for three AB magnitude intervals.}   
\label{tab:astro_photo}
\centering
\begin{tabular}{cccc}
data set & $\text{AB} < 25$  & $25 < \text{AB} < 27$ & $27 < \text{AB}$\\
\hline
 & \multicolumn{3}{c}{Average astrometric offsets (arcsec)} \\
\hline
MXDF & $0.12 \pm 0.04$ & $0.13 \pm 0.05$ & $0.13 \pm 0.06$ \\
UDF10 & $0.08 \pm 0.02$ & $0.10 \pm 0.05$ & $0.15 \pm 0.09$ \\
MOSAIC & $0.05 \pm 0.02$ & $0.08 \pm 0.05$ & $0.16 \pm 0.10$ \\
 & \multicolumn{3}{c}{Average AB magnitude differences} \\
\hline
MXDF & $0.08 \pm 0.04$ & $0.09 \pm 0.07$ & $0.25 \pm 0.19$ \\
UDF10 & $0.07 \pm 0.04$ & $0.23 \pm 0.13$ & $0.35 \pm 0.24$ \\
MOSAIC & $0.10 \pm 0.08$ & $0.24 \pm 0.16$ & $0.32 \pm 0.23$ \\
\end{tabular}
\tablefoot{The mean AB magnitude error reported by \textsf{SExtractor} for the HST convolved image amounts to 0.002, 0.011 and 0.030 in the $\text{AB} < 25$, $25 < \text{AB} < 27$ and $\text{AB} > 27$ intervals, respectively.}
\end{table}

\subsection{Spatial and spectral resolution}
\subsubsection{\mxdf\ spatial PSF}
\label{sec:fsf}
Spatial resolution is an important ingredient for most analyses. The spatial PSF was first estimated for each individual raw exposure to reject bad exposures before the final datacube was produced. Then, the final PSF was estimated on the combined datacube. 

Unfortunately, there is no point source bright enough in the \mxdf\ field of view to estimate the PSF in a single exposure of 25 min integration. Moreover, the seeing value reported by the Paranal seeing monitor is not sufficient by itself to estimate the PSF given the adaptive optics (AO) correction provided by the AOF GLAO system. We therefore used the tool \psfrec\ developed by \cite{Fusco2020} to derive an estimate of the PSF using AO telemetry. During an exposure, atmospheric parameters such as seeing, ground-layer turbulence fraction, turbulence outer scale, are derived from wavefront sensor telemetry and regularly recorded in a table in the raw FITS exposure. The algorithm uses this information to produce an estimate of the spatial PSF in the form of a circular 2D Moffat profile. %\citep{Moffat1969} with two parameters: the FWHM in arcsec and the $\rm \beta$ shape parameter\footnote{$\rm \beta = 1$ and $\rm \beta \gg 1$ for respectively a Lorentzian or a Gaussian distribution.}
\cite{Fusco2020} measure a standard deviation of 0.06 arcsec for the derived FWHM. We note that some systematic offsets were found at high spatial resolution \citep[Fig.~17 of][]{Fusco2020}. We also found a similar offset when we compared the derived value of \psfrec\ with a direct fit to the combined datacube (Table~\ref{tab:fsfcomp}).

\begin{figure}[htbp]
\begin{center}
\includegraphics[width=0.8\columnwidth]{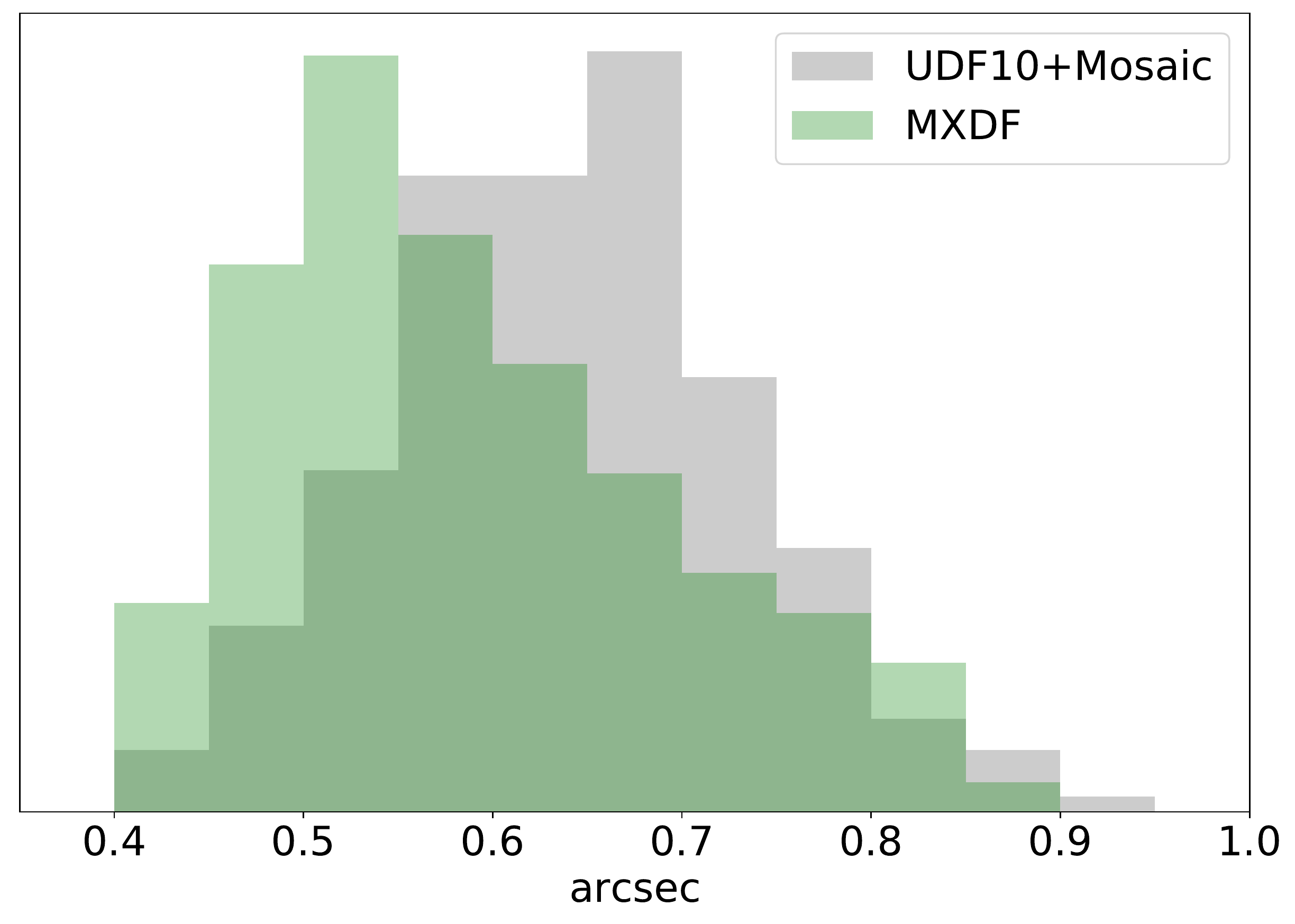}
\caption{Spatial resolution of individual \mxdf\  observations compared to the \udft\ plus \mosaic\ observations. The FWHMs (in arcsec) are derived from the Moffat PSF model at 7000 \AA. The normalized FWHM histograms of the \mxdf\ and \udft\ plus \mosaic\ observations are shown in green and gray, respectively. The FWHM \mxdf\ values have been corrected for a $\rm +0.06$ arcsec offset (see text).} %0.64 0.57 median values
\label{fig:iq_comp}
\end{center}
\end{figure}

\begin{figure}[htbp]
\begin{center}
\includegraphics[width=0.8\columnwidth]{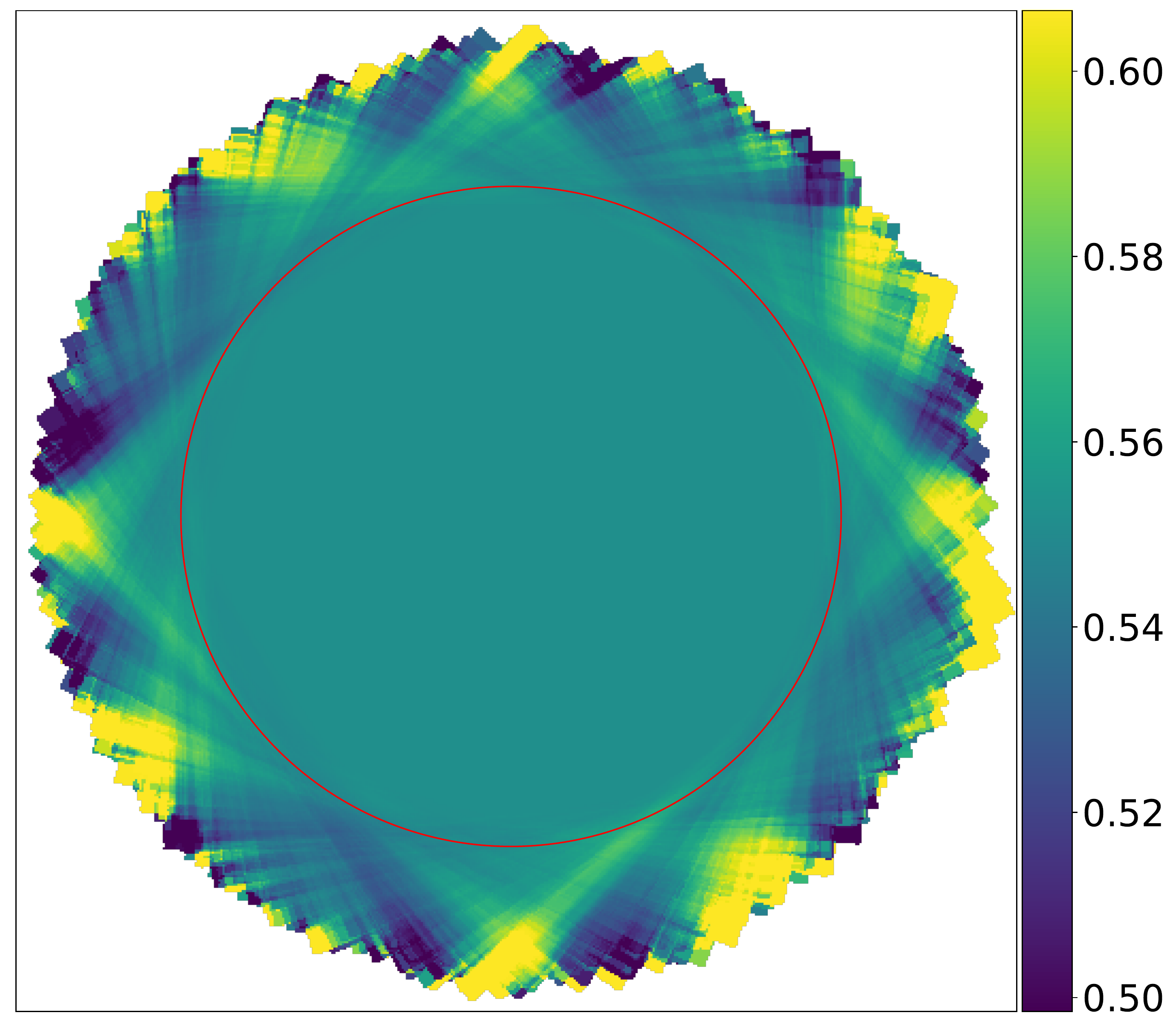}
\caption{Field evolution of the \mxdf\ spatial resolution. The colors indicate the mean value (in arcsec) of all individual Moffat FWHMs at 7000 \AA\ at each spaxel location as measured by the PSF AO reconstruction algorithm and corrected for an offset of $\rm +0.060$ arcsec. The red circle displays the 100-hour depth contour.} 
\label{fig:fsfvar}
\end{center}
\end{figure}

We show in Fig.~\ref{fig:iq_comp} the distribution of the computed \mxdf\ FWHMs\footnote{A value of 0.06 arcsec was added to the values of \mxdf\ \psfrec\ to account for the measured offset between \psfrec\ and the direct Moffat fit.}  at 7000 \AA\ for the 373 individual exposures. The benefit of the GLAO AOF system is evident when the histogram of \mxdf\ FWHMs is compared to the corresponding \mosaic\ plus \udft\ values. Due to the specific rotation scheme used for the \mxdf\ observations, the outer ring of the field results from the combination of different exposures and may thus  have a potentially different PSF than the central part. We show in Fig.~\ref{fig:fsfvar} that this is indeed the case with a variation of 0.10 arcsec FWHM. We note, however, that within the area with depth $>100$ hours there is no detectable spatial variation in the PSF.

\begin{figure}[htbp]
\begin{center}
\includegraphics[width=1\columnwidth]{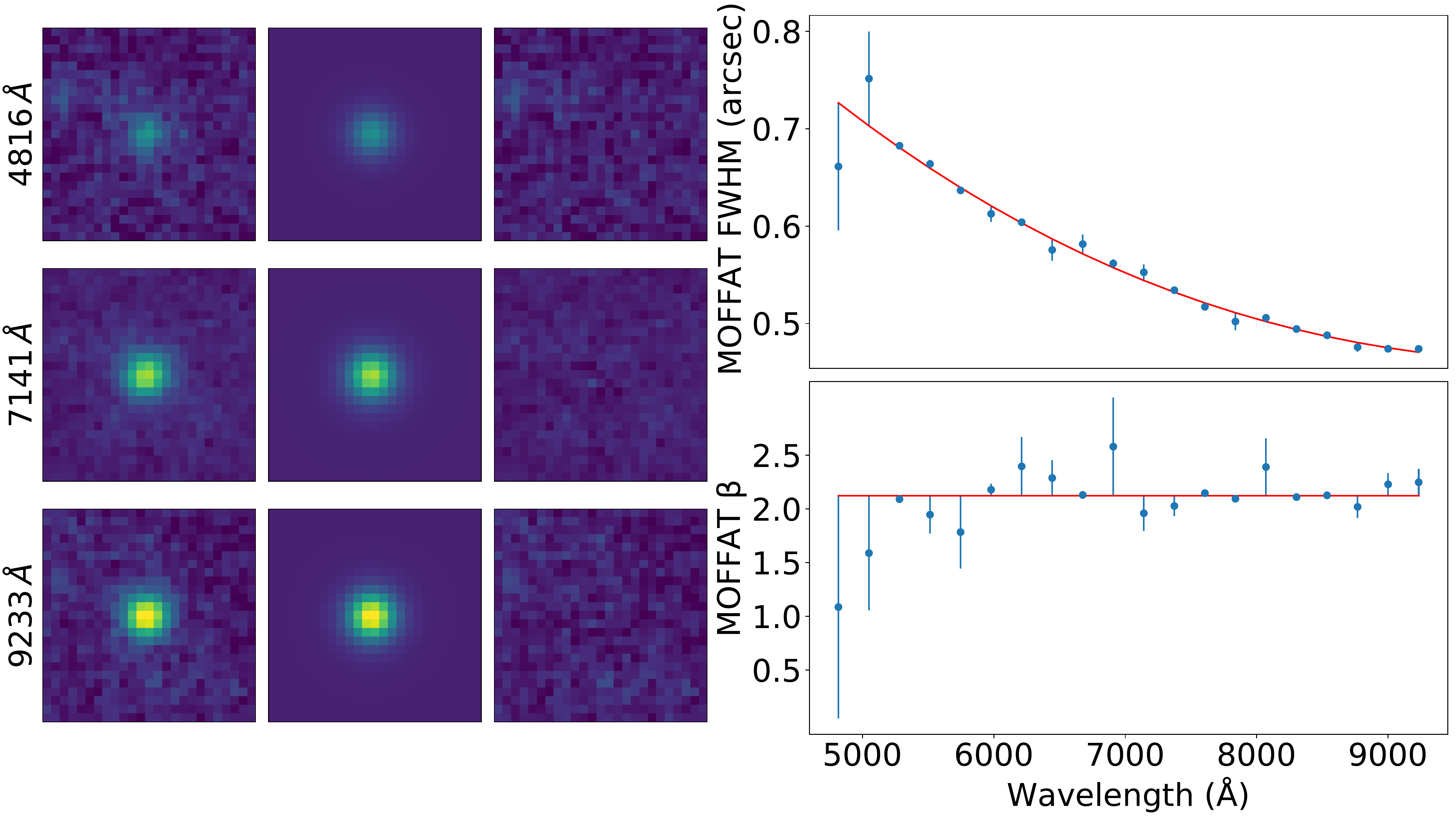}
\caption{\mxdf\ Moffat fit of the M star RID-5102. The fit is performed on 20 narrowband images of 232 \AA\ width, equally distributed along the \mxdf\ cube wavelength axis. Examples of the first, central, and last images (each box side is 5 arcsec in length) are displayed in the first column. The corresponding Moffat fit and residuals are shown in the two other columns. The FWHM and $\beta$ Moffat fitted values are shown in the upper and bottom right panels, together with their polynomial approximation.
}
\label{fig:fitstar}
\end{center}
\end{figure}

Although there is no bright star in the field of view, there is a fainter M star (ID 5102 in R15 catalog, F775W AB magnitude 24.7) located in the 100-hour depth area that can be used to measure the PSF. To take into account the evolution of the PSF with wavelength, we divided the final combined datacube into 20 wavelength slices of 232 \AA\ each and fit a circular Moffat function to each image. As shown in Fig.~\ref{fig:fitstar} (left panels), the circular Moffat function represents the data well. The figure also shows the evolution of the shape parameters FWHM and $\rm \beta$ as a function of wavelength. Except at the blue end of the MUSE wavelength range, where the M star becomes too faint, the shape parameter $\rm \beta$ is approximately constant. We adopt $\rm \beta = 2.123$ and perform the Moffat fit again, leaving the FWHM as a variable. The adopted third-order polynomial fitted to the evolution of the FWHM as a function of wavelength is shown in the upper right panel of Fig.~\ref{fig:fitstar}.
The final spatial PSF is then given by
\begin{align}
\label{eq:fsf}
\rm PSF(r, \lambda) &= (\beta - 1) / (\pi \alpha(\lambda)^2) \times (1 + r^2/\alpha^2 )^{-\beta} \\
\rm \alpha(\lambda) &= FWHM(\lambda) \times \left[ 2 \sqrt{2^{1/\beta}-1} \right] ^{-1} \nonumber \\
\rm FWHM(\lambda) &= a_3 \lambda'^3 + a_2 \lambda'^2  + a_1 \lambda' + a_0 \nonumber \\
\rm \lambda' &= ( \lambda - 4850 ) / (9350 - 4850) - 0.5. \nonumber
\end{align}
The wavelength ($\lambda$) and FWHM units are \AA\ and arcsec, respectively. The $\beta$ value and the FWHM polynomial coefficients ($a_n$) are given in Table~\ref{tab:fsfval}.

We compared this PSF estimate with the values derived by convolution of the HST images for the F606W, F775W, F814W, and F850LP broadband filters. The method is identical to that used in the paper I \citep[][Sect.~5.1]{Bacon2017} for the  \mosaic\ and \udft\ datacubes. For ease of comparison, a direct Moffat fit was performed on the M star for the four reconstructed HST broadband MUSE images. An additional comparison was performed by fitting a Moffat distribution to the average of all individual \psfrec\ PSFs. The Moffat FWHM and $\beta$ shape parameter for the three different estimates are shown in the Table~\ref{tab:fsfcomp}. Since the FWHM and $\beta$ parameters are not independent, we also provide the FWHM obtained by fitting a Gaussian profile to judge the difference in FWHM. However, we recall that the MUSE PSF is not well represented by a Gaussian function \citep{Husser2016, Weilbacher2020}.
As shown in Table~\ref{tab:fsfcomp}, all methods give similar results with the expected decrease in FWHM with increasing wavelengths. However, there is a systematic offset between the methods: the HST convolution method overestimates the FWHM by $\approx$+0.05 arcsec while the \psfrec\ method underestimates it by $\approx$-0.06 arcsec.

\begin{table*} 
\caption{Circular Moffat PSF models for the three MUSE data sets.}   
\label{tab:fsfval}
\begin{tabular}{ccccccccc}
\hline 
data set & $\beta$ & $a_0$ & $a_1$ & $a_2$ & $a_3$ & $F_{4850}$ & $F_{7000}$ & $F_{9350}$ \\ 
\hline 
\mxdf & 2.123 & 0.5465 & -0.2474 & 0.1975 & -0.0295 & 0.72 & 0.55 & 0.47 \\ 
\udft & 2.800 & 0.6179 & -0.1353 &  &  & 0.69 & 0.62 & 0.55 \\  
\mosaic & 2.800 & 0.6378 & -0.1515 &  &  & 0.71 & 0.64 & 0.56 \\ 
\hline 
\end{tabular} 
\tablefoot{ 
$\beta$ is the Moffat shape parameter and $a_n$ are the coefficients of the polynomial approximation to FWHM($\lambda$) defined in Eq.~\ref{eq:fsf}.
In the case of the \mosaic\ data set, we provide the median PSF for the nine fields. FWHM values in arcsec (e.g., $F_{4850}$) are given for the blue (4850 \AA), central (7000 \AA) and red (9350 \AA) wavelengths.}
\end{table*}

\begin{table*} 
\caption{Comparison of different spatial \mxdf\ PSF estimates for four HST reconstructed broadband images.}   
\label{tab:fsfcomp}
%\centering
\begin{tabular}{cccccccccc}
\hline 
Filter & \multicolumn{3}{c}{Moffat FWHM} & \multicolumn{3}{c}{Moffat $\rm \beta$} & \multicolumn{3}{c}{Eq. Gaussian FWHM}\\ 
& Star & HST & AO tel & Star & HST & AO tel & Star & HST & AO tel \\
\hline 
F606W & 0.60 & 0.64 & 0.54 & 2.10 & 2.41 & 2.01 & 0.72 & 0.75 & 0.65 \\
F775W & 0.52 & 0.56 & 0.45 & 2.17 & 2.25 & 1.88 & 0.61 & 0.66 & 0.55 \\
F814W & 0.50 & 0.55 & 0.43 & 2.19 & 2.35 & 1.84 & 0.59 & 0.64 & 0.53 \\
F850LP & 0.48 & 0.51 & 0.41 & 2.25 & 2.25 & 1.80 & 0.57 & 0.60 & 0.50 \\
\end{tabular}
\tablefoot{ 
FWHM (arcsec) and $\rm \beta$ values are given for the direct Moffat fit of the RID-5102 M star (Star columns), the comparison with HST (HST columns) and the mean of the PSF reconstruction values (AO tel columns). The Gaussian equivalent FWHMs are given in the last three columns.}
\end{table*}

\subsubsection{Line spread function}
\label{sec:lsf}

\begin{figure}[!tbp]
\begin{center}
\includegraphics[width=1\columnwidth]{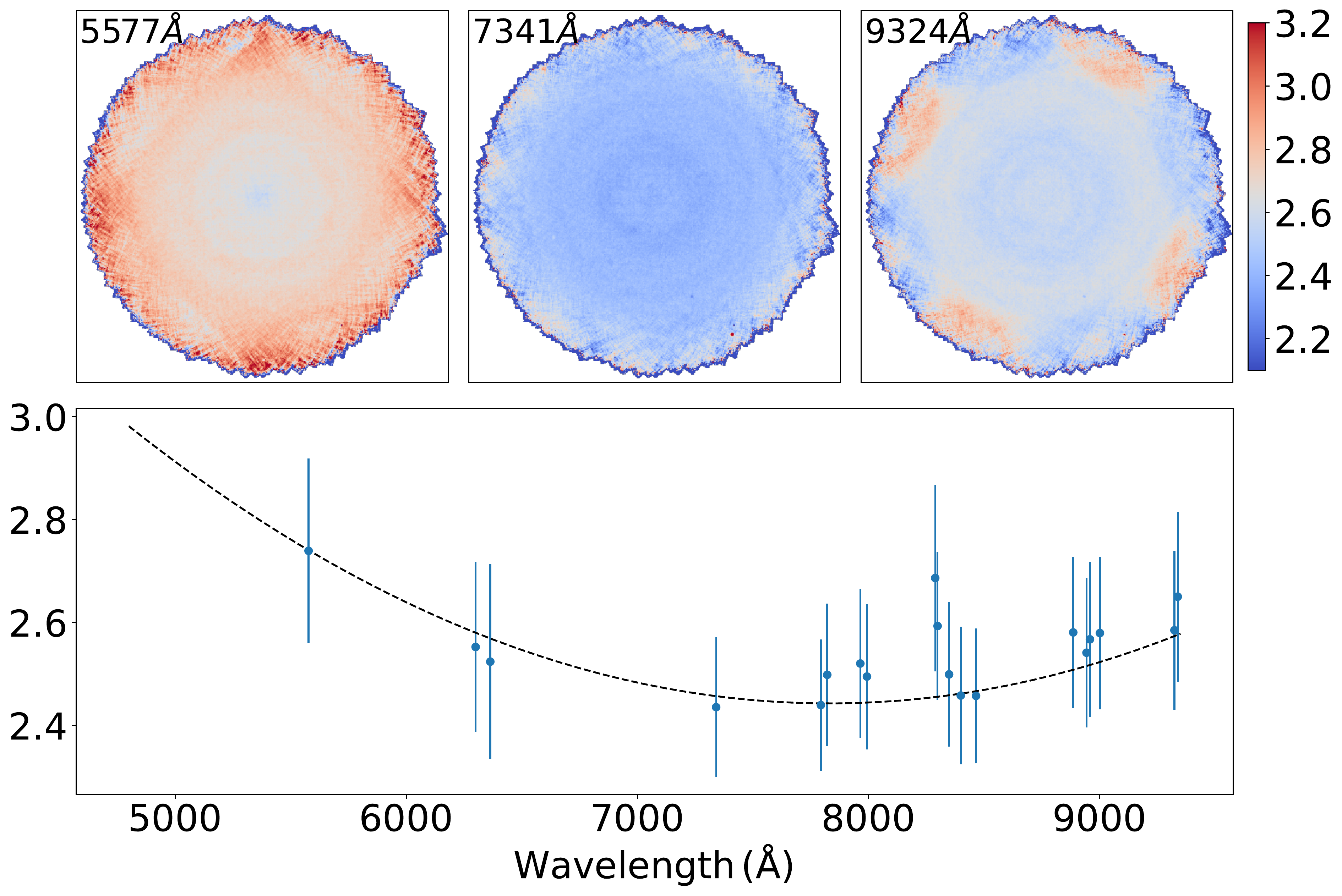}
\caption{Measurements of the \mxdf\ LSF in the combined datacube (not sky subtracted). The bottom panel displays the fitted Gaussian FWHM for a few bright sky lines. Variations with the field of view are given by the error bars (standard deviation). The DR1 \udft\ polynomial approximation is shown as a dashed line. Examples of FWHM spatial maps are given for three sky lines in the top panels. FWHM units is \AA.
}
\label{fig:lsf}
\end{center}
\end{figure}

As in DR1, we measure the line spread function (LSF) on a combined cube of data where the sky has not been subtracted. A Gaussian fit was performed on the bright sky lines taking into account the contribution of adjacent fainter lines. As shown in Fig.~\ref{fig:lsf}, the \mxdf\ LSF shows the expected instrumental evolution with wavelength. A good model of the median LSF FWHM is given by the \udft\  model: 
$\rm LSF(\lambda) = 5.866 \times 10^{-8} \lambda^2 - 9.187 \times 10^{-4} \lambda + 6.040$, with LSF and $\lambda$ in \AA.
The LSF is approximately constant in the field of view, except in the outer rings, which show larger variation. This is mainly due to the limited number of exposures that are combined together at the edges, which leaves the intrinsic variation between the 24 IFUs more visible.

\subsection{Noise property and limiting flux}
\label{sec:noise}

\begin{figure}[!tbp]
\begin{center}
\includegraphics[width=1\columnwidth]{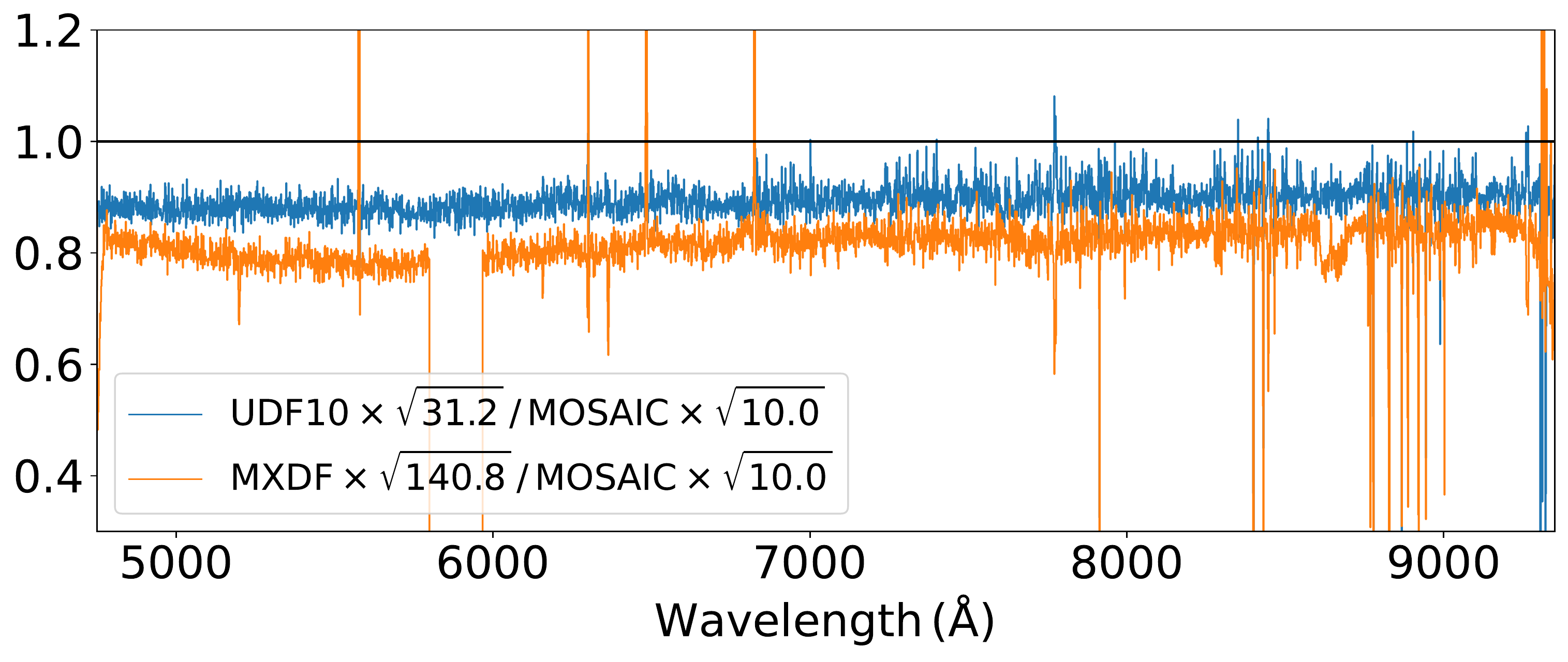}
\caption{Scaled raw noise comparison between the three data sets. The plot displays the  \udft\ and \mxdf\ to \mosaic\ scaled noise ratio. The scaled noise is computed as  
the median noise level for each data set in a common set of 1\arcsec\ empty apertures, after scaling by the square root of the depth in hours. We note that the noise level has not been corrected for correlations caused by the interpolation. 
}
\label{fig:noise_comp}
\end{center}
\end{figure}

To compare the noise level of the three data sets, we selected 214 one-arcsecond-diameter apertures located in empty locations and common to all data sets.  The location of the aperture was chosen to have a similar depth for each data set: $140.8 \pm 0.06$ hours, $31.2 \pm 0.2$ hours, and $10.0 \pm 0.3$ hours for the \mxdf, \udft, and \mosaic\ data sets, respectively. The corresponding noise level as a function of wavelength is calculated as the median of the 214 standard deviations. It should scale with exposure time ($t$) as $\rm \sqrt t$. In Fig.~\ref{fig:noise_comp} we display the \udft\ and \mxdf\ noise level scaled by the square root of the depth and divided by the \mosaic\ scaled value. The figure shows that the deviation from the theoretical expectation is indeed very small (i.e., 10-20\%). The slight improvement with depth results from a better control of the systematics due to the observation strategy (dithering and rotation; see Sect.~\ref{sec:obs}).

We estimated the 1$\sigma$ surface brightness limit for an unresolved lines ($\textit{SB}_{line}$) with the following formula:
\begin{equation}
\label{eq:sb}
\textit{SB}_{line}(\lambda) = \frac{\sqrt{V_{line}(\lambda)} \, \Delta \lambda}{\sqrt{25} \,\Delta s^2} \\ \mathrm{with} \\
V_{line}(\lambda) = \sum_{\lambda' = \lambda_0-k \sigma(\lambda)}^{\lambda_0+k \sigma(\lambda)} V_s(\lambda') 
,\end{equation}
where $V_{s}(\lambda)$ is the variance by spaxel, $\Delta \lambda$ the wavelength bin size (1.25 \AA), $\Delta s$ the spaxel bin size (0.2 arcsec), 25 the number of spaxels in 1 arcsec$^2$ and $\sigma(\lambda)$ is derived from the LSF approximation given in Sect.~\ref{sec:lsf} with $\sigma(\lambda) = \text{LSF}(\lambda)/2.355$. The wavelength interval ($k = 1.29$) was chosen to capture 90\% of the line flux.
The variance by spaxel was computed on the central \mxdf\ area with a depth larger than 120 hours after masking the bright sources and taking the median of the datacube variance. We note that the variance is already corrected for the correlated noise as described in Appendix~\ref{sec:postdrs}.

\begin{figure}[!tbp]
\begin{center}
\includegraphics[width=1\columnwidth]{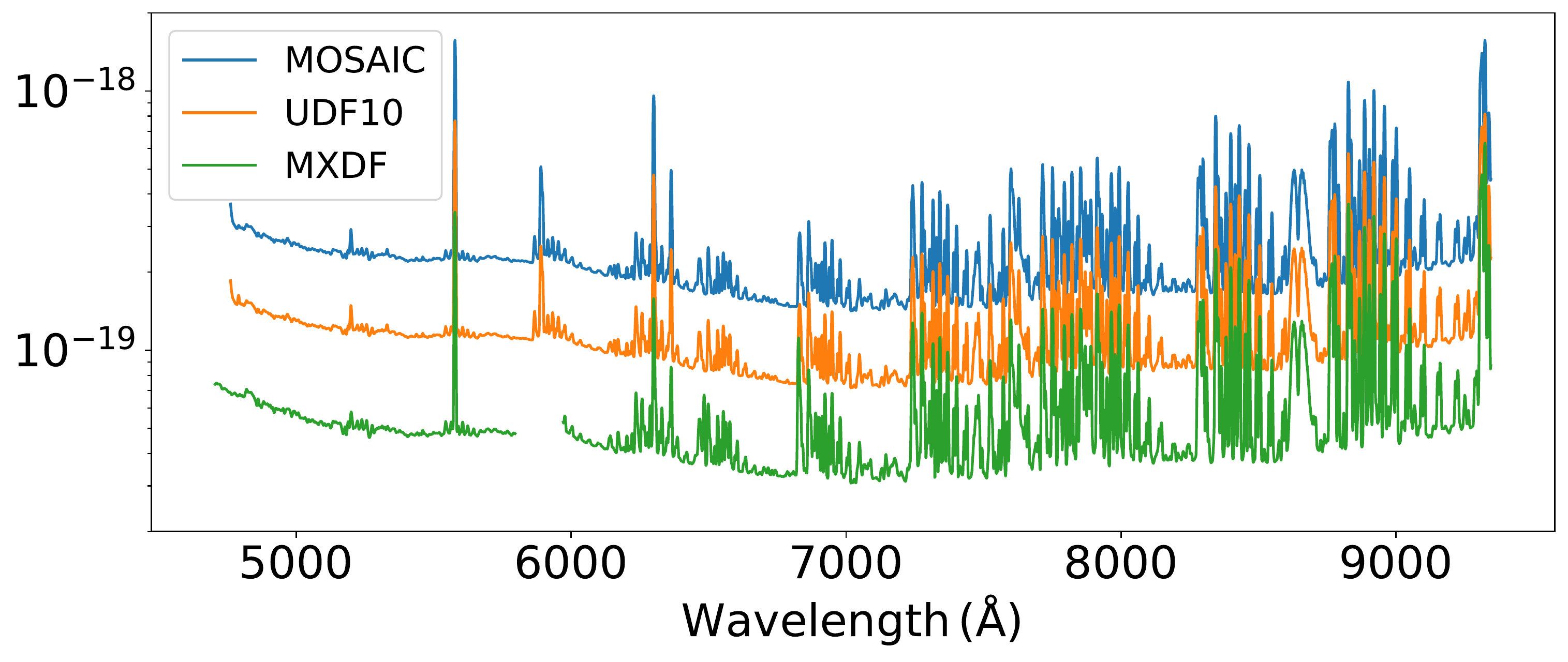}
\caption{Unresolved emission line median 1$\sigma$ surface brightness  limit for the three data sets in units of \ergsurfb.
}
\label{fig:sb_noise}
\end{center}
\end{figure}

The same computation was performed for the \udft\ data set, restricted to the area with depths greater than 30 hours, and the \mosaic\ data set. Results are shown in Fig.~\ref{fig:sb_noise}. The best sensitivity is achieved in the \mxdf\ deep area and in the 6800-7800 \AA\ wavelength range (outside sky lines) where an unresolved emission line surface brightness of \erglinesurf{1.0}{-19} will be detected at 3$\sigma$.

\begin{figure}[!tbp]
\begin{center}
\includegraphics[width=1\columnwidth]{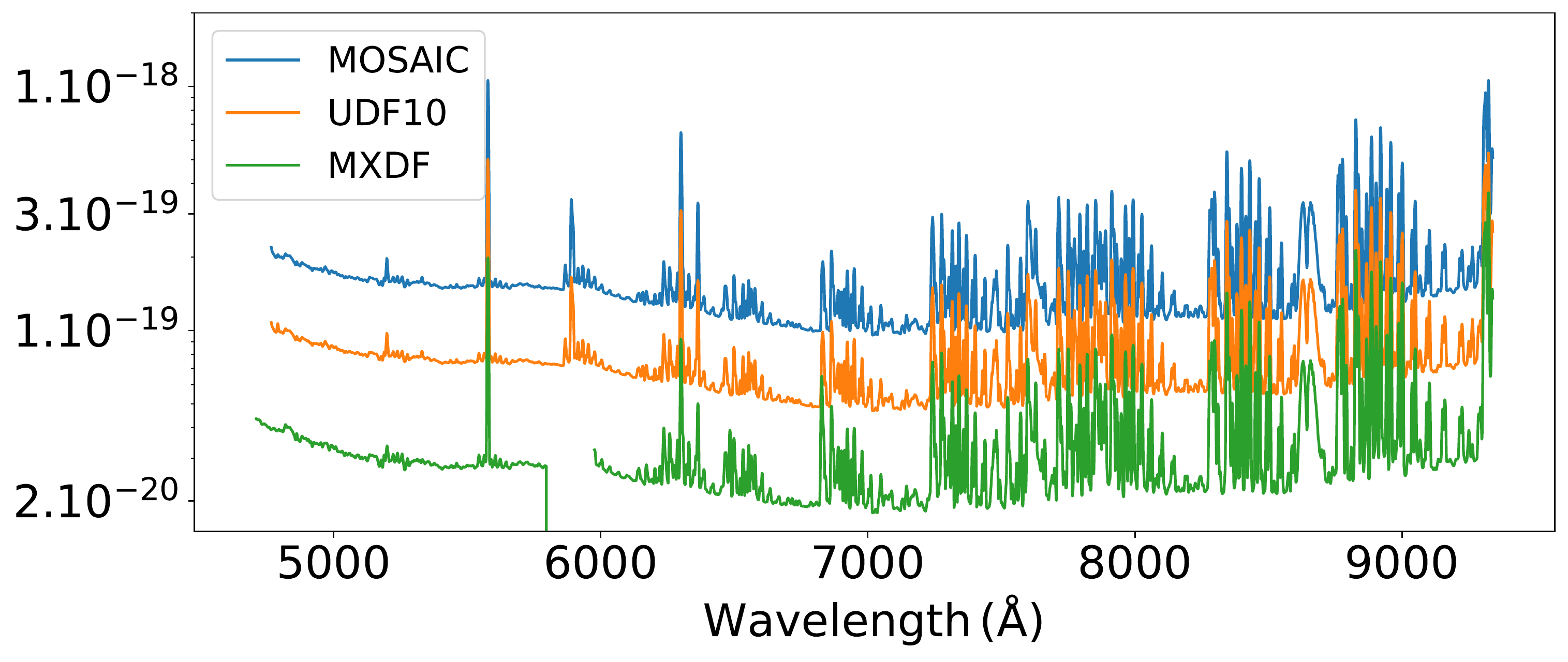}
\caption{Unresolved emission line median 1$\sigma$ point-like source limit for the three data sets in units of \ergsline.
}
\label{fig:pl_noise}
\end{center}
\end{figure}

We perform a similar estimation for a point-like source with an unresolved emission line, using the spatial PSF estimated in Sect.~\ref{sec:fsf} with the following formula: 
\begin{equation}
\label{eq:pl}
F_{pl}(\lambda) =  \Delta \lambda \, \sqrt{ \frac{\pi r^2}{\Delta s^2}  V_{line}(\lambda)}
,\end{equation}
where $V_{line}$ is defined in Eq.~\ref{eq:sb} and $r$ is the 80\% enclosed flux circular radius for the corresponding Moffat PSF model at 7000 \AA.
The best point-source sensitivity (Fig.~\ref{fig:pl_noise}) is achieved in the \mxdf\ central area and in the 6700-9300 \AA\ wavelength range (sky lines excluded) with a 3$\sigma$ detection limit of $6.3 \times 10^{-20} \ergsline$ for an unresolved emission line.

\section{Source detection and classification}
\label{sec:dr2process}
% intro_dr2process.tex

The next stage after the production of the final datacubes for the three data sets (\mosaic, \udft\ and \mxdf) is the  detection and classification of sources. This is a delicate task, involving many steps to ensure the data quality and homogeneity of the final product. An overview, followed by detailed descriptions of each of the step, is presented in the following subsections.

\subsection{Overview}
\label{sec:overview}
\begin{figure}[htbp]
\begin{center}
\includegraphics[width=0.7\columnwidth]{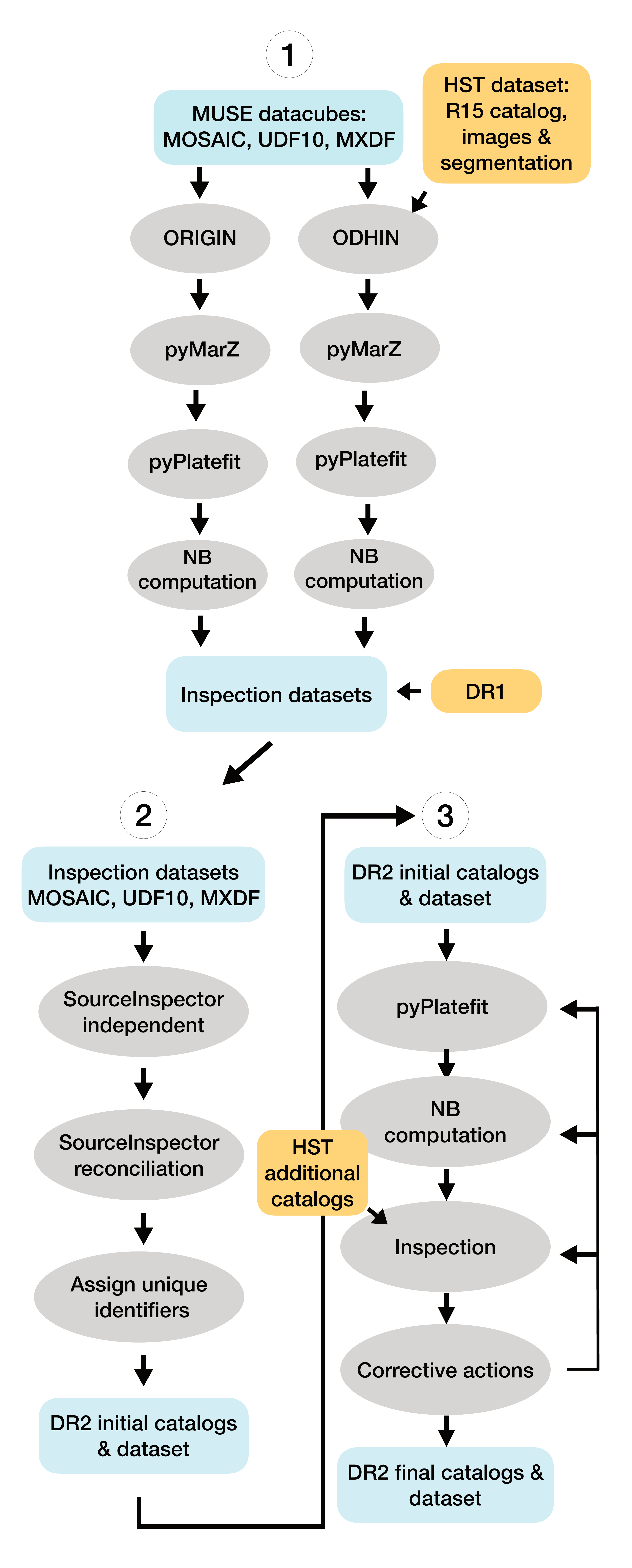}
\caption{Schematic of the processes involved in the data release production.
}
\label{fig:process}
\end{center}
\end{figure}

A schematic of the process is shown in Fig.~\ref{fig:process}.  The first phase is the production of inspection data sets (process chart one in Fig.~\ref{fig:process}). Candidate sources are produced in two parallel branches: (i) from blind detection using \origin\ software (Sect.~\ref{sec:origin}) and (ii) from de-blended extraction using the \odhin\ algorithm (Sect.~\ref{sec:odhinmain}) based on the R15 HST catalog and segmentation map. For each candidate source, the first five redshift solutions obtained with the \pymarz\ software are recorded (Sect.~\ref{sec:marz}). For each redshift solution, the corresponding emission and absorption lines are computed with \pfit\ (Sect.~\ref{sec:platefit}) and narrow bands for the highest S/N lines are derived (Sect.~\ref{sec:narrowband}). The data resulting from the different processes are organized into a database with a set of candidate source catalogs, one for each input datacube and each detection branch (i.e., \origin\ and \odhin).

The second phase (process chart two in Fig.~\ref{fig:process}, Sect.~\ref{sec:inspection}) consists of the visual inspection of a set of candidate sources by a group of experts who select the appropriate redshift solution, assigns confidence level and crossmatch the source with the HST R15 and \dro\ catalogs. Conflicts between the experts' solutions are identified and resolved in reconciliation meetings. Duplicates between \origin\ and \odhin\ sources are sorted out and the retained solutions are attributed a unique MUSE identifier\footnote{We use the following naming convention: MID-nnnn for MUSE identifiers and RID-nnnn for R15 identifiers}.

In the last step  (process chart three in Fig.~\ref{fig:process}, Sect.~\ref{sec:finaldp}), we selected the most appropriate extraction and refined the redshift solution, emission, and absorption line fits and narrow bands for each source. The results are inspected and corrective actions are identified and executed. This process is iterated until the results are deemed satisfactory. At the end of this last step, the final data products are produced, that is, a set of catalogs (Appendix~\ref{sec:catalogs}) and data organized in the MUSE Python Data Analysis Framework (\mpdaf) source format (Appendix~\ref{sec:sources}).

\subsection{\origin\ blind source detection}
\label{sec:origin}

The \origin\ software \citep{Mary2020} has been developed to automatically detect faint line emitters in MUSE datacubes. It is optimized for the detection of compact sources with faint spatial-spectral emission signatures and provides an automated and reliable estimate of the purity (i.e., one minus the fraction of false positives, \citealt{Benjamini1995, Mary2021}).
All three data sets were processed with the same input parameters. Below we present the example of the \mxdf\ data set.

\begin{figure}[htbp]
\begin{center}
\includegraphics[width=1\columnwidth]{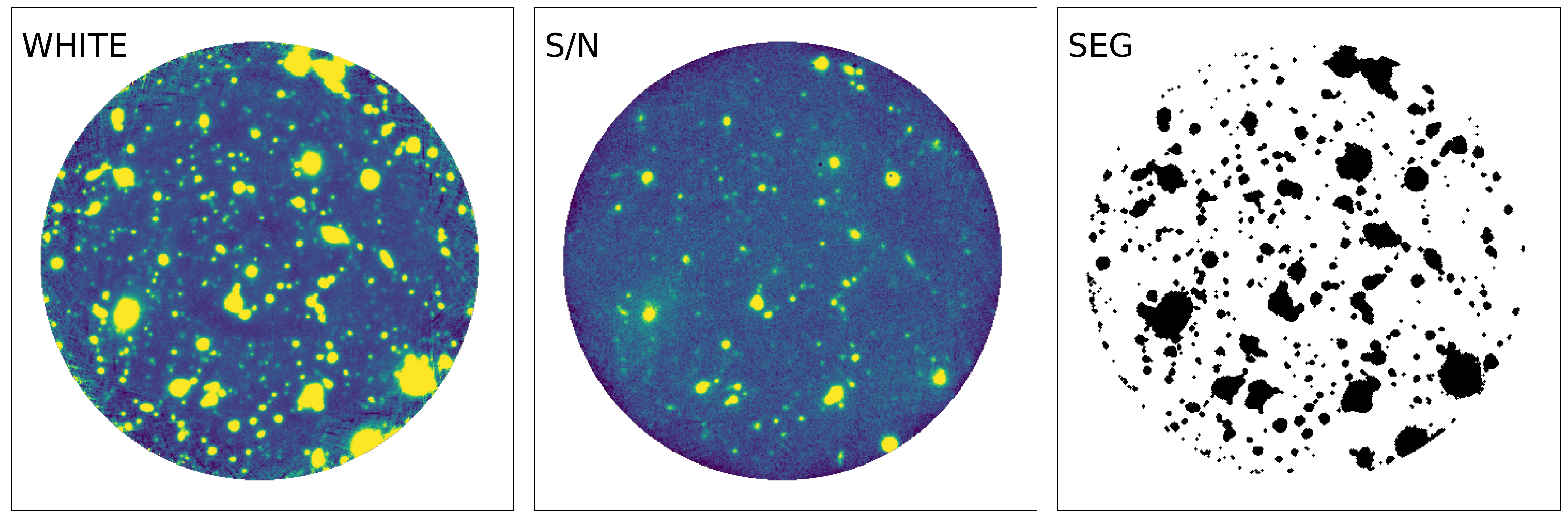}
\caption{\origin\ preprocessing steps. From left to right, the \mxdf\ MUSE white-light image, the S/N image and the segmentation map used to identify continuum sources and regions of low S/N.
}
\label{fig:orig1}
\end{center}
\end{figure}

The preprocessing steps (Fig.~\ref{fig:orig1}) perform a first pass continuum removal using a discrete cosine transformation of order ten along the datacube wavelength axis. The resulting white light continuum image is then segmented to mask the bright continuum sources. A second segmentation is also performed on the S/N residual image (i.e., the reconstructed continuum-free image of the datacube divided by the square root of the variance). The two segmentation masks are merged. The unmasked region defines the "faint region" that will be used in the following steps to avoid disturbance due to bright or noisy sources.

\begin{figure}[htbp]
\begin{center}
\includegraphics[width=1\columnwidth]{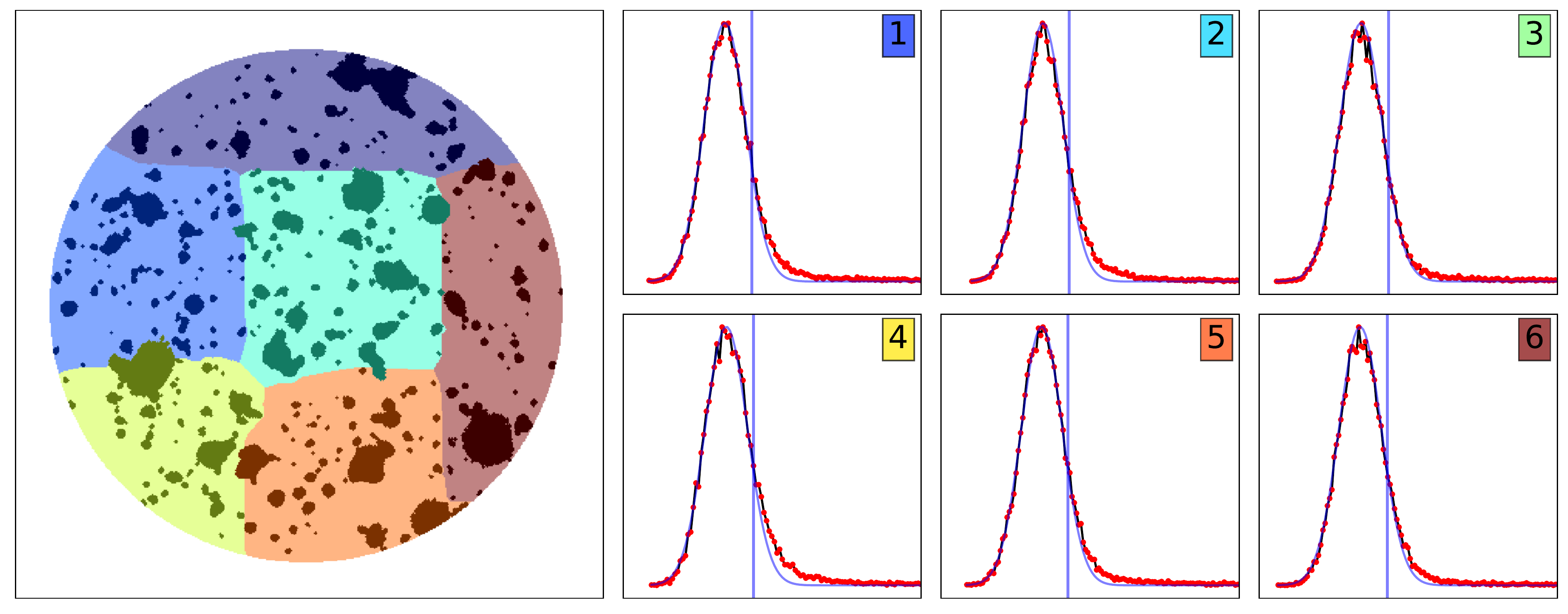}
\caption{\origin\ PCA thresholds. The left image displays the field segmentation in 6 zones. The 6 plots show the corresponding histograms of the PCA test statistics for each zone. The red curves exhibit the Gaussian fit of the distribution. The computed thresholds are displayed as blue vertical lines. They identify the minimum value above which the spectrum triggers the PCA cleaning process. This value corresponds to the 10\% upper quantile of the empirical distribution.
}
\label{fig:orig2}
\end{center}
\end{figure}

\begin{figure}[htbp]
\begin{center}
\includegraphics[width=1\columnwidth]{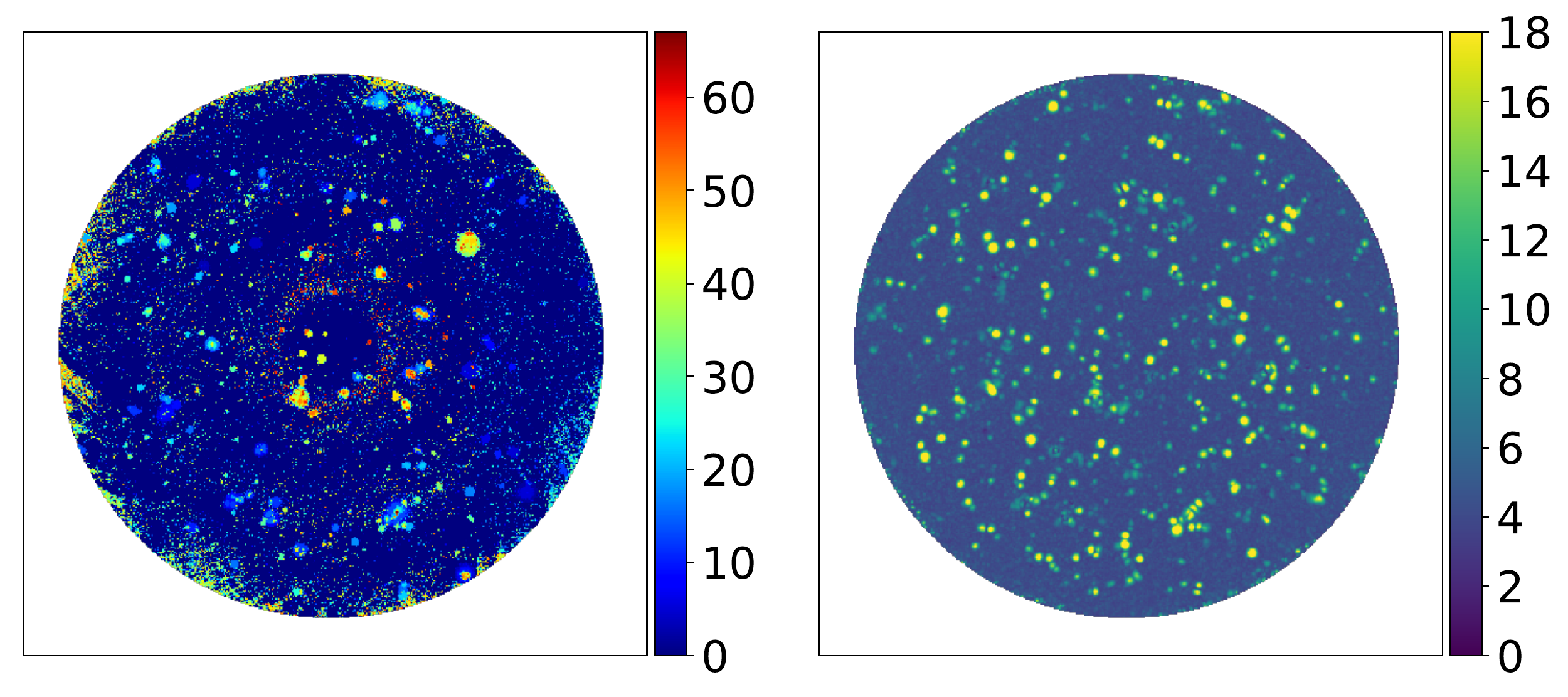}
\caption{Number of iterations where each spaxel was involved in the iterative PCA is shown in the left panel. The \origin\ maxmap, computed as the maximum of the General Likehood Ratio test datacube over the wavelength axis, is displayed in the right panel.
}
\label{fig:orig3}
\end{center}
\end{figure}

Although most of the continuum signal was removed in the preprocessing steps, there are still residuals in the resulting datacubes that will create too many false positives in the detection step, forcing the use of a higher detection threshold and preventing the detection of faint emission line sources. The continuum residuals, as well as the systematics present in the datacube, are subsequently removed with the iterative Principal Component Analysis (PCA) process described in Sect.~3.1.3 of \cite{Mary2020}. The \mxdf\ field of view is segmented into six zones (Fig.~\ref{fig:orig2} left panel; see Sect. 3.1.2 of \citealt{Mary2020}). In each segment, a threshold is calculated on the spaxel distribution to evaluate the deviation from the normal distribution\footnote{The noise distribution in the \mxdf\ is  well represented by a normal distribution, as are the \udft\ and \mosaic\ noise statistics; see Fig.~18 of \citealt{Bacon2017}.}. This threshold is then used by the iterative PCA process to clean the signal from nuisance sources. In the left panel of Fig.~\ref{fig:orig3}, we show the number of cleaning iterations at each spaxel. We note that while in most of the spaxels only a few iterations were needed to clean the data, many more were required at specific locations (e.g., at the edge of the field  or in the presence of complex bright sources residuals).

\begin{figure}[htbp]
\begin{center}
\includegraphics[width=1\columnwidth]{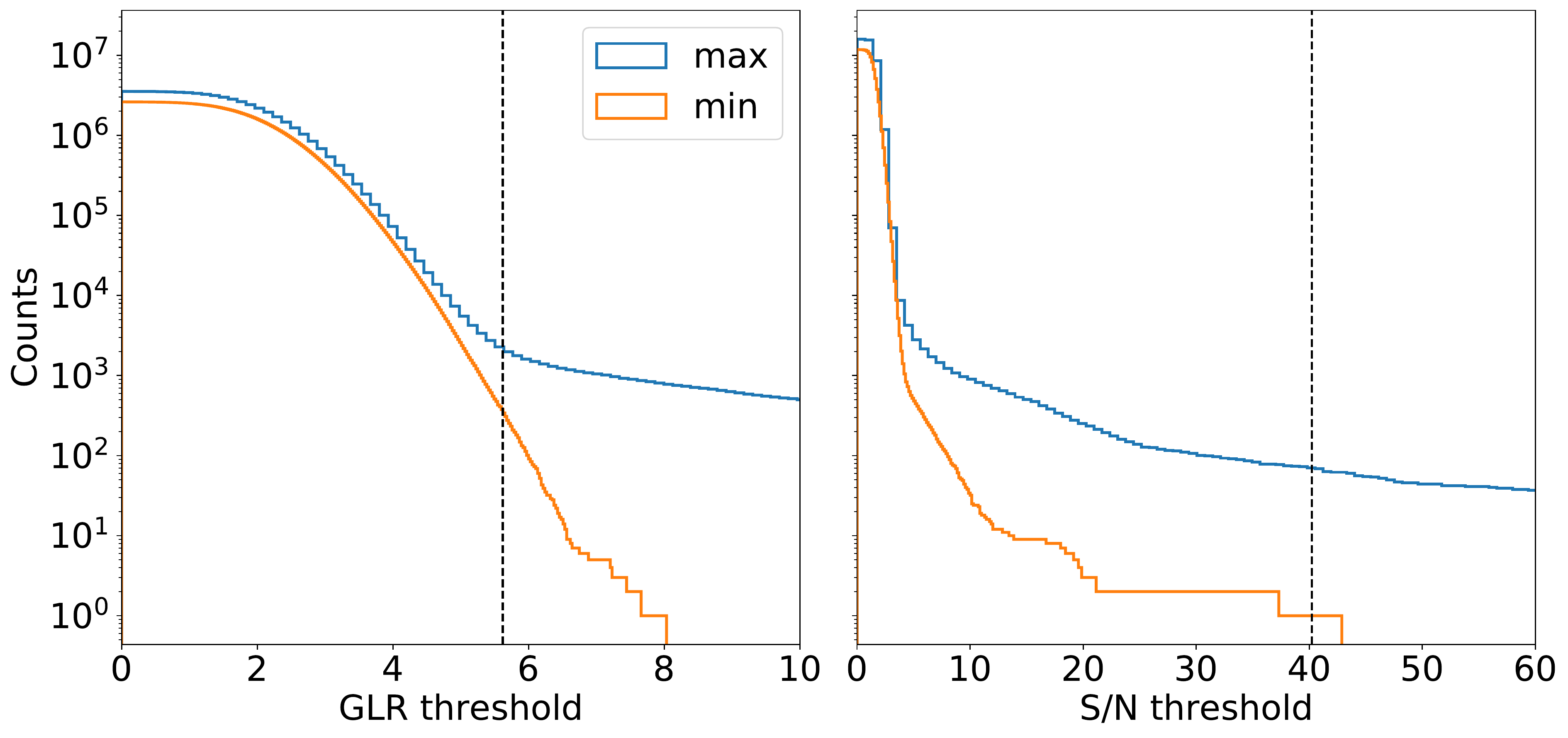}
\caption{Counts of positive and negative local maxima in the GLR datacube as a function of the threshold are shown in the left panel in blue and orange, respectively. The GLR threshold value computed to achieve the target purity value of 0.8 is displayed as a dotted vertical line. A similar plot for the S/N datacube and a target purity of 0.90 is shown on the right panel.
}
\label{fig:orig4}
\end{center}
\end{figure}

The datacube is then filtered using a library of possible 3D signatures, constructed as spectral line profiles spread spatially by the PSF and a generalized likelihood ratio (GLR) is computed for each voxel, as described in Sect.~3.2.1 of \cite{Mary2020}. The image of the maxima of the resulting GLR datacube is shown in Fig.~\ref{fig:orig3}, right panel. A comparison of the local maxima counts in the GLR datacube, restricted to the faint area, with the corresponding local minima counts is used to derive the purity as a function of the threshold applied to the cube of local maxima (Fig.~\ref{fig:orig4}, left panel). Using a purity of 0.8 for the GLR datacube, we identify 984 sources after spatial grouping of the detected emission lines. A similar operation is performed in the S/N datacube, prior to PCA subtraction, to recover possible bright sources affected by the PCA process.  In that case we use a higher purity of 0.9 to search for these additional bright sources. This yields 18 more sources.

For each source, an optimal extraction is performed on the MUSE datacube with the \cite{Horne1986} algorithm and using as a weighting  and segmentation image, the image obtained by summing the GLR datacube over a wavelength window centered on the detection peak wavelength and with a width of $\pm \,2 \times \text{FWHM}$, where FWHM is the width of the spectral template that provides the highest correlation peak (see Sect.~3.5 of \citealt{Mary2020}). 
We emphasize that the extraction is always performed on the original MUSE datacube, the \origin\ pseudo narrowband image being used only as weighting image.

The process was similar for the \udft\ and \mosaic\ data sets.
We compared our results with another blind source detection software program, \textsf{LSDCat}, developed by \cite{Herenz2017b} in the context of the MUSE-Wide survey \citep{Herenz2017a, Urrutia2019, Schmidt2021}. \textsf{LSDCat} is based on a 3D matched filter, which is, to a first approximation, similar to the mathematical basis of the \origin\ algorithm. However, there are  some important differences (e.g., the continuum subtraction or the use of GLR) and we should then be able to assess the robustness of our results by comparing the two detection catalogs. We use the same \mosaic\ datacube to perform this comparison. Excluding the \udft\ footprint to have a homogeneous depth, the \textsf{LSDCat} catalog is composed of 1190 sources. This must be compared to the 1287 \origin\ sources with a redshift confidence $\rm ZCONF > 0$ (see Sect.~\ref{sec:zconf}) using the same selection. 81 \textsf{LSDCat} sources (6\%) are unmatched in the \origin\ catalog. A detailed examination of these sources shows that most of them are low purity sources that were detected by \origin\ but not confirmed in the inspection process. Of the \origin\ sources,  246 (19\%) are not matched to the \textsf{LSDCat} catalog. Half of these sources are low confidence ($\rm ZCONF = 1$) sources that fall below the \textsf{LSDCat} detection threshold. Overall, the vast majority of high confidence sources (93\%) are detected by the two methods, giving confidence in the robustness of the results.

\subsection{\odhin\ HST-prior spectra extraction}
\label{sec:odhinmain}
For deep exposures, source blending increases, to a point where source confusion begins to be a problem. Source blending is already present in the 10-hour depth \mosaic\ field and of course has an increased impact in the ten times deeper \mxdf\ data. 
Fortunately, the 3D information content of MUSE helps to identify source mixing. For example, the location of the emission line can be measured to an accuracy of about half a spaxel ($0\farcs 1$) in reconstructed narrow bands when the S/N is sufficiently high. For blended sources with different emission lines it is then possible to separate the different contributions. Although this does not work with the continuum,  
we have another very valuable piece of information: the high spatial resolution broadband images from HST. 

The combination of data at different spatial and spectral resolutions is known as pansharpening in signal processing of hyper-spectral data \citep[see the review in][]{loncan2016}. The aim of these methods is to reconstruct a datacube with high spatial and spectral resolution. However, they are not directly applicable to our problem due to the very different spatial and spectral resolution in MUSE and HST\footnote{There is a factor of $\approx 5$ in spatial resolution and $\approx 1000$ in spectral resolution between MUSE and HST.} and the presence of noise. We have therefore developed a new and more robust method, nicknamed \odhin, based on linear source demixing.

The method assumes that at HST spatial resolution a source can be modeled as a single spectrum. This is obviously not valid for nearby galaxies that exhibit spatially resolved spectral variations. This is also not strictly valid at high redshift where some line emission may be more extended than the broadband morphology \citep[e.g.,][]{Leclercq2017}. \odhin\ is similar to the \textsf{TDOSE} method \citep{Schmidt2019} developed for the MUSE-Wide survey \citep{Urrutia2019, Schmidt2021} but it differs in three aspects: it is nonparametric, it uses multiple broadband HST images, and it implements a regularization process to avoid noise amplification for very close sources. The details of the algorithm are presented in Appendix~\ref{sec:odhin}. 
 
The inputs to the algorithm are the 3 HST broadband images (F606W, F775W, and F850LP) that fall within the MUSE spectral range (Fig.~\ref{fig:specrange}), the HST source catalog and its segmentation map and the MUSE datacube. We ran \odhin\ for three HST catalogs: our primary catalog R15 \citep{Rafelski2015}  but also the \candels\ \citep{Whitaker2019} and \tdhst\ \citep{Skelton2014} catalogs. Although these extractions use the same HST broadband images \citep[XDF,][]{Illingworth2013}, they are based on different segmentation maps and may thus provide different results.

\begin{figure}[htbp]
\begin{center}
\includegraphics[width=1\columnwidth]{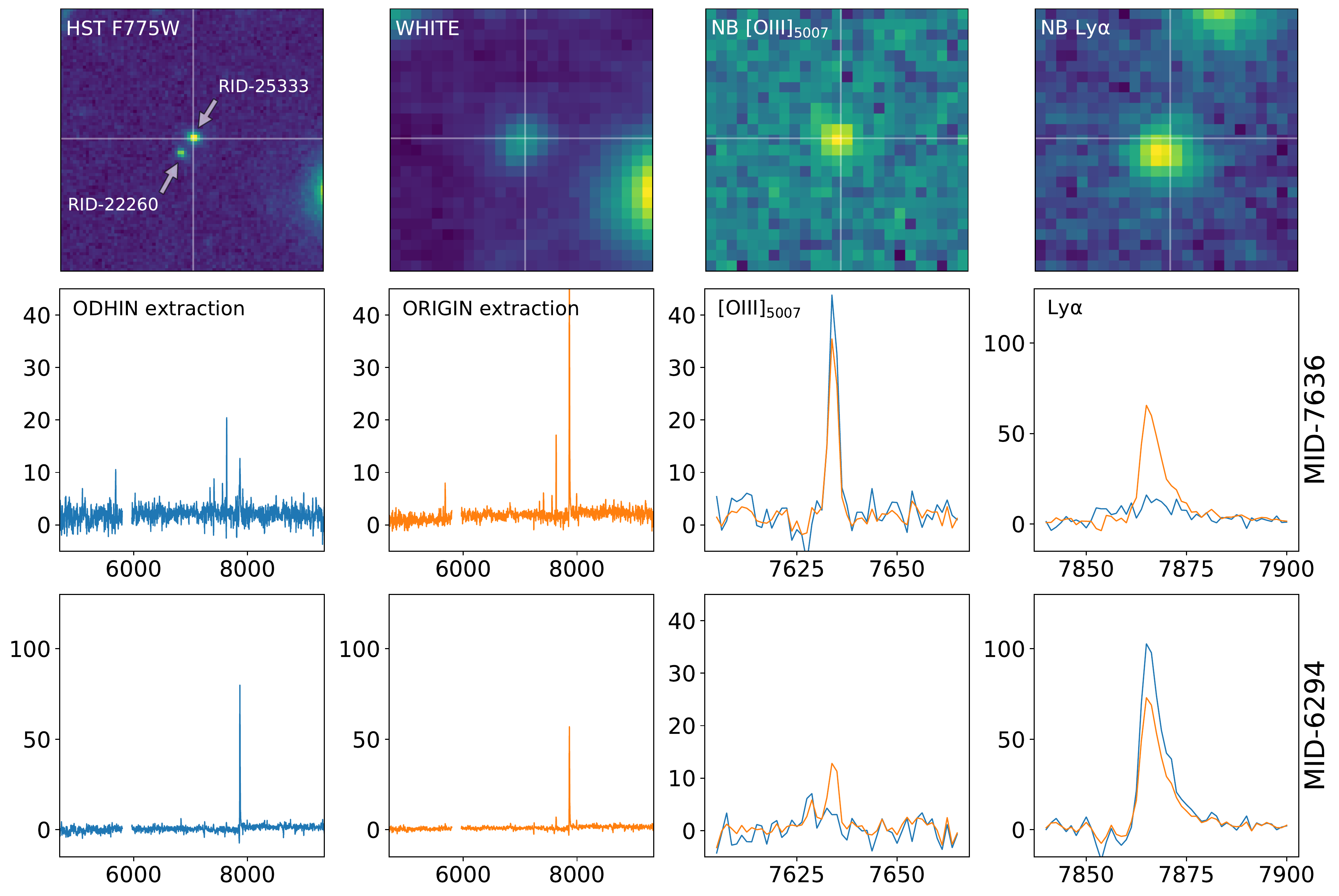}
\caption{Example of \odhin\ de-blending of two close sources MID-7636 (RID-25333), a $z=0.5$ star forming galaxy, and MID-6294 (RID-22260), a $z=5.5$ \lae. The separation of the sources is 0.4 arcsec. The top row of panels shows, from left to right: the HST F775W, MUSE white-light, [OIII]5007\AA\ and \lya\ narrow band images ($\rm 5 \times 5 \, arcsec^2$ size). The central and bottom panels display the spectra derived from \odhin\ (in blue) and \origin\ extraction (in orange) for each source. Fluxes and observed wavelengths are expressed in units of \ergsa{-20} and \AA, respectively.}
\label{fig:deblend_example}
\end{center}
\end{figure}

An example of source de-blending is given in Fig.~\ref{fig:deblend_example}. Two close sources can be identified in the HST image (upper left panel) separated by 0.4 arcsec, but they are blended at the MUSE spatial resolution as shown in the MUSE white light image (second upper panel). The brightest source (F775W AB 27.2) MID-7636 (RID-25333) is a nearby star-forming galaxy at $z=0.5$, with multiple emission lines: for example, \oiid, \oiiia, \hb\ and \oiiib (central row of Fig.~\ref{fig:deblend_example}). The second source (F775W AB 27.8) MID-6294 (RID-22260) is a $z=5.5$ \lae\ with a strongly asymmetric \lya\ line (last row of Fig.~\ref{fig:deblend_example}). The fact that the narrow-band image of the \lya\ is slightly offset from the \oiiib\ narrow-band image indicates that the \lya\ emission is indeed related to RID-22260 and not RID-25333. However, the overlap is too large to de-blend it from the narrowband segmentation at the spatial resolution of MUSE. Therefore, for each source, the spectra provided by the \origin\ extraction (shown in orange in Fig.~\ref{fig:deblend_example}) show strong contamination due to the nearby source. In contrast, the \odhin\ extraction (shown in blue) is able to de-blend the contributions from each source.

One can see that, despite its intrinsic limitations\footnote{We assume that the broadband light distribution is representative of any light emission and do not allow for spectral variation within a source.}, \odhin\ is capable of de-blending closed sources with good accuracy. We note, however, that in the presence of emission that extends much beyond the HST broadband, the method will miss part of the flux (see Fig.~\ref{fig:ori_emi_ext} for an example). Furthermore, the method is by construction blind to any source undetected by HST.

\subsection{Redshift determination with \pymarz}
% marz.tex
\label{sec:marz}
The software \pymarz\ is a Python implementation of the redshift fitting software \textsf{MarZ} \citep{Hinton2016} originally developed for multi-object fiber spectroscopy with OzDES \citep{Lidman2020}. \textsf{MarZ} itself is based on a modified version of the cross-correlation algorithm \textsf{AUTOZ} \citep{Baldry2014}. It is written in JavaScript with a dedicated user interface. The development of the Python version was motivated by the need to integrate it more easily into the full data analysis chain where all parts are written in Python. We have also developed our own interface \se\ (Sect.~\ref{sec:si}), which is run separately from \pymarz.

Compared to the version used in DR1 \citep{Inami2017}, the main change, apart from the Python language, is the update of the spectral templates based on a set of higher S/N spectra of selected sources in \udft\ and \mosaic\ datacubes. We tested it on a selection of \udft\ high confidence isolated sources and found that the first ranked solution returned by \pymarz\ is correct 84\% of the time. We note that the probability of finding the correct solution among the first two \pymarz\ solutions is 95\%. In the previous version used for DR1, the probability of finding the correct solution was only 71\% and 87\% for the first and first two returned solutions, respectively. Although the improvement is significant, the selection of the correct redshift solution requires additional information and cannot be performed automatically based only on the top ranked \pymarz\ solution. For this reason, we retain the top five \pymarz\ solutions for the manual inspections presented in Sect.~\ref{sec:inspection}.
The software, its documentation and the templates used for this work are made publicly available (Appendix~\ref{sec:soft}).

\subsection{Line flux measurements with \pfit}
\label{sec:platefit}
In the past years, the increasing number of galaxy surveys with wide field or highly multiplexed IFUs (e.g., SAMI, MANGA, and MUSE) has been accompanied by the development of software tools to efficiently fit emission and/or absorption lines in spectra or datacubes (e.g., pPXF \citealt{Cappellari2017, Sarzi2017}, ALFA \citealt{Wesson2016}, LZIFU \citealt{Ho2016}, pipe3D \citealt{Sanchez2021}). Most of these tools have been developed for the spatially resolved study of nearby galaxies and are not optimized for the population of medium to high redshift galaxies present in our observations. We have therefore developed a specific fitting code \pfit, inspired by the \textsf{Platefit} IDL code developed for the SDSS survey \citep{Tremonti2004, Brinchmann2004}.

\pfit\ is a python module to perform emission line fitting of astronomical spectra. Although it was specifically developed for this data release\footnote{We used the \textsf{Platefit} IDL code for the DR1 release.}, it should work for spectra delivered by other instruments, provided that the input spectrum is in MPDAF format \citep{Piqueras2019}.
The program takes as input a spectrum and an input redshift and returns a list of emission line parameters: for example, flux, velocity dispersion, equivalent width and their errors. The program performs a continuum fit, using a simple stellar population model (\citealt{Bruzual2003}, see \citealt{Brinchmann2013}), and then fits the emission lines after subtracting the continuum. In addition, the program can also perform a basic fit of the absorption lines. Lines are grouped into families, each of which is fitted independently. All lines belonging to the same family share the same velocity (or redshift) and velocity dispersion. The default lines table (Appendix~\ref{sec:linename}) has 3 different families: the Balmer series, the non-Balmer emission lines and the ISM absorption lines. In addition, some major resonant lines like \lya\ or \mgii\ are fitted separately as distinct families. 

All emission and absorption lines are modeled as Gaussian with 3 parameters: central wavelength ($\lambda_0$), wavelength dispersion ($\sigma$) and flux ($F_0$). The only exception is the \lya\ line, which is fitted as a skewed Gaussian \citep{Birnbaum1950}, with an additional unit-less parameter ($\gamma$) measuring the asymmetry of the line:
\begin{equation}
\label{eq:asymline}
F(\lambda) = F_0  \left[ 1 + \erf \left(\gamma \frac{\lambda - \lambda_0}{\sqrt{2} \sigma} \right) \right]
\exp \left( -\frac{\left(\lambda - \lambda_0 \right)^2}{2 \sigma^2} \right)
.\end{equation}

Optionally, a double peaked \lya\ line profile can be fitted as the sum of two skewed Gaussians.
The program uses the front-end lmfit minimization package\footnote{\href{https://lmfit.github.io/lmfit-py/}{lmfit.github.io/lmfit-py}} \citep{Lmfit2014} based on scipy optimize nonlinear least-squares minimization \citep{Scipy2020}. The default minimization routine is the trust region reflective least squares algorithm, which is more robust and gives a better error estimate than the classical Levenberg–Marquardt algorithm. Other minimization algorithms can be selected (see the lmfit documentation for details). The Markov chain Monte Carlo emcee package \citep{Foreman2013} is also used to improve error computation.
Information about the public released version of the software and its documentation is given in Sects.~\ref{sec:soft} and \ref{sec:linename}.

\subsection{Narrowband measurements}
\label{sec:narrowband}
For each redshift solution, we compute the narrowband images associated with each detected emission or absorption line, as provided by \pfit. These narrowband images are important to confirm the detection of the line and to verify its origin. For example, spectral pollution due to nearby sources will be more easily spotted by looking at the location of the corresponding narrowband image (e.g., Fig.~\ref{fig:deblend_example}). Another example is the case of a low S/N spectrum when a noise pattern due to sky subtraction residuals is fitted by \pfit\ because its spectral location matches the expected wavelength of a line. In this case, the narrowband image should not present a coherent peak at the source location, thus helping discard the line measurement.

The narrowband process selects each emission and absorption line with $\rm (S/N)_{line} > 2$ plus all major\footnote{e.g., \lya\ or \oiid; see \pfit\ input line table in Appendix~\ref{tab:linename}.} emission lines, regardless of their S/N. The continuum is then subtracted from the datacube using a median filter size of 101 \AA\ total width (81 pixels). A narrowband image is finally obtained using a simple summation of the continuum-subtracted datacube over the wavelength range defined by the fitted center of the line and a spectral width of $\rm \pm 2 \times FWHM$, where FWHM is the fitted line width at half maximum. For emission line doublets like \oiid, the wavelength range is extended to cover both lines. 
For each narrowband image  with $\rm (S/N)_{line} > 2$, we perform a \sex\ segmentation and derive its center, flux and signal-to-noise ratio $\rm (S/N)_{nb}$.

We also derived composite emission and absorption narrowband images obtained by optimally adding  all emission or absorption narrow bands. The process starts with the highest $\rm (S/N)_{nb}$ of the emission or absorption narrow bands and adds additional narrowband images until the $\rm (S/N)_{nb}$ of the composite narrowband stops increasing.

\subsection{Visual inspections and reconciliations}
\label{sec:inspection}
The material produced in the previous steps, that is, for each candidate source, the five redshift solutions provided by \pymarz\ as well as their emission and absorption lines fits and their associated narrow bands, must be organized and evaluated by a group of experts. Once the conflicts have been resolved during the reconciliation meetings, one can proceed to the creation of the final catalog by assigning a unique identifier to the detected sources. The details of the process are presented in the following sections. The figures and statistics quoted are given for the \mosaic\ largest data set, but the process was similar for the \udft\ and \mxdf\ data sets.

\subsubsection{Inspection sample}
The \mosaic\ parent sample is composed of 10450 sources divided into 7977 \odhin\ sources derived from the R15 catalog (Sect.~\ref{sec:odhinmain}) and 2473 \origin\ sources (Sect.~\ref{sec:origin}). 
Previous investigation has shown that all DR1 \origin\ sources to which a secure redshift could be assigned have a purity greater than 0.9. We therefore restrict the \origin\ sample to the 1946 sources with purity greater than 0.9. 

The \origin\ subsample contains all line sources that have at least one detectable emission line, regardless of their continuum flux. The remaining sources of interest are those whose continuum flux is bright enough to identify other spectral features (e.g., absorption lines, breaks). After some trial, we adopted a continuum S/N cut\footnote{The continuum S/N is computed as the maximum of the S/N spectrum after running a median filter of 100 pixels width to remove the impact of emission lines.} of \ergsb{0.8}{-20}  per spectral pixel, which corresponds roughly to an AB F775W magnitude of 27.6. This reduces the \odhin\ sample to 782 sources\footnote{Our experience has taught us that galaxies with fainter continuum cannot be attributed an aborption line redshift and that galaxies with emission lines are already detected with \origin.}. 

We further reduced the \mosaic\ inspection sample by removing all sources located in the \udft\ area, taking a margin of 1 arcsec. The final \mosaic\ inspection sample is composed of 2412 sources: 1727 \origin\ and 685 \odhin\ sources. This sample was randomly distributed to four groups of three experts. In addition, a randomly selected common subsample of 50 sources was given to the four groups. This control sample will be used to assess the homogeneity of the classification between the groups.
We proceeded likewise for the \udft\ and \mxdf\ data sets and produced final (i.e., \odhin, \origin, and control) inspection samples of 1234 and 1599 sources, respectively.

\subsubsection{\se}
\label{sec:si}
\begin{figure*}[htbp]
\begin{center}
\includegraphics[width=0.8\textwidth]{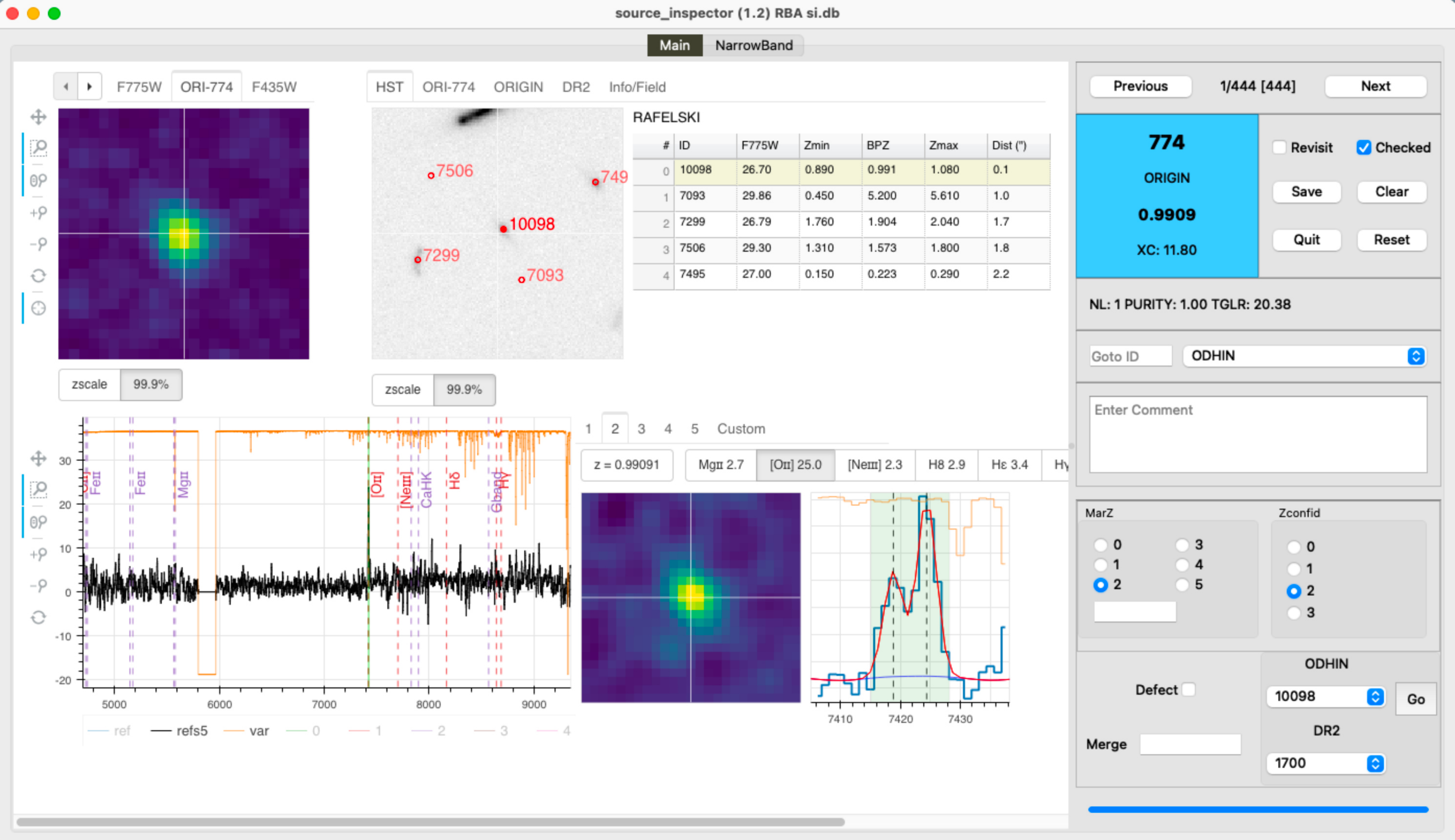}
\caption{\se\ interface in {evaluation} mode. The left side of the interface displays the interactive html file for the current source. It shows the five redshift solutions proposed by \pymarz\ plus their corresponding \pfit\ line fits and associated narrow bands in the bottom panels. The html upper panels display MUSE and HST images, and the locations of neighboring sources from HST catalogs. The right side of the interface allows the user to navigate in the source list, select the redshift solution or assign a new redshift, assign confidence and provide crossmatch information.
}
\label{fig:se_eval}
\end{center}
\end{figure*}

The inspection package provided to each expert consists of a small database, a set of interactive html files and the \se\ software. \se\ is a PyQt tool that allows the expert to display each candidate source, to select the redshift solution or to provide a new one, to give a confidence to the redshift assignment and to match the source to other sources (e.g., \origin\ to \odhin) and to HST catalogs. All operations are performed locally. An example of the \se\ interface is shown in Fig.~\ref{fig:se_eval}. The inspection results are exported at the end of the process to be used later for the reconciliation step.

\subsubsection{Redshift confidence}
\label{sec:zconf}
The expert must assign a confidence to the redshift solution. This redshift confidence (called ZCONF in the tables) is defined as follows.

A confidence level of 0 means no redshift solution can be found. It should be noted that this is  decoupled from the detection status. An emission line can be well  detected with high purity by \origin, even though no redshift solution can be found.

A confidence level of 1 indicates a low confidence. A tentative redshift solution has been found. Reasons for assigning $\rm ZCONF = 1$ include low S/N of the lines (e.g., $\rm S/N_{line} < 3$), poor fit, existence of other valid redshift solutions, noisy associated narrow bands, additional lines present in the spectrum with reasonable S/N but unexplained by the proposed redshift solution.

A confidence level of 2 indicates a good confidence. For non \laes\ (i.e., $z < 2.9$) sources, we assign this confidence level to sources with multiple lines detected with good S/N ($\rm S/N_{line} > 5$). For example, a resolved \oiid\ doublet with a clear narrowband would be sufficient to assign $\rm ZCONF = 2$. For \laes, the requirement is to have a \lya\ line with good S/N (i.e., $\rm S/N_{line} > 5$) and a width and asymmetry compatible with \lya\ line shapes.

Finally, a confidence level of 3 indicates a high confidence. For non-\laes, the criteria are similar to the $\rm ZCONF = 2$ requirements, but we expect more lines and higher S/N, as well as high S/N narrowband images. For \laes, if there is no other line than \lya, we require the \lya\ line to achieve high S/N ($\rm S/N_{line} > 7$) with the expected line shape: that is a pronounced red asymmetrical line profile\footnote{$\gamma$ is the asymmetric Gaussian parameter used by \pfit\ (Sect~\ref{sec:platefit}).} ($\rm \gamma > 2$) and/or a blue bump or double peaked line profile\footnote{Double peaked \lya\ profiles are easily differentiated from \oiid\ or \ciiid\ using the line spacings and widths.}.

In the case of \origin\ detections and if the source can be matched to an HST counterpart, it adds confidence to the detection. If, in addition, the R15 photometric redshift is reliable and matches well with the MUSE redshift (see Eq.~\ref{eq:photoz}), this adds confidence to the redshift assignment and could lead to a higher ZCONF than in the case where the \origin\ source is not matched with an HST source. On the other hand, photometric redshifts for faint sources are often not very reliable and thus disagreement between the MUSE and photometric redshifts is not considered a strong negative constraint.

We note that in the case of low S/N spectra with a single emission line, we are not always able to determine the nature of that line with a high degree of certainty. In many cases, the combination of line shape and photometric redshift argues in favor of \lya, but it is desirable to have a handle on the amount of misclassifications in this subsample. We focus on \oii\ and \lya\ emitters here as other emission lines are much less likely because of the small accessible volume.

In order to estimate the expected fraction of \lya\ and \oii\ emitters as a function of magnitude, we need LFs and rest equivalent-width distribution functions (EWDFs) for both. We take the \lya\ LF from  \cite{Herenz2019} and the \lya\ EWDF from \cite{Hashimoto2017}. For \oii\ we adopted the evolving LF from \cite{Saito2020}, which agrees fairly well with our UDF data. We also need an EWDF for \oii, which we prefer to derive from our high quality ($\rm ZCONF \ge 2$) data. To do that we adopt the point-source selection function from \cite{Herenz2019}, which we derived for \lya\ emitters but which reproduces the minimum \oii\ flux as a function of wavelength fairly well for the present sample when the selection function is adjusted for difference in exposure time by  $\rm \sqrt{t}$. We then use a simple Vmax estimator for the EWDF after verifying that the \oii\ line luminosity and equivalent width are approximately independent, in good agreement with previous work \citep{Blanton2000, Ciardullo2013}. The distribution is found to be well fitted by a Gaussian in log EW(\AA) with mean = 1.61 and sigma = 0.17.

Taking these together with the assumption that the spectrum of our galaxies is approximately flat in flux density ($\rm f_{\nu}$) we can calculate the expected number of \oii\ and \lya\ emitters with observed magnitude F775W fainter than a particular value. When we do that, we find that at magnitudes $\rm F775W > 28.5$ the number of \laes\ is expected to be $10$---$100$ times that of the \oii\ emitters at all redshifts of relevance down to a flux limit of \ergsa{-18}. Thus, we conclude that the likelihood of misclassified lines is less than 1/10 for galaxies fainter than 28.5.

\subsubsection{Cross identifications}
\label{sec:crossmatch}
For \origin\ sources, the expert must also identify the possible HST counterparts and, likewise, must find the possible \origin\ counterparts for \odhin\ sources. Crossmatching \origin\ and HST sources can be difficult when the field is overcrowded or if the \origin\ location is offset with respect to HST (see an example in Fig.~\ref{fig:offset_lya}). The use of the photometric redshift can help with the HST association, but not always. At a later stage we perform an evaluation of the HST crossmatch confidence (Sect.~\ref{sec:crosshst}).

Occasionally a source is visually identified in one of the HST images at the \origin\ location, but is not present in the HST R15 catalog. This might occur when the source has been incorrectly merged with a nearby source by \sex, but in other cases the source is simply too faint and escapes detection. We identify these cases by setting "missed, faint, or undetect" in the corresponding HST crossmatch flag (see Sect.~\ref{sec:nohst}). 

The \origin\ process identifies individual emission lines and merges them into a single source when they are close (see Sect. 3.5 in \citealt{Mary2020}). However, sometimes the process fails to merge emission lines when the source is very extended, or, on the contrary, it merges lines that belong to two different sources (i.e., superpositions). The expert is then responsible for evaluating if the \origin\ sources should be split or merged with another \origin\ source. Such actions are then performed in the final step (Sect.\ref{sec:finaldp}).

Cross-identification with the \dro\ catalog is also performed during this evaluation process. When a source is found at the location of a \dro\ source but is assigned a different redshift, we mark it as a \dro\ unmatched source. On the other hand, if only the redshift confidence is different, we mark the source as \dro\ matched.

\subsubsection{Reconciliation}
\label{sec:recon}
When all experts in the group have completed their evaluation, the group leader performs the reconciliation process. All sources that disagree on any of the assigned items (i.e., redshift value, redshift confidence, HST, DR1, or \origin\ crossmatch) are flagged for the reconciliation meeting. In addition, any source flagged with "{revisit}" by an expert is added to the review list. The only exception to this process is when the only discrepancy is due to redshift confidences between two and three, in which case the source is not flagged for reconciliation and its confidence is set to the average confidence rounded to the nearest integer.

An example of the statistics returned by \se\ in automatic checking mode is given in Table.~\ref{tab:recon}.
Disagreements are then discussed and resolved in a face-to-face reconciliation meeting with the help of the reconciliation mode of \se. On average, between 30\% and 40\% of the original sample had to be revisited. 

Most of the sources to be re-examined are weak, single-line sources for which experts have a different assessment of the reality of the measured redshift. For example, some less experienced experts would be inclined to identify a \lae\ for a low S/N line, even if there is no additional information to support this assumption. In most cases, however, when reconciling, the redshift solution will be ruled out.

\begin{table}
\caption{Example of automatic reconciliation.}   
\label{tab:recon}
\centering
\begin{tabular}{lr}
\hline 
\mosaic\ group 1 reconciliation\\
\hline
Total number of sources & 790 \\ 
Sources autochecked with $\rm ZCONF=0$ & 57 \\ 
Sources autochecked with $\rm ZCONF > 0$ & 430 \\ 
Sources to be manually reconcilied & 303 \\ 
With revisit flag & 24 \\  
With discrepant redshift & 185 \\ 
With discrepant redshift confidence & 45 \\ 
With discrepant matching information & 36 \\ 
With discrepant split or merge info & 13 \\ 
\hline 
\end{tabular} 
\tablefoot{
"autochecked" means that all experts agree.
}
\end{table}

\subsubsection{Control sample}
\label{sec:control}
A control sample of 50 randomly selected \mosaic\ sources was given to the same four experts groups. The same experiment was done for the \mxdf\ evaluation. After reconciliation, the control sample was analyzed for remaining disagreements. The 50 sources consisted of 36 \origin\ and 14 \odhin\ sources. All groups agreed on the redshift assessments. Eight sources did not obtain a redshift solution. Of the 42 sources with a redshift solution, nine sources were credited with different redshift confidence: five sources were ranked with low and good/high confidence and four sources were ranked with good and high confidences. Four sources were also crossmatched to different HST sources.

The results for this control sample demonstrate that we achieved overall excellent agreement among all experts, with no discrepancies in redshift solutions. Redshift confidence is more subjective and we do identify 20\% of disagreements between groups. We note that we do not consider the difference between confidence two and three to be critical. Nevertheless, we have 10\% of sources that were considered low confidence by some groups and good or high confidence by others. 

Similar results were obtained on a control sample of 49 \mxdf\ sources given to three different groups. All groups agreed on the redshifts, but there were ten sources with different confidence assignments. In addition, some groups assigned a low confidence level to 10 sources, while the others classified them as having no redshift (confidence 0).  

As we discuss in the next section, a review of all sources with redshifts is performed as a final step by a single expert. This helps to homogenize the results, but given the amount and complexity of information that goes into the confidence evaluation (Sect.~\ref{sec:zconf}), it currently seems difficult to have a more objective criterion for redshift confidence assignment.

\subsection{Creation of final data products}
% finaldp.tex
\label{sec:finaldp}
In this final step, we create the final catalog by giving a unique MUSE identifier to all sources with an assigned redshift. We start by splitting the catalog in \origin\ only, \odhin\ only
and crossmatched \origin-\odhin\ sources. All sources already matched to the \dro\ catalog keep the same MUSE identifier, all others are given a unique new identifier.

Split and merge of \origin\ sources defined during the inspection step was implemented. 
A final review of all sources with redshift was then completed. During this review the reference extraction (Sect.~\ref{sec:ext}) and the reference center (Sect.~\ref{sec:narrowband}) were chosen. Emission or absorption line fits were fine-tuned with \pfit\ when needed and double asymmetric \lya\ fits implemented for double peaked profiles. The computed narrowband $\rm S/N_{nb}$ was used as an additional guide for the redshift confidence (Fig.~\ref{fig:zconf}), which was updated when requested.

% extract.tex
\subsubsection{Selection of reference extraction}
\label{sec:ext}

For each source, we derive the spectra from the MUSE datacube using different extraction schemes (e.g., \origin\ or \odhin). There is no single way to extract the signal: for example, the stellar continuum is not necessarily superimposed on the location of the flux peak of the emission lines, and its spatial extend is often different. The most obvious cases are those of \laes, which show extended \lya\ emission that is on average ten times larger than their continuum counterparts (e.g., \citealt{Wisotzki2016, Leclercq2017}). In this case, a spectrum based on the continuum surface brightness will miss a large faction of the \lya\ flux, and vice versa, a spectrum optimally extracted for the \lya\ emission will be suboptimal for the continuum S/N.

In the \origin\ extraction scheme (Sect.~\ref{sec:origin}), the pseudo-narrowband image is derived from an improved continuum subtracted datacube using an iterative principal component (PCA) scheme and thus has the advantage of being mostly free from continuum residuals. It also has a high S/N,  as it results from a process similar to optimal filtering (see Sect.~3.2 of \citealt{Mary2020}). 
However, in the case of a very bright emission line, the PCA process may produce  artifacts by removing part of emission line. In this case, we perform an optimal extraction based on the narrowband composite image and its segmentation (Sect.~\ref{sec:narrowband}).
This narrowband extraction method is called \nbext\ in the rest of the document. 

\begin{figure*}[!tbp]
  \begin{subfigure}[b]{0.5\textwidth}
    \includegraphics[width=1\textwidth]{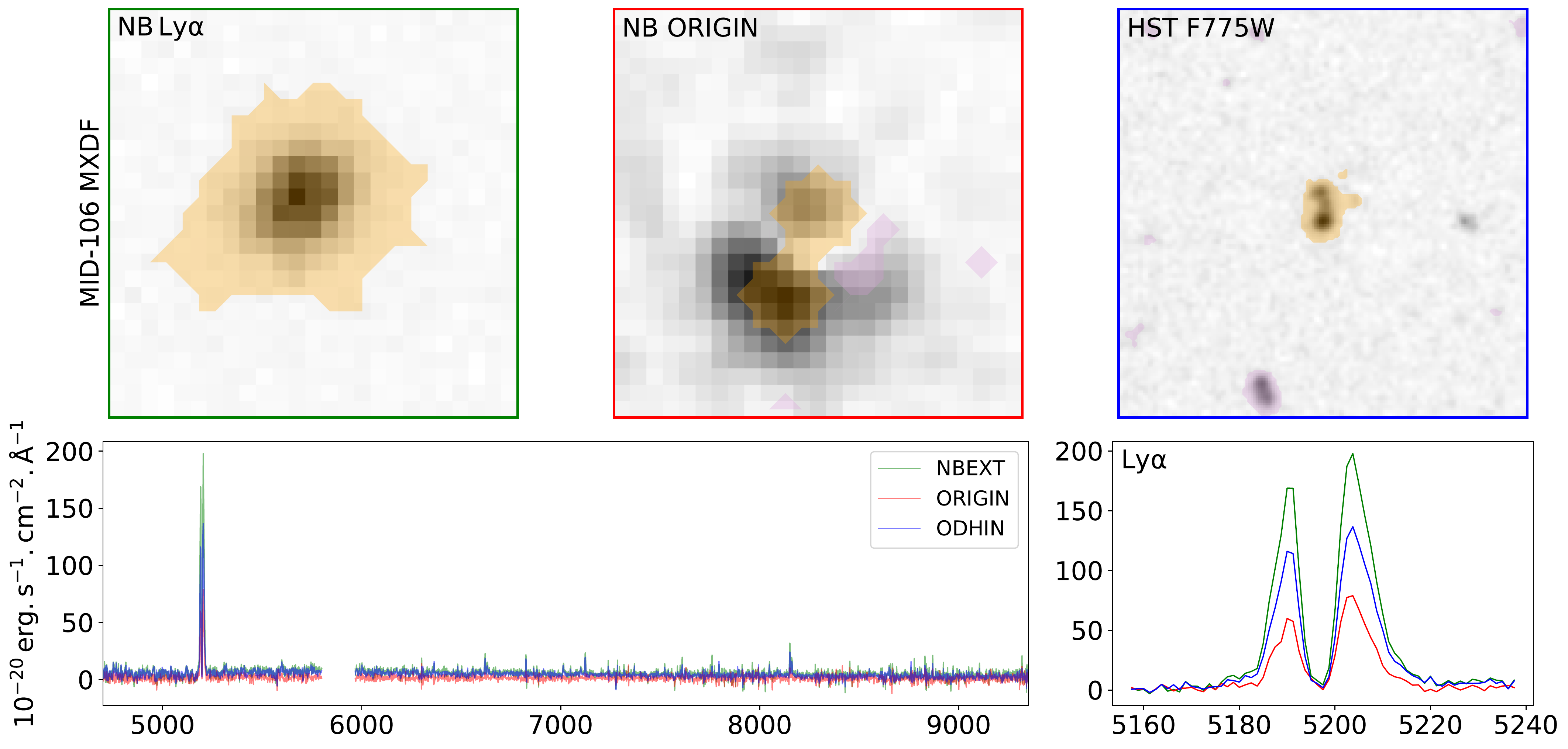} 
    \caption{MID-106.}
    \label{fig:src106}
  \end{subfigure}
  \hfill
  \begin{subfigure}[b]{0.5\textwidth}
        \includegraphics[width=1\textwidth]{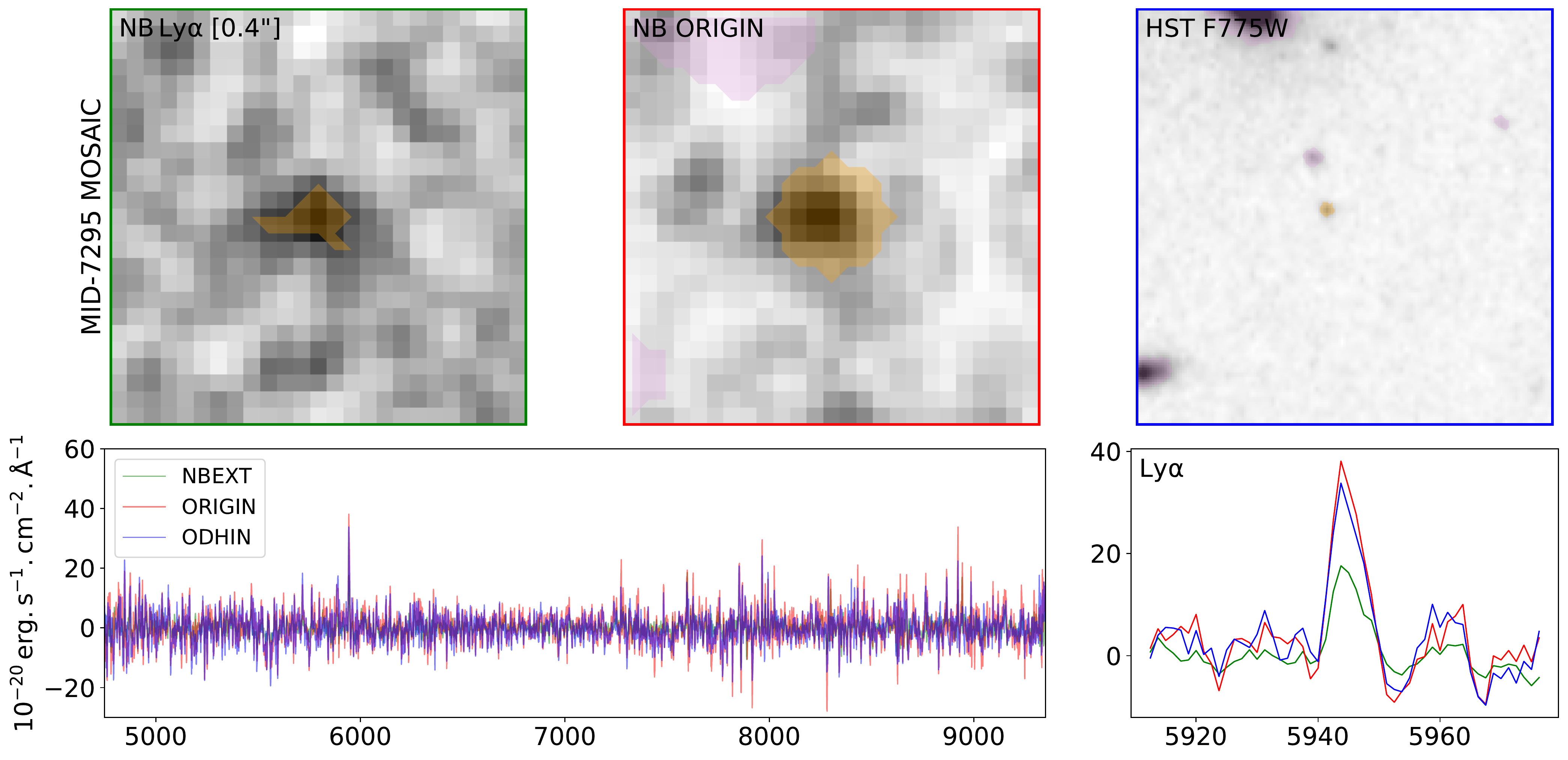}
        \caption{MID-7295.}
        \label{fig:src7295}  
  \end{subfigure}
  \caption{Example of sources where spectral extraction gives different results. Three different extraction schemes are shown for the \laes\ MID-106 (a) and  MID-7295 (b). For each subfigure we show the full spectrum (lower left panel) and a zoomed region around the \lya\ line (lower right panel). In green we display the spectrum derived from the weighted extraction using the narrowband \lya\ weighted map (\nbext) and its segmentation (top left panel), in red the \origin\ weighted extraction (top central panel) and in blue the \odhin\ extraction based on the R15 segmentation map (upper right panel). The colored area in the left and central top panels indicate the source (orange) and masked objects (magenta) segmentation maps. In the case of the low S/N source MID-7295 (b), the spectra have been smoothed with a 5 pixels width kernel and a Gaussian filter of 0.4 arcsec FWHM has been applied to the \lya\ narrowband image (top left panel of subfigure b). Observed wavelengths are in \AA\ units.}
  \label{fig:ori_emi_ext}
\end{figure*}

We illustrate the difference between the two methods with two examples in Fig.~\ref{fig:ori_emi_ext}. 
Figure~\ref{fig:src106} shows the results of the \origin\ and \nbext\ weighted extractions for a bright \lae\ in which this effect is particularly strong. It can be seen that in the case of the \origin\ pseudo narrowband image (upper central panel), the bright central \lya\ emission has been partially removed by the PCA, leaving only the outskirts. As a result, only a small portion of the \lya\ flux is recovered by the weighted extraction (shown in red in the lower panels). We note that the \odhin\ extraction (Sect.~\ref{sec:odhinmain}) based on the R15 segmentation map (blue line in bottom panels) is more efficient but still misses some \lya\ flux compared to the narrow band \lya\ extraction (green line in bottom panels).
The second example shown in Fig.~\ref{fig:src7295} presents the same extraction methods applied to a faint \lae. We observe that the narrowband image is too noisy to capture the \lya\ flux and that the \origin\ weighted extraction performs better. We stress that in most cases, however, the different extractions methods give very similar results.

For each source, we assigned a reference extraction. In general, we favored \origin\ or \nbext\ for \laes\ and \odhin\ for low-$z$ sources when the continuum was not too faint. In some cases where the contamination is strong, \odhin\ was preferred for \laes, even if it misses some \lya\ flux. The selection was made on a case-by-case basis. However, the reference extraction may not be the best for a given science case and so we provide all extractions in the delivered data products (Appendix~\ref{sec:sources}).

\subsubsection{Crossmatch with HST source catalogs}
\label{sec:crosshst}

During the inspection step (Sect.~\ref{sec:inspection}) the \origin\ sources were matched to the HST R15 catalog. However, since the publication of \cite{Rafelski2015}, new HST source catalogs in the CANDELS area have been published. These catalogs are based on an improved data reduction of HST photometry \citep{Whitaker2019}, additional HST GRISM information as in \tdhst\ \citep{Skelton2014}, or additional ground-based  photometric bands \citep{Merlin2021}. In addition, the detection image\footnote{The \cite{Rafelski2015} detection image is the average of the four optical ACS images with the four WFC3 images. 
\cite{Whitaker2019} perform a noise-equalized combination of one ACS (F850LP) with three WFC3 bands (F125W, F140W, F160W). The \cite{Merlin2021} catalog is based on the \cite{Guo2013} catalog, which uses the F160W as detection image.}
 and the segmentation input parameters were different, resulting in different segmentation maps. The biggest difference is for closely packed faint galaxies, which can be merged in a single source in some catalogs, whereas they are split into subcomponents in others. In our investigation, we found many cases where the HST source should be split into two different sources (e.g., Fig.~\ref{fig:hst_missed}) or, on the contrary, where two close HST sources should be considered as a single one. 

In this final step, we therefore crossmatched our results with the \candels\  \citep{Whitaker2019}, \tdhst\ \citep{Skelton2014} and \astrod\ \citep{Merlin2021} catalogs. These catalogs and their corresponding segmentation maps were also used to produce the corresponding \odhin\ extraction (Sect.~\ref{sec:odhinmain}).

For all sources, we calculated an HST environment index to quantify whether the source can be considered  spatially isolated or has close or distant neighbors. This index is the percentage of the area covered by neighboring sources measured in the R15 segmentation map\footnote{We use the \candels\ segmentation map when the HST source is not present in R15.} in a given aperture centered on the MUSE source location. We used aperture radii of 0.5 and 1.0 arcsec to define the close and distant environment index, respectively. At the MUSE spatial resolution, an object with a nonzero close environment index will overlap with its neighbors, at least partially, while an object with zero close environment index and nonzero distant environment index will be resolved by MUSE.
A source is considered isolated if its distant environment index is zero.

The presence or absence of close or distant neighbors is coded in the IFLAG column of the main catalog (Sect.~\ref{sec:catalogs}). It is defined as follows:
(i) Sources with a close HST environment index of zero and a distant environment index of less than 5\% are considered isolated with respect to HST. If, in addition, there is no other non-HST DR2 sources within 1 arcsec radius, they are defined as fully isolated. Their isolation indicator ICONF is set to three. (ii) Sources with $\rm ICONF = 2$ have only distant neighbors, either from HST or from DR2 non-HST sources. (iii) All sources with at least one close HST or non-HST neighbor have ICONF set to one.

To assess the quality of the provided matching information, we assigned a matching confidence indicator (hereafter MCONF) to each source.  Its values are given in Table~\ref{tab:mconf} and range from 0 (ambiguous) to 3 (high confidence). MCONF combines the neighborhood information as defined above with additional redshift and narrowband offset information.

An HST source is matched in redshift with a MUSE source when its photometric redshift is considered reliable and the MUSE redshift is within the 95\% confidence interval, that is,

\begin{equation}
\label{eq:photoz}
\frac{z_{max} - z_{min}}{1 + z_{best}} < 0.22 \\
 z_{min} < z_{MUSE} < z_{max}
,\end{equation}where $ z_{best}, z_{min}, z_{max}$ are respectively the photometric redshift solution and its bounds given by the 95\% error probability. In the case of \tdhst\ spectroscopic redshifts we use  $ | z_{MUSE} - z_{spec} | / (1 + z_{MUSE})  < 0.01$. 

The last criterion is the offset between the MUSE absorption or emission composite narrowband (Sect.~\ref{sec:narrowband}) from the center of the matched HST source. This offset is compared to the size of the HST source and is considered acceptable when it satisfies the following condition: $ \rm offset < \max{( 2 \times size , 0.4\arcsec) }$ where size is the radius of the  circular HST aperture  (in arcsec) enclosing half of the total flux, as given in the catalog. 

All isolated sources that do not have an HST counterpart in any catalog are assigned a high matching confidence ($\rm MCONF = 3$). There are 99 sources in this category. All isolated sources matched to a single HST counterpart are also assigned a MCONF value of three if all reliable photometric redshifts of the two used HST catalogs can be matched to the MUSE redshift and if there is no significant offset between the HST source and the narrowband location. If the offset is too large and/or there is at least one reliable photo-z discrepant with respect to MUSE, MCONF is set to two.

When sources cannot be considered isolated, in addition to the previous criteria, we search if any neighbor, not already assigned to any MUSE sources, has a reliable photo-z that could be matched to the MUSE redshift. If so, MCONF is lowered to two or one, depending on if the neighboring source is distant or close to the MUSE source. 
Finally, sources with multiple HST counterparts assigned during the manual evaluation are assigned a zero value of MCONF. There are only eight sources in this category.

The counts of the matching confidence are given in Table.~\ref{tab:mconf}. Most MUSE sources have a high HST matching confidence (68\%), either because the MUSE source can be assigned to a unique identifier in the HST catalogs (53\%) or because they are, instead, isolated and have no HST counterpart in the catalog (15\%).

\begin{table*}
\caption{Isolation flag (IFLAG) and HST crossmatch confidence (MCONF) definition and counts.}   
\label{tab:mconf}
\centering
\begin{tabular}{crr||crr}
\hline 
IFLAG & Description & Counts & MCONF & Description & Count\\
\hline 
& & & 0 &  Ambiguous & 8\\ 
1 & Close neighbors & 679 & 1 &  Low confidence & 61\\ 
2 & Distant neighbors & 154 & 2 &  Good confidence & 644\\ 
3 & Isolated & 1388 & 3 & High confidence & 1508\\ 
\hline 
\end{tabular}
\end{table*}

\subsubsection{Systemic redshift estimation for \laes}
\label{sec:zsyslae}

For most \laes\ the measured redshift is based on the peak of the \lya\ line. However, due to the resonant scattering properties of \lya\ photons in the interstellar medium, the \lya\ redshift is systematically different from the systemic redshift \citep[e.g.,][]{Shapley2003, McLinden2011, Rakic2011, Song2014}. Typical velocity offsets are $\approx$ 200 \kms\ for \laes\, with larger values ($\approx$ 500 \kms) for Lyman-break galaxies \citep{Shibuya2014, Muzahid2020}. 

We used the empirical recipes provided by \cite{Verhamme2018} to estimate the velocity offset between the \lya\ and systemic redshift. In their paper, \cite{Verhamme2018} supply two types of corrections. The first correction is based on the separation between the two peaks of the \lya\ line (i.e., the blue bump separation). The center of the two peaks is a good estimate of the systemic redshift with a rest-frame scatter of 53 \kms. The asymmetric double peaked line profile used in \pfit\ (Sect.~\ref{sec:platefit}) provides this value directly. We simply add the 53 \kms\ standard deviation derived by \cite{Verhamme2018} to the redshift error budget in quadrature.

The second type of correction is based on the FWHM of the \lya\ line profile. The rest-frame velocity offset ($\rm \Delta V$ in \kms) is inferred from the following linear relation with the rest frame measured FWHM in \kms\ (uncorrected for the MUSE LSF): $\rm \Delta V = 0.9 \times FWHM - 34 \, \kms$.
It has a larger scatter of 72 \kms\ and therefore we use the first correction when a double \lya\ peak was detected and successfully fitted.

\begin{figure}[htbp]
\begin{center}
\includegraphics[width=0.8\columnwidth]{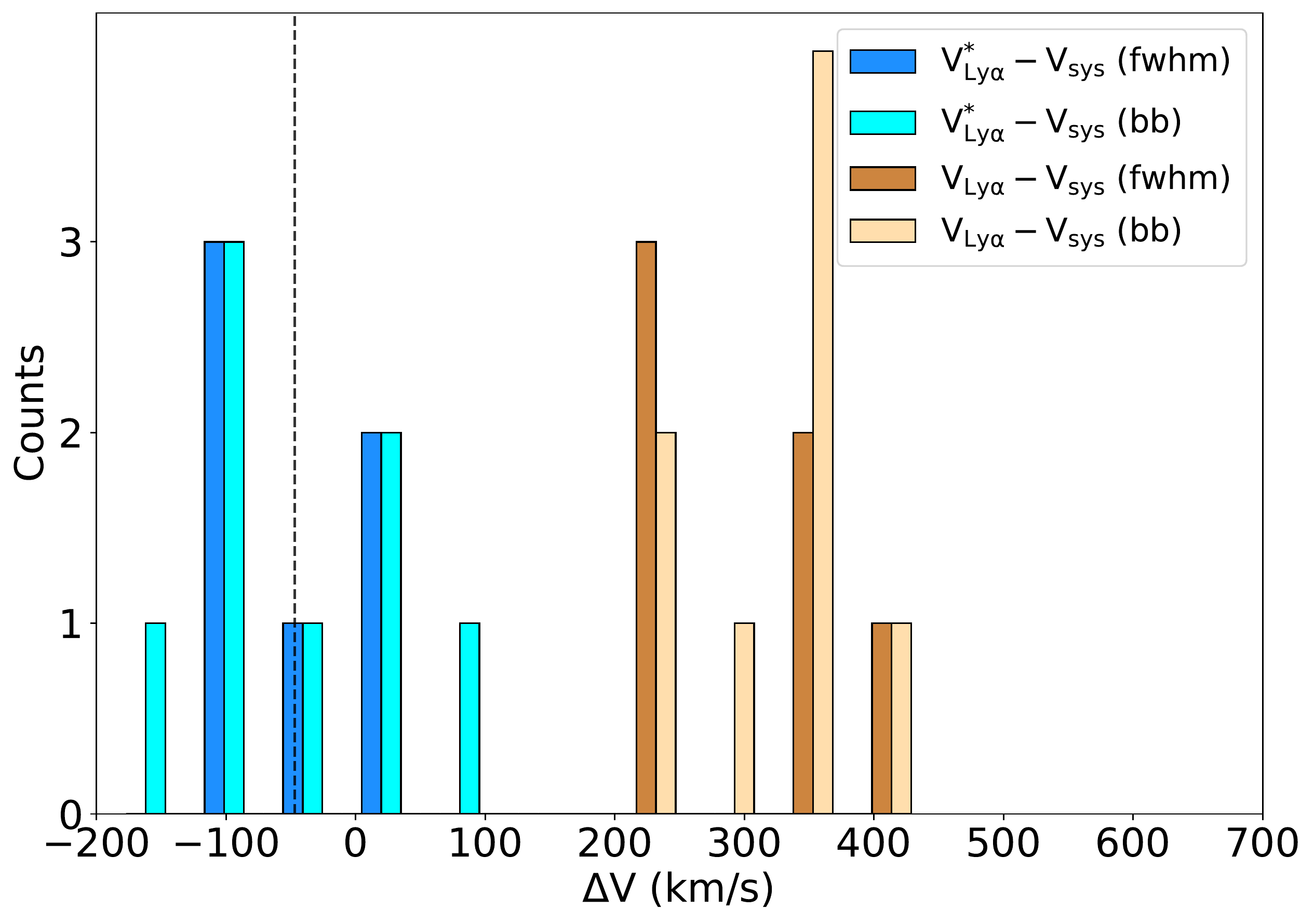}
\caption{Check of systemic redshift correction for a subsample of DR2 \laes\ with systemic redshifts measurements. The histogram of the rest-frame velocity difference between the measured \lya\ red peak location ($\rm V_{\lya}$) and the systemic velocity ($\rm V_{sys}$) is shown in orange. The same histogram after systemic velocity correction ($\rm V_{\lya}^{*}$) is shown in blue. The type of correction, that is, blue bump separation (bb) or \lya\ line FWHM (fwhm), are identified in respectively darker and lighter colors. The dashed vertical line displays the median value of the  velocity offset distribution after correction ($\rm -47 \, \kms$). 
}
\label{fig:zsys}
\end{center}
\end{figure}

We checked the accuracy of the correction by selecting the few \lya\ emitters with high redshift confidence, high S/N \lya\ line and additional high S/N nonresonant emission or absorption lines. This selection resulted in a sample of 14 \laes: six with single \lya\ line profiles and eight with double peaked profiles. Systemic velocities were measured from forbidden emission lines (mostly \ciiid) or strong absorption lines (e.g., \citealt{Boogaard2021}). The comparison of the velocity offsets before and after correction is presented in Fig.~\ref{fig:zsys}. We can see that the systematic velocity offsets of about 300 \kms\ are reduced to a few tens of \kms. The measured scatter (77 \kms\ RMS) is comparable to the values measured by \cite{Verhamme2018}.

\subsubsection{Final catalogs}
In Table~\ref{tab:from} we display the source count by extraction method and data set. We find that the majority of sources (67\%, 1499/2221) detected with \origin\ have also an HST counterpart, but we also observe that the fraction of \origin\ sources without HST counterparts increases with depth, from 13\% (202/1505) in the \mosaic\ to 29\% (203/691) in the \mxdf\ data sets.

When a source is detected in multiple data sets (e.g., \udft\ and \mxdf), we retain the data set with the greatest depth at the source location. The only exception is MID-171, where we selected the \udft\ instead of \mxdf\ data set, despite its lower depth (31 versus 136 hours). The reason for this is the \lya\ line that falls into the \mxdf\ masked wavelength range because of the sodium laser guide star notch filter (Fig.~\ref{fig:specrange}). When available, the other data sets are also recorded in the delivered data product (Appendix~\ref{sec:sources}).

The final catalog contains 2221 sources with redshifts.
The statistics and properties of these sources are discussed in the following sections.

\begin{table}
\caption{Source counts with respect to detection method.}   
\label{tab:from}
\centering
\begin{tabular}{lrrrr}
\hline 
data set & \textsf{ORI}\tablefootmark{a} & \textsf{ODH}\tablefootmark{b} & \textsf{ORI} \& \textsf{ODH} & Total \\ 
\hline 
\mosaic & 202 & 220 & 1083 & 1505 \\ 
\udft & 78 & 48 & 225 & 351 \\ 
\mxdf & 203 & 78 & 410 & 691 \\ 
\drt & 428 & 294 & 1499 & 2221 \\
\hline 
\end{tabular} 
\tablefoot{
\tablefoottext{a}{\origin} 
\tablefoottext{b}{\odhin}
}
\end{table}

\section{Results}
\label{sec:results}

\subsection{Redshifts}
% redshift.tex
\label{sec:redshift}

One of the most visible results of the survey is the number of galaxies for which we were able to assign high quality redshifts. A detailed presentation is given in the following subsections.

\subsubsection{Redshift distribution and confidence}
\label{sec:zstat}

\begin{figure*}[!tbp]
  \begin{subfigure}[b]{0.5\textwidth}
    \includegraphics[width=1\textwidth]{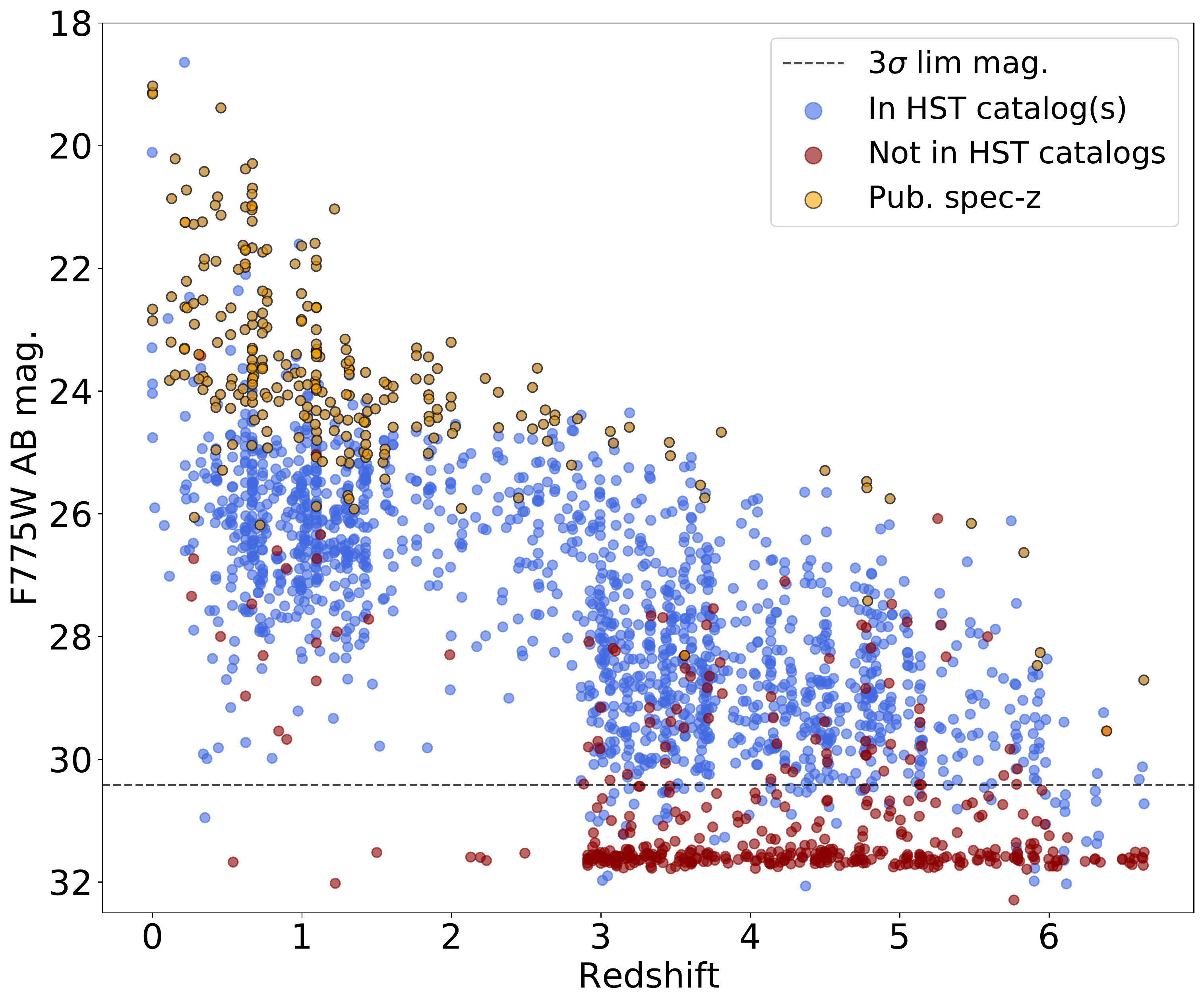} 
    %\caption{Source id-106.}
    \label{fig:zmag}
  \end{subfigure}
  \hfill
  \begin{subfigure}[b]{0.5\textwidth}
        \includegraphics[width=1\textwidth]{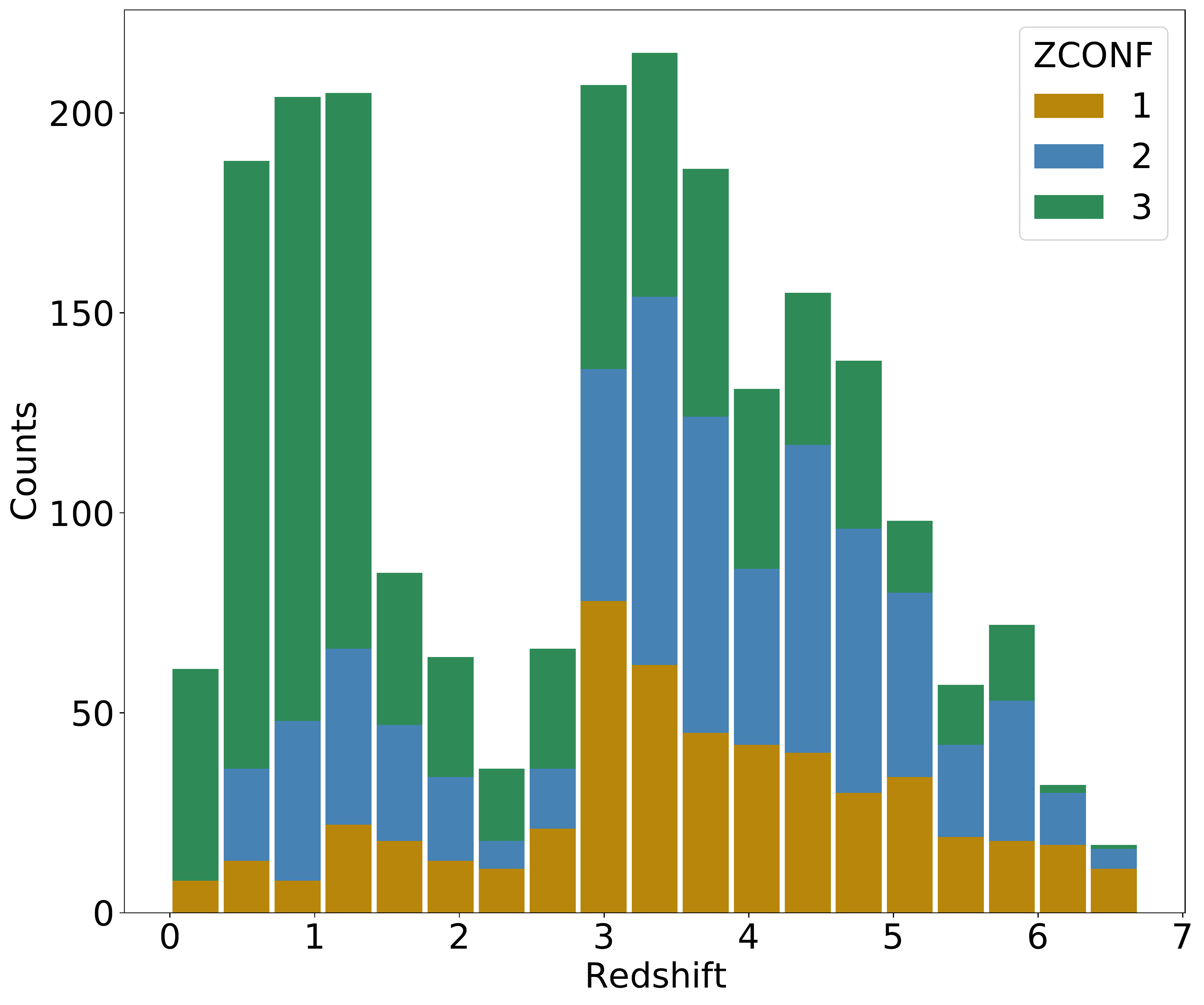}
        %\caption{Source id-7295.}
        \label{fig:zhist}  
  \end{subfigure}
  \caption{Redshift distribution of the MUSE data release II. The magnitude - redshift scatter plot is displayed on the left panel. All MUSE sources with an HST counterpart in at least one of the HST catalogs are shown as blue symbols. We note that objects with low S/N in F775W could still be in HST catalogs if they are detected in another filter (e.g., NIR filters for high-z objects). 
The matched spectroscopic redshifts previously published (Sect.~\ref{sec:specz}) are shown as yellow circles.
For sources without catalog HST counterparts, the brown circles show the 0\farcs{4} aperture magnitude as measured on the HST F775 image (c.f Sect.~\ref{sec:nohst}). 
The horizontal dashed line indicates the median F775W $\rm 3 \sigma$ limiting magnitude.
The redshift histogram, colored by redshift confidence, is presented in the right panel. 
  }
  \label{fig:zstat}
\end{figure*}

\begin{figure*}[htbp]
\begin{center}
\includegraphics[width=0.9\textwidth]{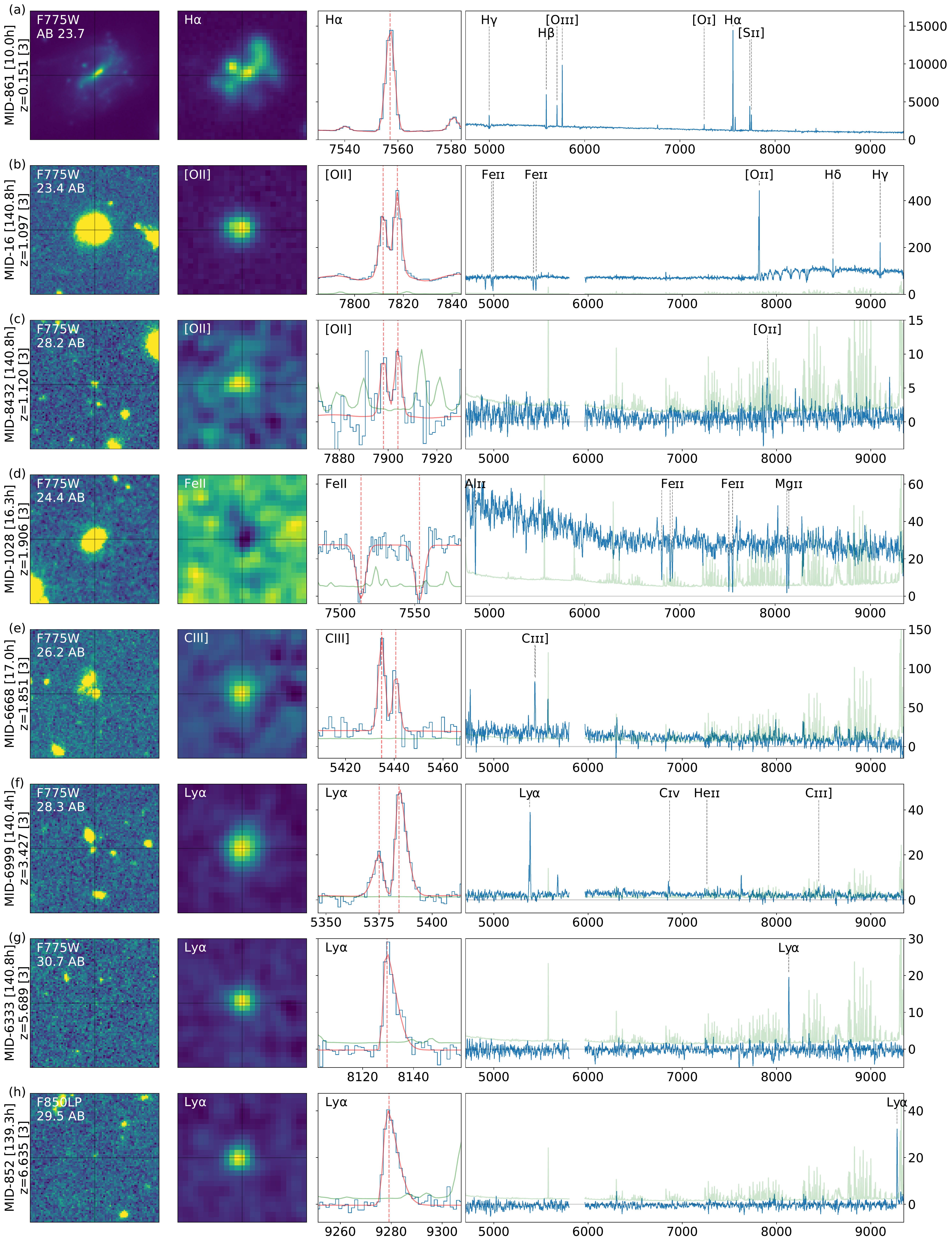}
\caption{Examples of sources. From left to right the columns display (1) the HST image in the F775W or F850LP filter (2) the emission or absorption narrowband for the main line (3) the spectrum zoomed-in on the main line (in blue) and its fit (in red) (4) the full spectrum (in blue). The green curves show the noise standard deviation. Except for the first two sources, the narrowband images and the full spectra have been filtered with respectively a Gaussian of 0.5\arcsec\ FWHM and a box filter of 5 pixels. Image sizes are $5\arcsec \times 5\arcsec$. Observed wavelengths are in \AA\ and the flux unit is \ergsa{-20}. For each source the MUSE identifier, depth in hours, redshift value and confidence, and AB HST magnitude are indicated.
}
\label{fig:ex_gal}
\end{center}
\end{figure*}

The distribution of redshifts is shown in Fig.~\ref{fig:zstat}. Table~\ref{tab:ztable} presents the breakdown of the 2221 redshifts in five categories. Given the small area (8.3 arcmin$^2$) and the Galactic pole location, we identified only eight stars, mostly M types. Likewise, the number of nearby galaxies ($\rm z < 0.25$) is limited (25) in this small volume (280 \vmpc). For this latter  category, dedicated wide field spectroscopic surveys (e.g., SDSS \citealt{Blanton2017} and MANGA \citealt{Bundy2015}) are much more efficient. Nevertheless, for these few bright galaxies (median AB F775W of 23.3), we obtain spatially resolved spectroscopy with exquisite S/N up to large galactocentric distance. An example is given in Fig.~\ref{fig:ex_gal} row a.

In the redshift range 0.25--1.5, the \oiid\ doublet falls in the MUSE wavelength range. At the MUSE spectral resolution ($\rm R \approx 3000$), the \oii\ doublet is resolved in most cases, making line identification easier. With 677 galaxies, this population is the second most populated category, after the \laes. These galaxies cover a broad range in F775W AB magnitude with 23.3, 25.7 and 27.6 for the 10\%, 50\% and 90\% percentiles, respectively. Examples are given in Fig.~\ref{fig:ex_gal}, rows b and c. We note that a small fraction  (5\%) of the galaxies in this wavelength range are not star forming \oii\ emitters, but passive galaxies detected through their absorption lines. 

The redshift range 1.5--2.8 corresponds to the MUSE redshift desert, a region where \oii\ is redshifted out of the MUSE wavelength range and \lya\ is not yet observable. The number of galaxies drops to 201. In this redshift range, galaxies are detected either through their UV absorption lines (50\%) or by fainter emission lines, mainly the \ciiid\ or \mgiid\ doublets (see examples in Fig.~\ref{fig:ex_gal}, rows d and e). 

The last category comprises the \laes, covering the redshift range 2.8--6.7. With 1308 \laes\ or 60\% of the complete sample, it is the most numerous and also the faintest population (29.4 median AB F775W). The vast majority of these galaxies (98\%) are identified by their high equivalent width asymmetric \lya\ emission line but a few galaxies (3\%) exhibit in addition fainter emission lines such as \civd\ or \ciiid. A small percentage (2\%) are \lya\ absorbers with broad absorption and little or no detectable \lya\ emission. We note that a significant fraction of the identified \laes\ have no HST counterpart (see Sect~\ref{sec:nohst}). A few examples of \laes\ are displayed in Fig.~\ref{fig:ex_gal}, rows f to h.

It can be seen in Table~\ref{tab:ztable} and in the right panel of Fig.~\ref{fig:zstat} that the fraction of good to high redshift confidence ($\rm ZCONF > 1$) is high (87\%) for the low and intermediate redshift populations (i.e., $z < 2.8$). This ratio is slightly lower (70\%) for the \laes\ for the reason that in most cases the \lya\ line is the only line identified. 

\begin{figure}[htbp]
\begin{center}
\includegraphics[width=1\columnwidth]{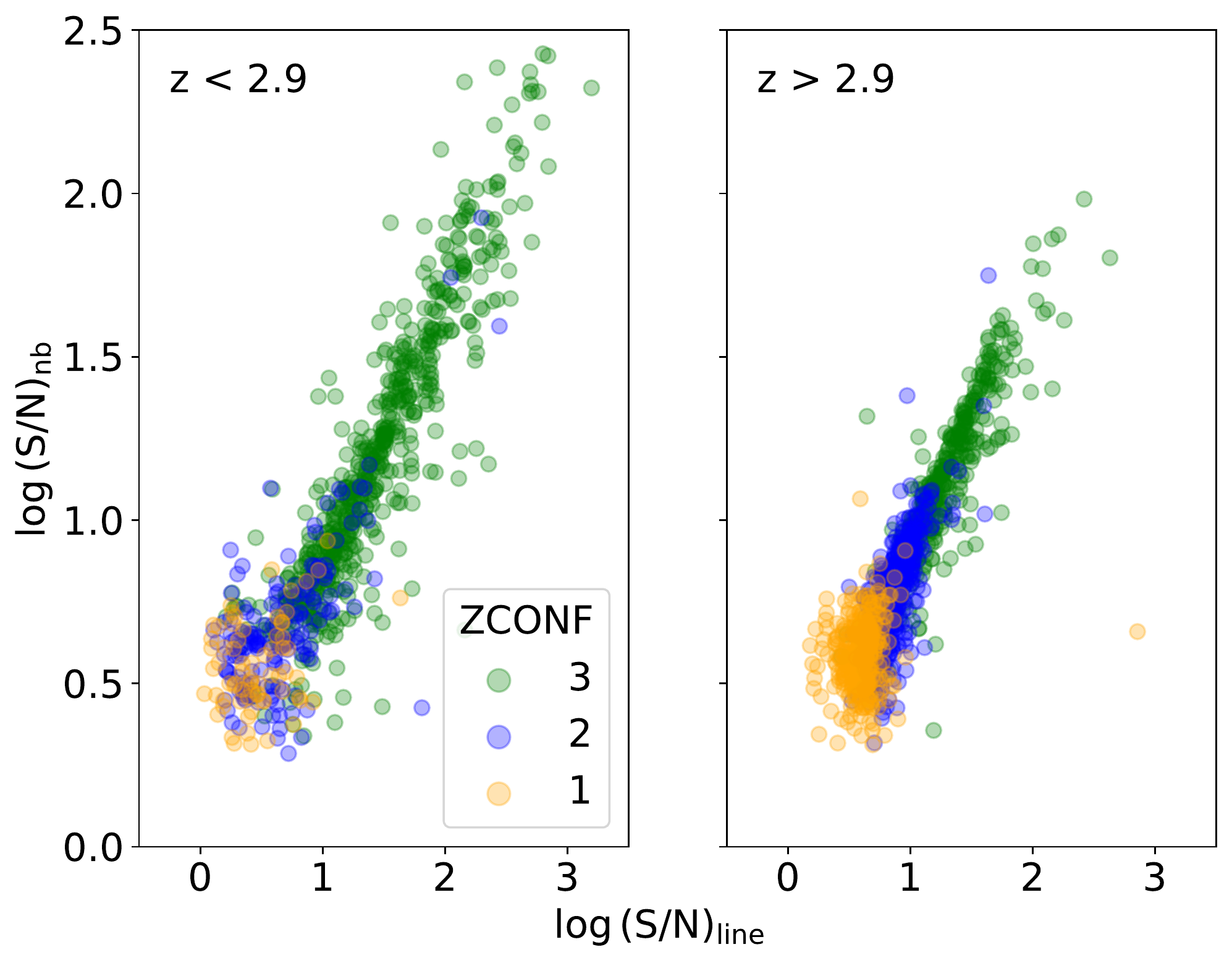}
\caption{S/N estimates as a function of redshift confidence. The log of the S/N emission or absorption composite narrowband is shown as a function of the log of the leading line S/N derived from \pfit. 
Low (1), good (2) and high (3) redshift confidences are shown in respectively orange, blue and green colors, while  low-z ($z<2.9$) and high-z ($z>2.9$) sources are displayed in the left and right panels, respectively.
%\todo{Jarle: Given the shape of the distributions might it be less confusing to have two adjacent panels, one for low and one for high?}\ans{Roland: done}
}
\label{fig:zconf}
\end{center}
\end{figure}

In Sect.~\ref{sec:zconf} we presented the redshift confidence definition and assessment. Although it is a complex process based on multiple parameters and data sets, including the expert's judgement, emission line S/N is one of the primary factors in the decision process. This can be seen in Fig.~\ref{fig:zconf} where we show the confidence values as a function of the main emission line S/N derived from the extracted spectra and the corresponding narrowband S/N. 

\begin{figure}[htbp]
\begin{center}
\includegraphics[width=1\columnwidth]{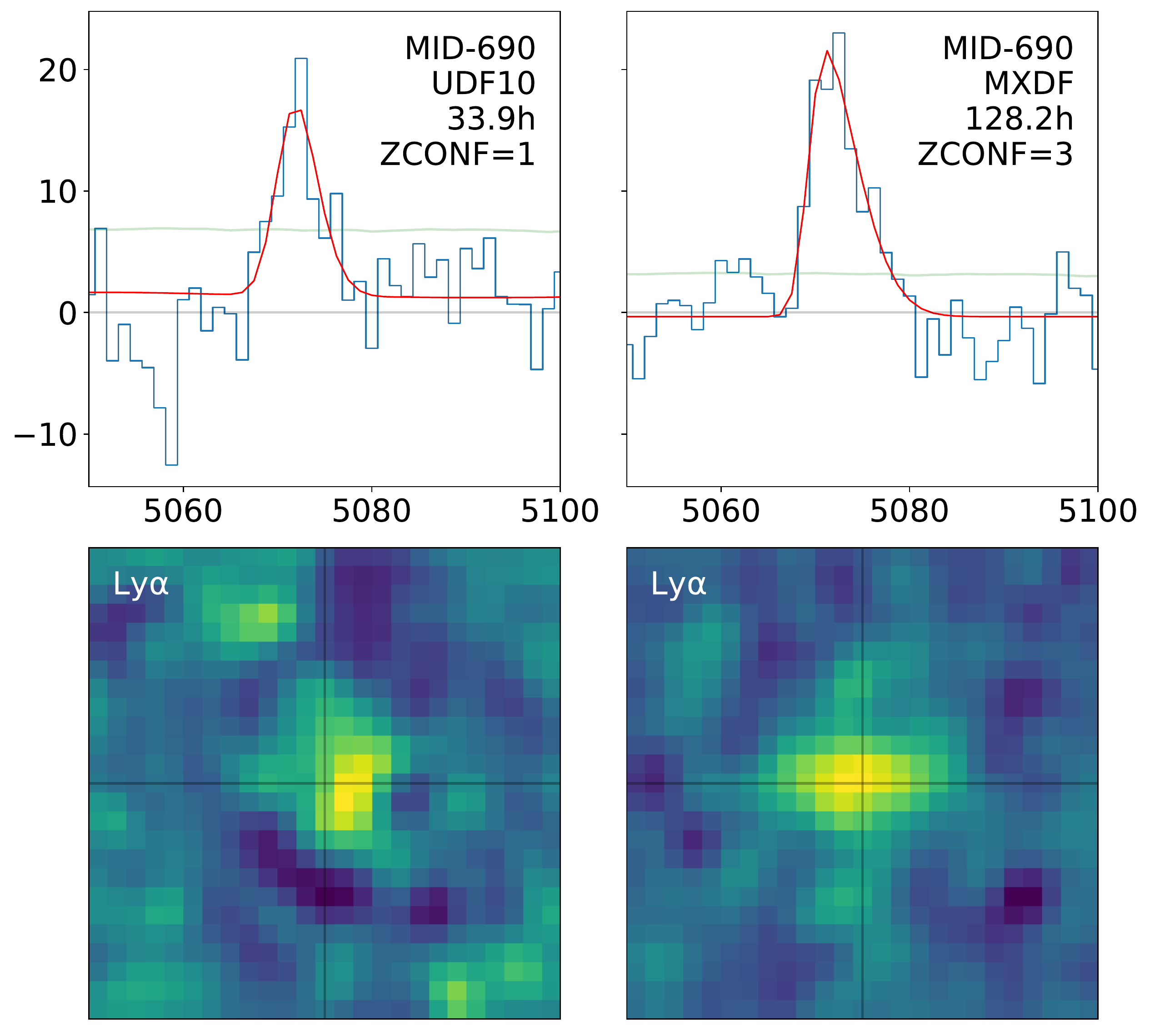}
\caption{Impact of depth on the redshift confidence assignment. The \lya\ line (data in blue, fit in red) and the corresponding narrowband image of the \lae\ MID-690 are shown for the \udft\ (left column) and the \mxdf\ (right column) data sets. The green curves display the noise standard deviation. The narrow bands have been smoothed with a Gaussian of 0.5\arcsec\ FWHM. Image sizes are $5\arcsec \times 5\arcsec$. Observed wavelength and flux units are respectively \AA\ and \ergsa{-20}.
}
\label{fig:zconfcomp}
\end{center}
\end{figure}

We point out that sources with low redshift confidence ($\rm ZCONF = 1$) have nevertheless a clear detection signal, but in general the S/N is too low to assign a redshift with high confidence. An example is given in Fig.~\ref{fig:zconfcomp} where the same source is shown for the \udft\ and \mxdf\ data sets with 34- and 128-hour depths, respectively. In the \udft\ data set (upper left panel in Fig.~\ref{fig:zconfcomp}), the emission line is detected with an S/N of 5.3 and does not appear to be very asymmetric ($\rm \gamma = 1.5 \pm 2.6$). The corresponding narrowband (lower left panel) shows a clear peak, although with a low S/N of 3.5. The measured redshift is 3.17, which is not far from the R15 photometric redshift of 2.74 estimated for the matched HST source. We note, however, that the photometric redshift ($\rm zp$) measurement is not very reliable with $\rm (zp_{max} - zp_{min})/(1+zp) =  0.27$. Taking all these points into consideration, the expert gave a low redshift confidence  to the source. The same source with the deepest \mxdf\ observations (right column in Fig.~\ref{fig:zconfcomp}) obtains a higher S/N of 10.7 and 7.8 for the emission line and the corresponding narrowband, respectively. The line profile is now clearly asymmetric with $\rm \gamma = 3.6 \pm 2.6$. Given these measurements, a high redshift confidence ($\rm ZCONF = 3$) was given to the source.

\begin{table}
\caption{Distribution of redshifts by confidence and for the different classes of sources.}
\label{tab:ztable}
\begin{tabular}{lcrrrrr}
\hline
Class & \multicolumn{5}{c}{ZCONF} \\
 & 1 & 2 & 3 & \multicolumn{2}{c}{1-3} \\
\hline
%\begin{table}
%\begin{tabular}{cccccc}
%type & zc1 & zc2 & zc3 & zcall & zcfrac \\
Stars & 0 & 0 & 8 & 8 & 0.4\% \\
Nearby ($\rm z<0.25$) & 4 & 0 & 21 & 25 & 1.1\% \\
\oii\ ($\rm 0.25 < z < 1.5$) & 55 & 122 & 500 & 677 & 30.5\% \\
Desert ($\rm 1.5 < z < 2.8$) & 55 & 57 & 89 & 201 & 9.0\% \\
\lya\ ($\rm z > 2.8$) & 396 & 538 & 374 & 1308 & 58.9\% \\
All & 510 & 717 & 994 & 2221 & 100.0\% \\
\hline
%\end{tabular}
%\end{table}
%\hline
\end{tabular}
\tablefoot{
ZCONF range from 1 (less secure redshift) to 3 (most secure redshift); see Sect.~\ref{sec:zconf}.
}
\end{table}

\subsubsection{Comparison with DR1}
\label{sec:dr1}

We recall that the DR1 catalog \citep{Inami2017} was based on an early data reduction of the \udft\ and \mosaic\ data sets and did not include the new  \mxdf\ data set. Furthermore, although the overall detection and inspection process was similar, there are a few important differences. First, the data reduction was significantly improved with the use of super-flatfield, improved sky-subtraction and self-calibration (Appendix~\ref{sec:datared}). Second, the HST basic aperture extraction was replaced by the de-blending \odhin\ procedure (Sect.~\ref{sec:odhinmain}). Third, we use new templates for \pymarz\ (Sect.~\ref{sec:marz}) and the improved line fitting software \pfit\ (Sect.~\ref{sec:platefit}). So we expect some differences between DR1 and the new catalog, even in the area outside the \mxdf\ footprint.
A summary of the differences is given in Tables~\ref{tab:dr1_unmatch} and \ref{tab:dr2_new}.

The DR1 catalog contains 1574 sources. While most of the DR1 sources are present in the new catalog, a small fraction (9\%) were not confirmed. The reasons for the failure are given in Tab.~\ref{tab:dr1_unmatch}. The rejected sample is mainly composed of low-confidence DR1 sources that were rejected or assigned a different redshift in the new deeper (\mxdf) or improved (\mosaic, \udft) data sets. For traceability purposes, new MUSE identifiers were given to DR1 sources that had been assigned a different redshift. %The detailed list of DR1 rejected sources is given in appendix~\todo{\ref{subsec:dr1}}. 

\begin{figure}[htbp]
\begin{center}
\includegraphics[width=1\columnwidth]{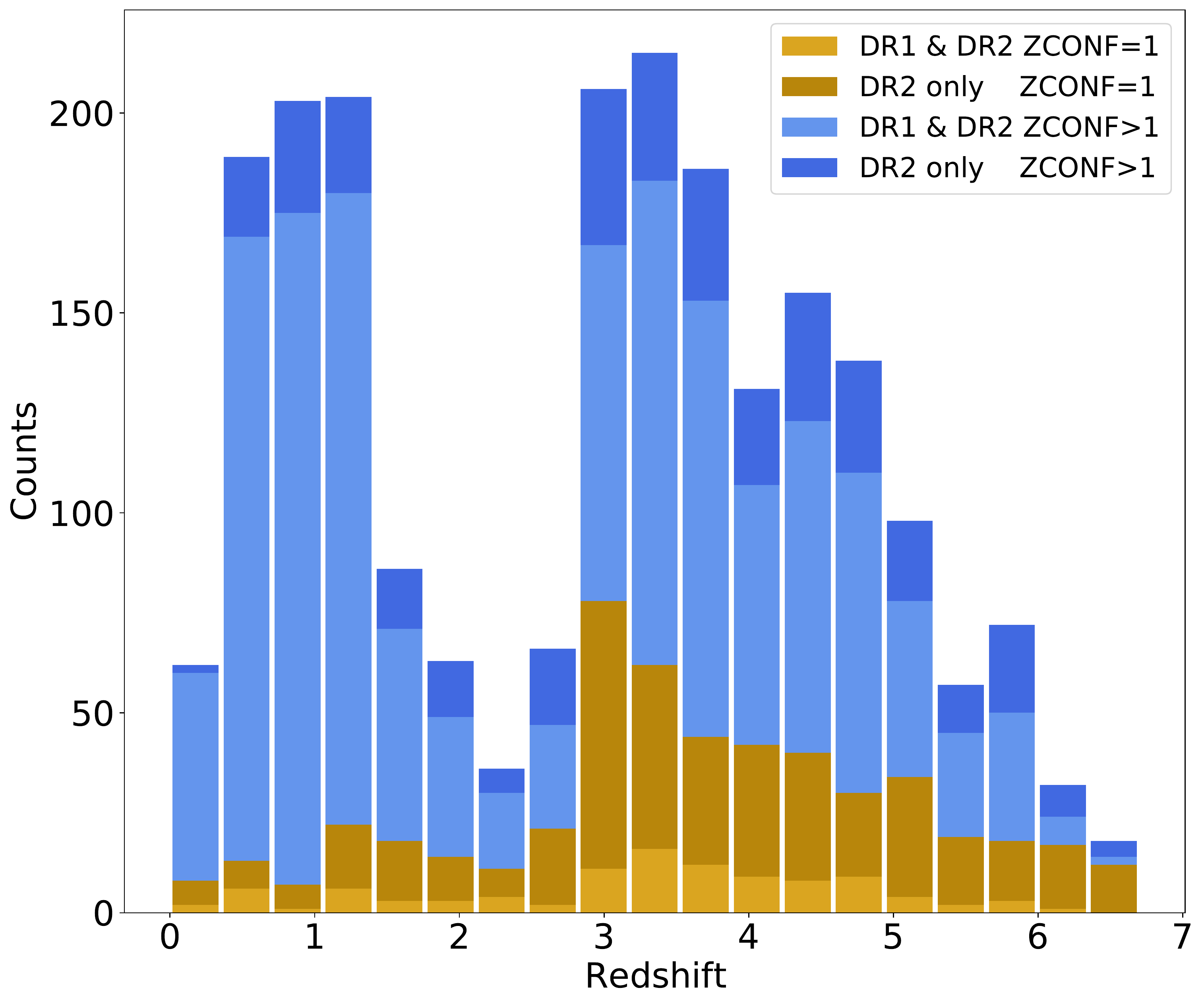}
\caption{Comparison of redshift statistics between the DR1 and DR2 data releases. Light and dark colors are for low ($\rm ZCONF=1$) and high ($\rm ZCONF>1$) redshift confidence, repectively. 
}
\label{fig:dr1}
\end{center}
\end{figure}

In Fig.~\ref{fig:dr1} we display the quantitative improvement between DR1 and DR2.
As shown in Tab.~\ref{tab:dr2_new}, the majority of the 792 new DR2 sources (53\%) are provided by the \mxdf\ data set. This was expected given the improved depth, that is, a factor of 12 and 4 with respect to \mosaic\ and \udft, respectively. However, there is still a significant gain of about  25\% for the \udft\ and \mosaic\ data sets.  This demonstrates that the effort spent on improving the entire process (i.e., data reduction, source detection and classification) has paid off.

\begin{table}
\caption{Comparison of DR2 and DR1 catalogs.}
\subcaption{DR1 sources not matched in DR2}   
\label{tab:dr1_unmatch}
\centering
\begin{tabular}{lrrrrrr}
\hline 
Reason & \multicolumn{2}{c}{$\rm ZCONF=1$} & \multicolumn{2}{c}{$\rm ZCONF>1$} & \multicolumn{2}{c}{Total} \\ 
\hline 
New Z & 25 & 1.6\% & 5 & 0.3\% & 30 & 1.9\% \\ 
No Z & 88 & 5.6\% & 17 & 1.1\% & 105 & 6.7\% \\ 
Duplicate & 1 & 0.1\% & 5 & 0.3\% & 6 & 0.4\% \\ 
New Match & 1 & 0.1\% & 0 & 0.0\% & 1 & 0.1\% \\ 
All & 115 & 7.3\% & 27 & 1.7\% & 142 & 9.0\% \\ 
\hline 
\end{tabular} 
\subcaption{New DR2 sources with respect to DR1} 
\label{tab:dr2_new}  
\begin{tabular}{lrrrrrr}
\hline 
data set & \multicolumn{2}{c}{$\rm ZCONF=1$} & \multicolumn{2}{c}{$\rm ZCONF>1$} & \multicolumn{2}{c}{Total} \\ 
\hline 
\mosaic & 214 & 15.6\% & 110 & 8.0\% & 324 & 23.6\% \\
\udft & 35 & 19\% & 17 & 9.2\% & 52 & 28.3\% \\
\mxdf & 159 & 23.9\% & 257 & 38.7\%  & 416 & 62.7\% \\
All & 408 & 18.4\% & 384 & 17.3\% & 792 & 35.7\% \\
\hline 
\end{tabular} 
\end{table}

\subsubsection{Comparison with published spectroscopic redshifts}
\label{sec:specz}
We also compared our redshifts with the compilation of all published spectroscopic redshifts in the CDFS given by \cite{Merlin2021} prior to 2021\footnote{We note that the MUSE DR1 redshifts published in \cite{Inami2017} were not included in the \cite{Merlin2021} compilation.}. The compilation has 4951 entries, of which 292 are within the area of MUSE HUDF, which is 14 times smaller than the CDFS area.

A large majority (255 or 87\%) of the published spectroscopic redshifts were successfully matched to MUSE redshifts, with the exception of 26 sources (9\%) that were assigned a different redshift. Excluding the three galaxies with a low MUSE redshift confidence (MID-7951, MID-8025 and MID-8038) for which the proposed published solution is a viable alternative, we are confident that the MUSE assigned redshift is the correct solution. In addition, 11 galaxies (4\%) of the published sample did not get any redshift in MUSE. We verify the proposed redshift solution and confirm the absence of clear features in the MUSE wavelength range.
For these galaxies, the main feature (e.g., \lya\ emission or absorption, \oiiid\ emission) lies outside the MUSE wavelength range. 
%Details of the discrepant redshifts is given in \todo{Annex, Table~\ref{tab:bad_specz}}.

The scaled deviation of the published redshift with respect to MUSE, that is, $\rm \Delta z = (z_{MUSE} - z_{Pub})/(1+z_{MUSE})$, has a standard deviation of $\rm \Delta z = 0.014$ after clipping 25 sources at $5 \sigma$.
We note that the redshift errors reported in the literature are often greatly underestimated: on average, they are 6 times smaller than the estimated error based on the MUSE reference redshifts. 
 
%In Fig.~\ref{fig:zspec_comp}, we plot $\rm \Delta z = (z_{MUSE} - z_{Pub})/(1+z_{MUSE})$, the scaled deviation of the published redshift with respect to MUSE. The standard deviation of $\rm \Delta z$ is 0.014 after clipping out 25 sources at $5 \sigma$.

\begin{figure}[htbp]
\begin{center}
\includegraphics[width=1\columnwidth]{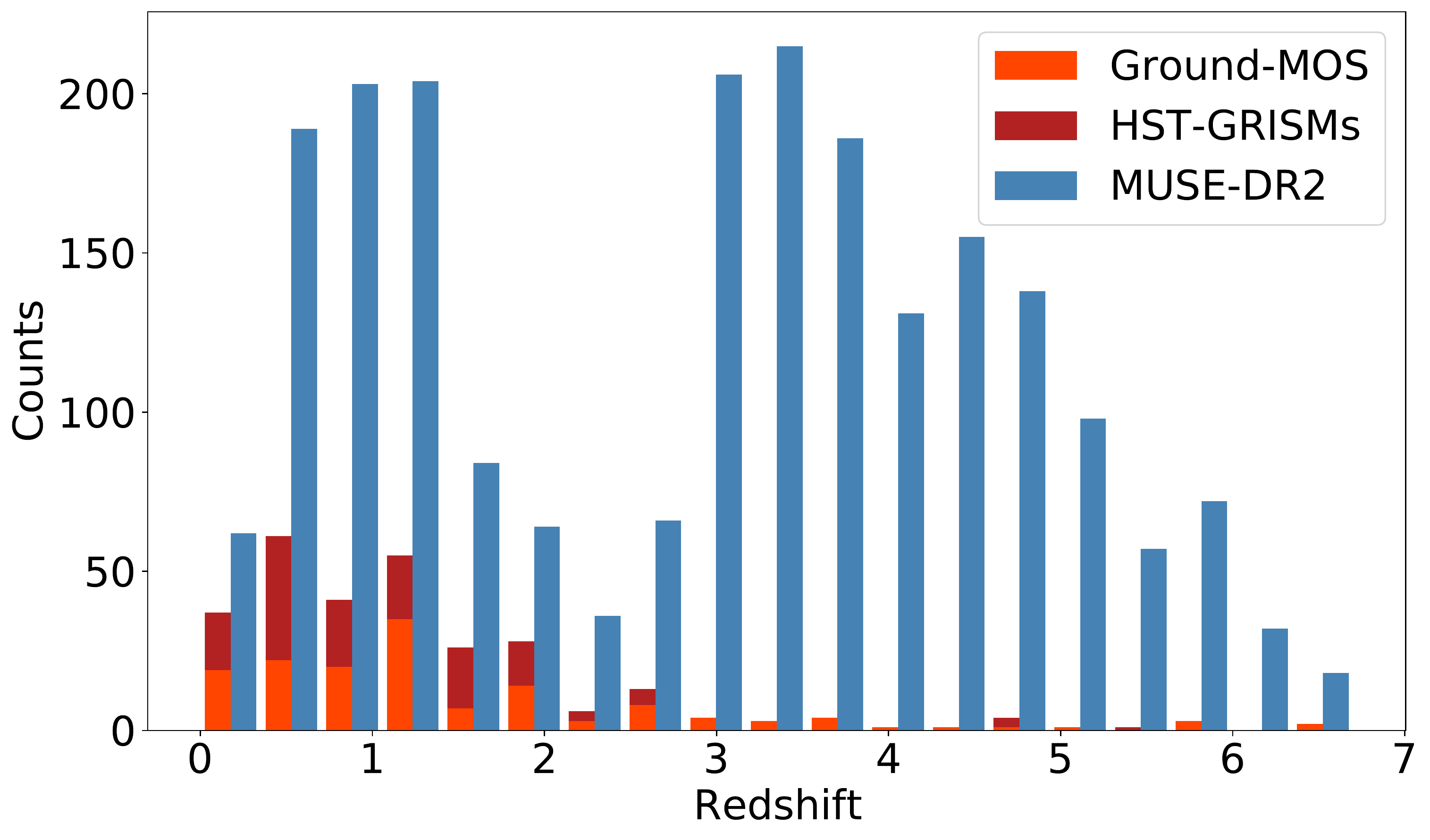}
\caption{Comparison of the spectroscopic redshift distributions of ground-based MOS and HST slitless GRISM published spectroscopic surveys in the HUDF area and the MUSE DR2 redshifts.
}
\label{fig:zspec}
\end{center}
\end{figure}

Just looking at the raw numbers (i.e., 2221 versus 292 spectroscopic redshifts), one can appreciate the dramatic progress achieved by MUSE. As shown in Fig.~\ref{fig:zspec}, the difference is even more striking at high redshift ($z > 2.9$) where only 20 sources out of 1291 have previously been identified. The same holds for faint magnitudes:  most of the published spectroscopic redshifts are from galaxies brighter than AB 24 in F775W, while the magnitudes of MUSE sources extend to AB 30 and beyond (Fig.~\ref{fig:zstat} left panel).

\subsubsection{Comparison with photometric redshifts}
\label{sec:photz}

A comprehensive comparison of photometric redshifts with MUSE DR1 redshifts was performed by \cite{Brinchmann2017}. Photometric redshifts are essential ingredients for on-going and upcoming cosmological surveys (e.g., KiDS, \citealt{Hildebrandt2017}; LSST, \citealt{Ivezic2019}; \textit{Euclid}, \citealt{Laureijs2011}; Roman, \citealt{Spergel2015}). 
Some surveys have very stringent requirement in photo-z accuracy; for example, for weak lensing, the mean redshift in each redshift bin must be constrained to better than $2 \times 10^{-3} (1+z)$ \citep{Newman2015}.
The goal of the \cite{Brinchmann2017} study was to test the performance of photometric redshifts for galaxies in the \textit{Hubble} Ultra Deep Field down to the 30th magnitude. 

One of the main conclusions of this study was the existence of systematic biases at intermediate ($z = 0.4-1.5$) and high redshift ($z > 3$), and a $\approx$10\%  proportion of outliers. We do not intend to repeat such a detailed analysis here, but we are interested in whether these conclusions based on DR1 are still valid for the more complete DR2 catalog.  

For this comparison we use three different photometric redshifts catalogs. The first catalog is from \cite{Rafelski2015} (hereafter BPZ-R15), which was already used in \cite{Brinchmann2017}.  We use the Bayesian photometric redshift values derived with the \textsf{BPZ} code \citep{Benitez2000}. The second catalog contains the 3D-HST redshifts from \cite{Momcheva2016}. These redshifts were computed from a simultaneous fit of multiband HST photometry and 2D spectra extracted from the HST WFC3 G141 GRISM. Although it is not a pure photometric redshift catalog, we included it in the comparison given that the additional GRISM data should provide improved redshifts with respect to purely photometric redshifts. We note, however, this is only true in the wavelength range covered by G141 and only for specific redshifts for which spectral features (especially breaks) are sampled by the G141 wavelength range. It is also restricted to object bright enough (i.e., with  F140 JH magnitudes brighter than 26). 
The last one is the recent \cite{Merlin2021} ASTRODEEP-GS43 spectro-photometric catalog. It is based on HST CANDELS data complemented with additional ground-based photometric data, reaching a total of 43 different photometric wide and narrow bands. Their photometric redshifts are derived with three different codes (\textsf{LePhare}; \citealt{Ilbert2006}, \textsf{EAzY}; \citealt{Brammer2008}, \textsf{z-phot}; \citealt{Fontana2000}), using the median of the three derived values. 

The input catalog is composed of 1711 sources with good and high confidence redshifts ($\rm ZCONF > 1$). The BPZ-R15, 3D-HST, and ASTRODEEP catalogs have respectively 1404, 1088, and 1009 matched sources. As in \cite{Brinchmann2017}, we compute the normalized redshift error as 
\begin{equation}
\Delta z = \frac{z_{\rm MUSE} - z_{\rm phot}}{1+z_{\rm MUSE}}.
\end{equation} 
We evaluated the evolution of $\Delta z$ with redshift and magnitude by computing the median error and the outlier fraction in bins of about 100 elements. We note that the distributions of redshift and magnitude are different for the three catalogs. For example, BPZ-R15 has more sources than the other two catalogs, it extends to fainter magnitudes and has a larger number of sources at high-z. Consequently, the redshift and magnitude bins  differ between catalogs. In each bin, the outlier fraction is computed as the fraction of sources with $\Delta z > n \, \sigma_{MAD}$ where $\sigma_{MAD}$ is the median absolute deviation around the median, adjusted to match a standard deviation for a Gaussian distribution (Eq.~2 of \citealt{Brinchmann2017}). A value of $n = 3$ is used for simple outliers and $n = 10$ for catastrophic outliers.

\begin{figure}[htbp]
\begin{center}
\includegraphics[width=1\columnwidth]{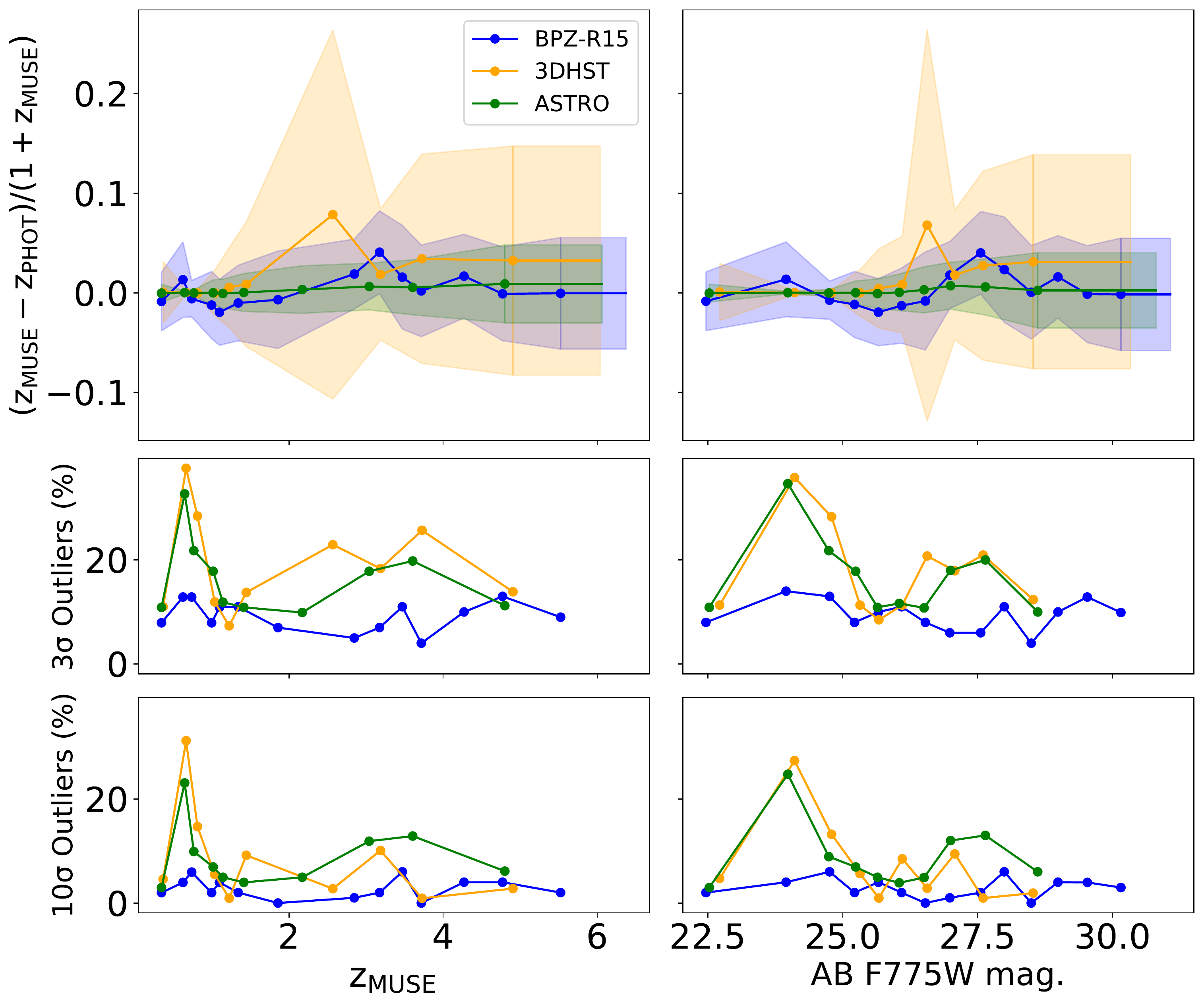}
\caption{Redshift offﬀset between spectroscopic redshifts from MUSE and the photometric redshifts from \cite{Rafelski2015} BPZ-R15 (blue), \cite{Momcheva2016} 3D-HST (orange) and \cite{Merlin2021} ASTRODEEP (green) catalogs, all normalized by $\rm 1 + z_{MUSE}$. The top left panel show the median (solid line) as a function of redshift and the scatter on the median is shown by the shaded areas. The middle and bottom left panels show the fraction of outliers at respectively $\rm 3 \sigma$ and $\rm 10 \sigma$. The right panels display the same quantities as a function of HST F775W AB magnitude. 
}
\label{fig:zphot_comp}
\end{center}
\end{figure}

The results are shown in Fig.~\ref{fig:zphot_comp}. The BPZ-R15 curve (blue curve in top left panel) displays the same behavior with redshift as observed in DR1 (Fig.~5 left panel of \citealt{Brinchmann2017}), with a systematic negative offset ($\rm \Delta z \approx -0.03$) at $\rm z \approx 1.2$ and a positive offset ($\rm \Delta z \approx +0.05$) at $\rm z \approx 3$. We note, however, that the increased offset at $\rm z > 4$ observed in DR1 is not observed in our larger sample. The median standard deviation measured in BPZ-R15 is 0.04, only slightly higher than the DR1 estimation. The evolution as a function of magnitude (blue curve and shaded area in the top right panel) displays a similar trend as for DR1: the standard deviation increases from 0.03 at $\rm AB < 25$ to 0.06 at $\rm AB > 28$. The observed fraction of outliers is 9.4\% and 2.8\% for the catastrophic outliers. These figures are consistent with the DR1 measurements.

The comparison with the ASTRODEEP photometric redshifts is shown in green in Fig.~\ref{fig:zphot_comp}. The additional photometric bands have improved the accuracy of the photometric redshifts. There are almost no detectable systematic offset with either redshift or magnitude. The scatter is also much reduced compared to BPZ-R15. It is less than 0.02 at AB 26 and then increases to 0.04 at AB 28.5. We observe, however, that the average fraction of outliers increases,  with 16\% and 9\%  for the $\rm 3 \sigma$ and $\rm 10 \sigma$  outliers, respectively. We note that these numbers (scatter and outlier fraction) are higher than those quoted by \cite{Merlin2021}. Their underestimation probably results from the limited spectroscopic sample used (see Sect.~\ref{sec:specz}) in their tests.

Surprisingly, the comparison with the 3D-HST catalog (orange curve and shaded area in Fig.~\ref{fig:zphot_comp}) reveals a net degradation of performance  at $\rm AB > 26$. We note that the \cite{Momcheva2016} catalog
only includes grism information for sources with JH F140 magnitudes brighter than 26. The increased scatter must therefore be due to the photometric redshift estimate.

\subsection{Sources without HST counterparts}
\label{sec:nohst}

\begin{figure}[!tbp]
\begin{center}
\includegraphics[width=1\columnwidth]{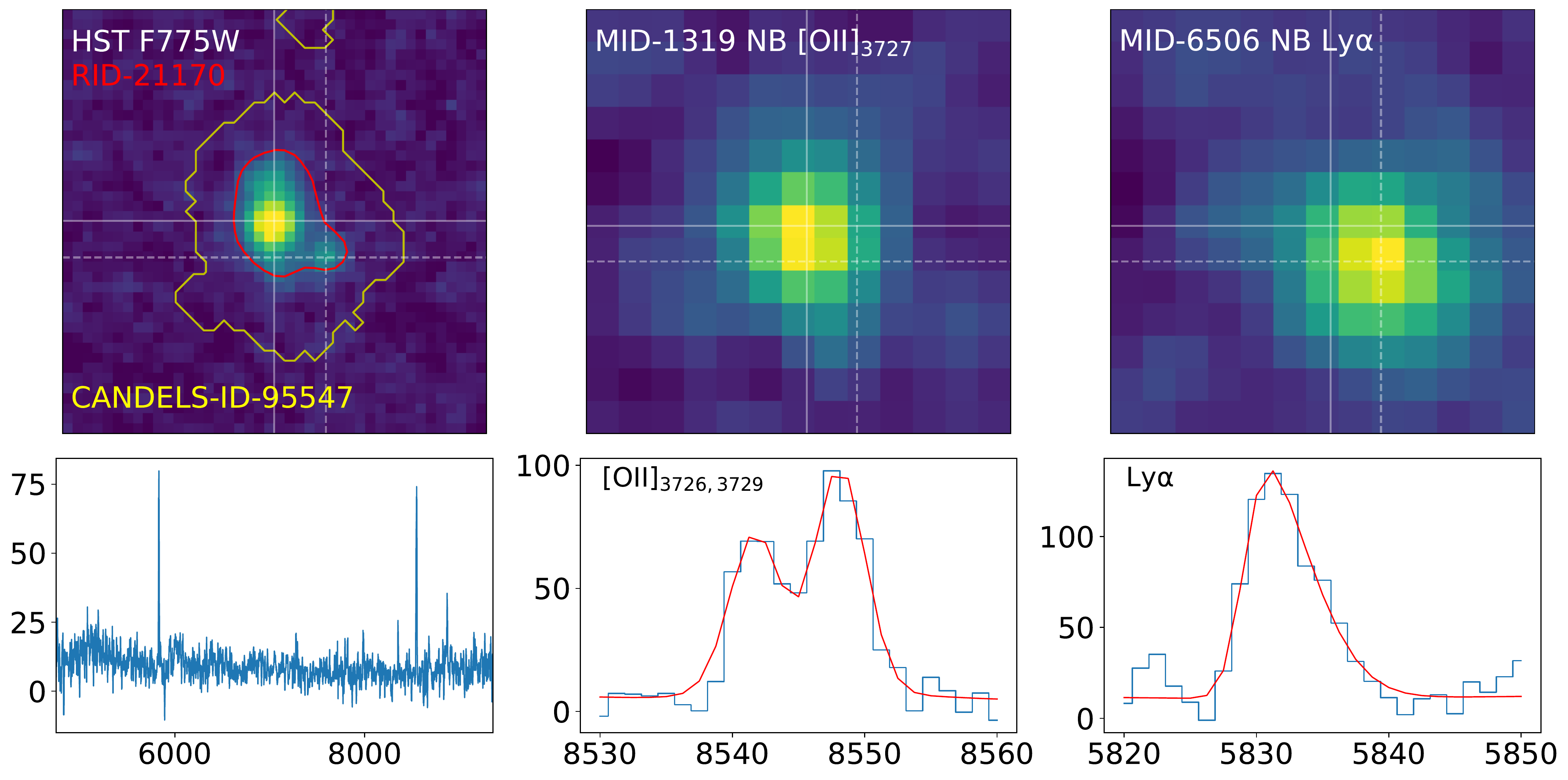}
\caption{Example of a MUSE source missed in HST catalogs. The upper left panel display the HST F775W image ($2\farcs5 \times 2\farcs5$) at the location of RID-21170. The red line shows the R15 segmentation contour, while the yellow line displays the \candels\ segmentation contour of the corresponding source (ID 95547) in \cite{Whitaker2019}.
Although all HST catalogs identify a single source, MUSE unambiguously identifies two sources: MID-1319, an \oii\ emitter at $ z=1.292$ (bottom central panel) and MID-6506 (bottom right panel), a \lae\ at $ z=3.798$. The MUSE reference spectrum is shown in the bottom left panel. From the corresponding \oii\ and \lya\ narrowband images, one can see that the MID-1319 \oii\ emitter matches the brighter part of the HST image, while the MID-6506 \lae\ coincides with the southeast secondary HST peak. Observed wavelength and flux units are respectively \AA\ and \ergsa{-20}.
}
\label{fig:hst_missed}
\end{center}
\end{figure}

During the inspection process (Sect.~\ref{sec:crossmatch}) we identified 424 (20\%) MUSE sources that cannot be matched to any of the four HST catalogs: R15 \uvudf\ \citep{Rafelski2015}, \candels\ \citep{Whitaker2019}, \tdhst\ \citep{Skelton2014, Momcheva2016}, \astrod\ \citep{Merlin2021}. In a some cases (86), the source can still be clearly seen in the HST image. An example of such "missed" sources is shown in Fig.~\ref{fig:hst_missed}. This demonstrates the limitations of source de-blending based solely on morphology and broadband colors. 

The vast majority of sources not found in the HST catalogs were classified as "faint" or "undetect" by the experts. In order to give more quantitative ground for this classification, we performed our own HST photometry at the locations of MUSE sources unmatched to the HST catalogs.

As in \cite{Maseda2018}, we measure the S/N ratio for objects using the HST/ACS mosaics from the XDF survey \citep{Illingworth2013}.  We determine the fluxes in apertures of 0.4 arcsec diameter centered on the DR2 catalog positions. This aperture corresponds to a physical size of 3.1 kpc (2.2 kpc) at $\rm z = 2.9$ (6.6). The local background level is calculated by measuring the standard deviation of the fluxes in 250 identical apertures spread randomly within a $10\arcsec \times 10\arcsec$ cutout centered on the object, with other objects masked according to the R15 segmentation maps. If the aperture flux is greater than three times the local background level in an HST band, then we consider the object “detected” in that band.  Otherwise we use the 3-$\sigma$ upper-limit to the flux in that band.

\begin{figure}[!tbp]
\begin{center}
\includegraphics[width=1\columnwidth]{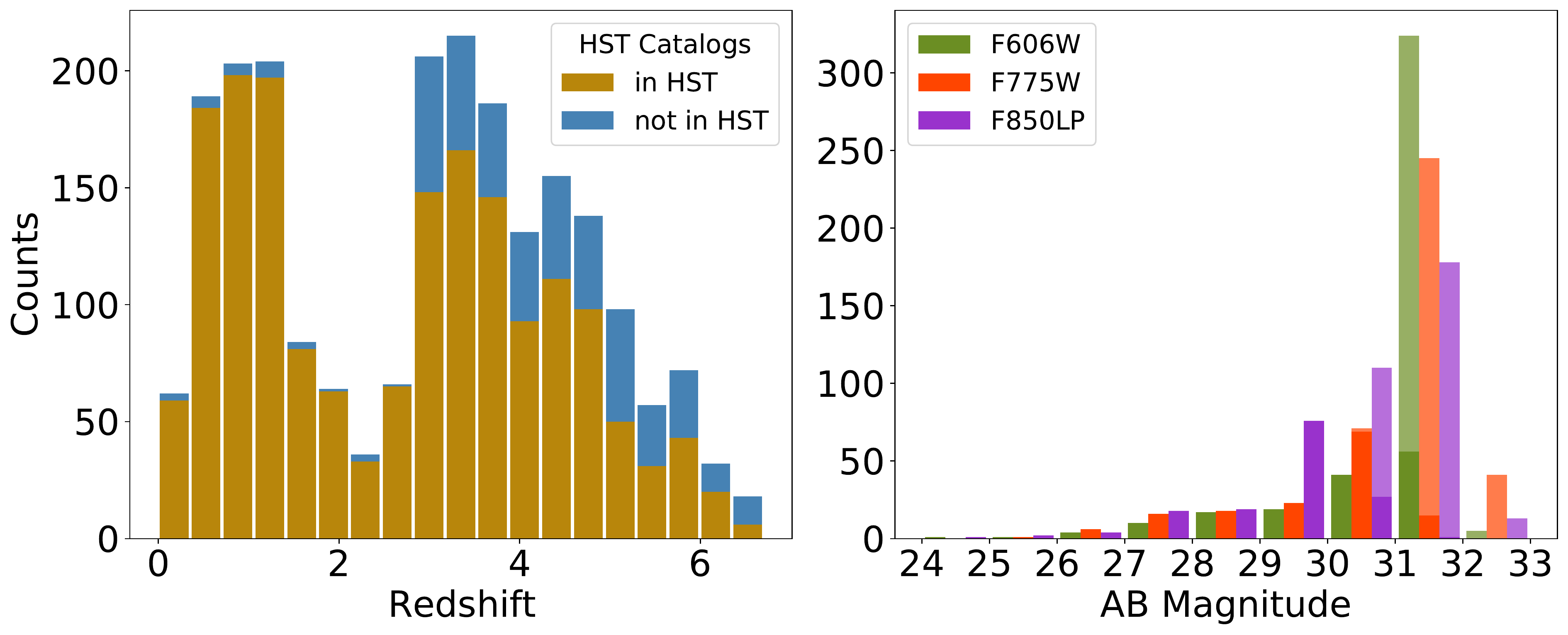}
\caption{Statistics of MUSE sources without HST counterpart. Left panel: comparison of the redshift distributions of MUSE sources with (brown color) and without (blue color) HST counterparts. Right panel: estimated AB aperture magnitude for MUSE sources without HST counterparts. The histogram is colored according to the HST filters. Successful photometric measurements ($\rm S/N > 3$) are displayed in dark colors, while measurements with $\rm S/N < 3$ are displayed with light colors. In the latter case the reported magnitude is the measured noise RMS standard deviation.
}
\label{fig:nohst_histo}
\end{center}
\end{figure}

The computed magnitudes of these sources are shown in the right panel of Fig.~\ref{fig:nohst_histo}. Using the $\rm 3 \sigma$ S/N detection limit, we have 189 sources that fall below the detection limit in all HST bands. We note that the majority of this sample  (90\% or 380 sources) have AB magnitude greater than 29. In practice, as shown in Sect.~\ref{sec:photz}, no reliable photometric redshifts can be obtained for these faint sources, even at the UDF HST depth (Fig.~\ref{fig:zphot_comp}).

We observe that almost all sources without HST counterparts are \laes\ (left panel of Fig.~\ref{fig:nohst_histo}). The nature of these sources has been discussed by \cite{Maseda2018, Maseda2020}. The authors conclude that they constitute the high equivalent-width  tail  of  the  \lae\  population and that they have on average a high ionizing photon production efficiency. Although their study is based on 103 faint sources only and does not include the deepest part of the DR2 catalog related to \mxdf\ observations, these conclusions are not invalidated by the present data set.

\subsection{Completeness}
\label{sec:completeness}

The completeness measures how complete our sample is with respect to one of its observational or physical properties (e.g., source magnitude or line flux). It is an essential measurement to infer the LF of a given galaxy population. To estimate the completeness, we have to evaluate the number of missed sources at a given luminosity and therefore capture all the biases introduced by the whole data chain (i.e., sample selection, observation, data reduction, and analysis).

In this paper, the objective is to maximize the number of reliable redshifts in the HUDF area. We therefore have used multiple detection methods: blind detection with \origin\ and HST prior extraction with \odhin. We also used  additional available information (e.g., photometric redshifts) to choose between various redshift solutions. Each detection method has its own bias. While HST prior \odhin\ detection is based on continuum detection and will therefore be biased toward galaxies with stellar continuum, \origin\ blind detection will  preferentially select high equivalent-width emission line galaxies. Inferring the completeness of our catalog  is therefore a challenging task and beyond the scope of this paper.

\begin{figure}[htbp]
\begin{center}
\includegraphics[width=1\columnwidth]{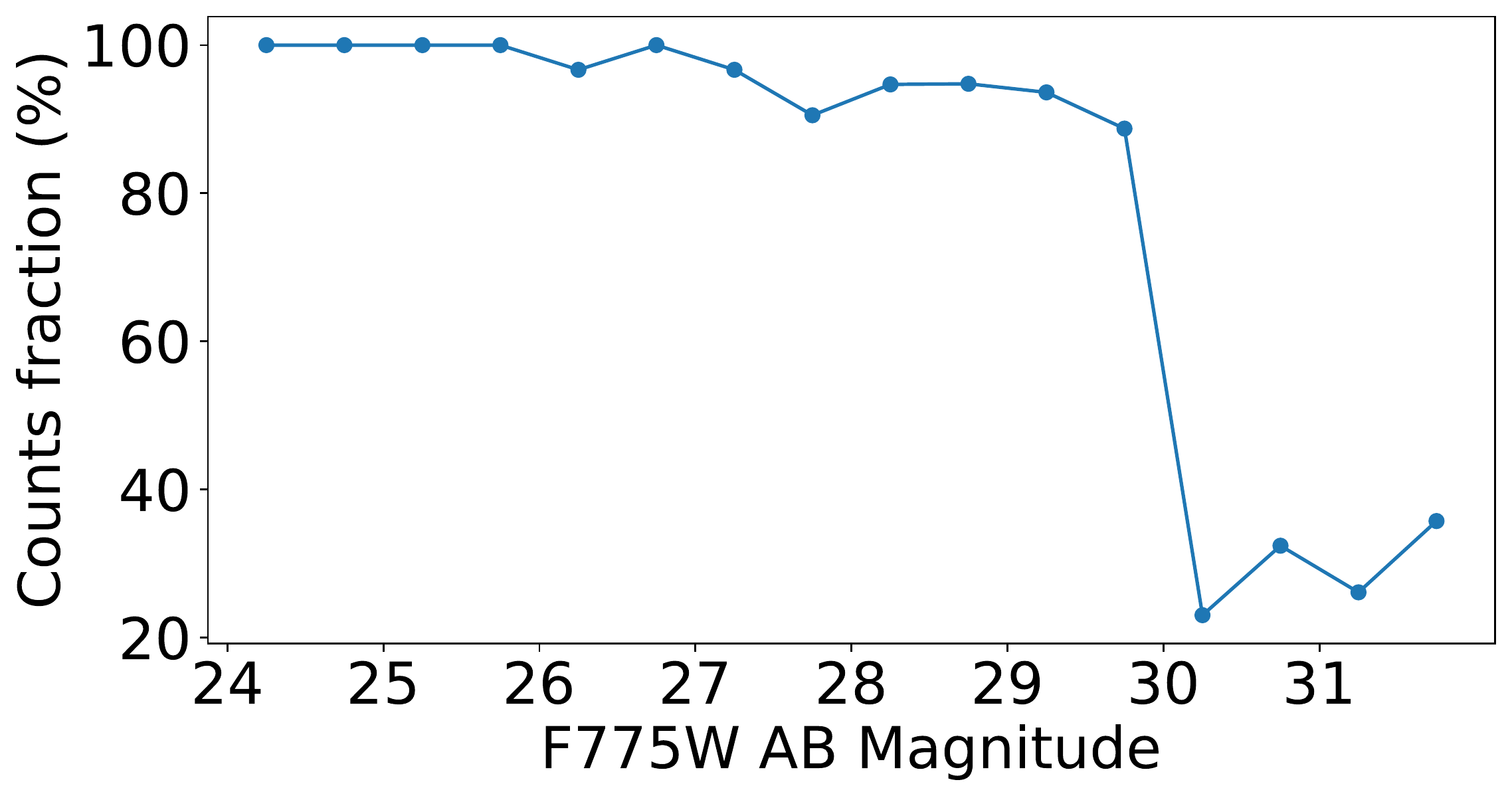}
\caption{
Fraction of MUSE \laes\ with HST counterparts as a function of AB F775W magnitude in bins of 0.5 magnitude.}
\label{fig:lae_hst}
\end{center}
\end{figure}

A useful estimate of completeness can nevertheless be made by comparing our catalog with the HST spectro-photometric catalogs. Given the great depth reached by Hubble in the UDF (i.e., $\rm 5 \sigma$ 29.5 AB F775W magnitude), the HST catalogs should be quite complete and could be used as an unbiased parent sample of broadband-selected galaxies. A rough estimate of HST \lya\ completeness can be obtained for the \lae\ population by counting the fraction of MUSE sources with HST counterparts with respect to the full \laes\ sample. As shown in Fig.~\ref{fig:lae_hst}, the HST catalogs can be considered as complete (90\%) up to AB 29.5. At fainter magnitudes, the HST catalogs have a significant drop. 
 
\begin{figure}[htbp]
\begin{center}
\includegraphics[width=1\columnwidth]{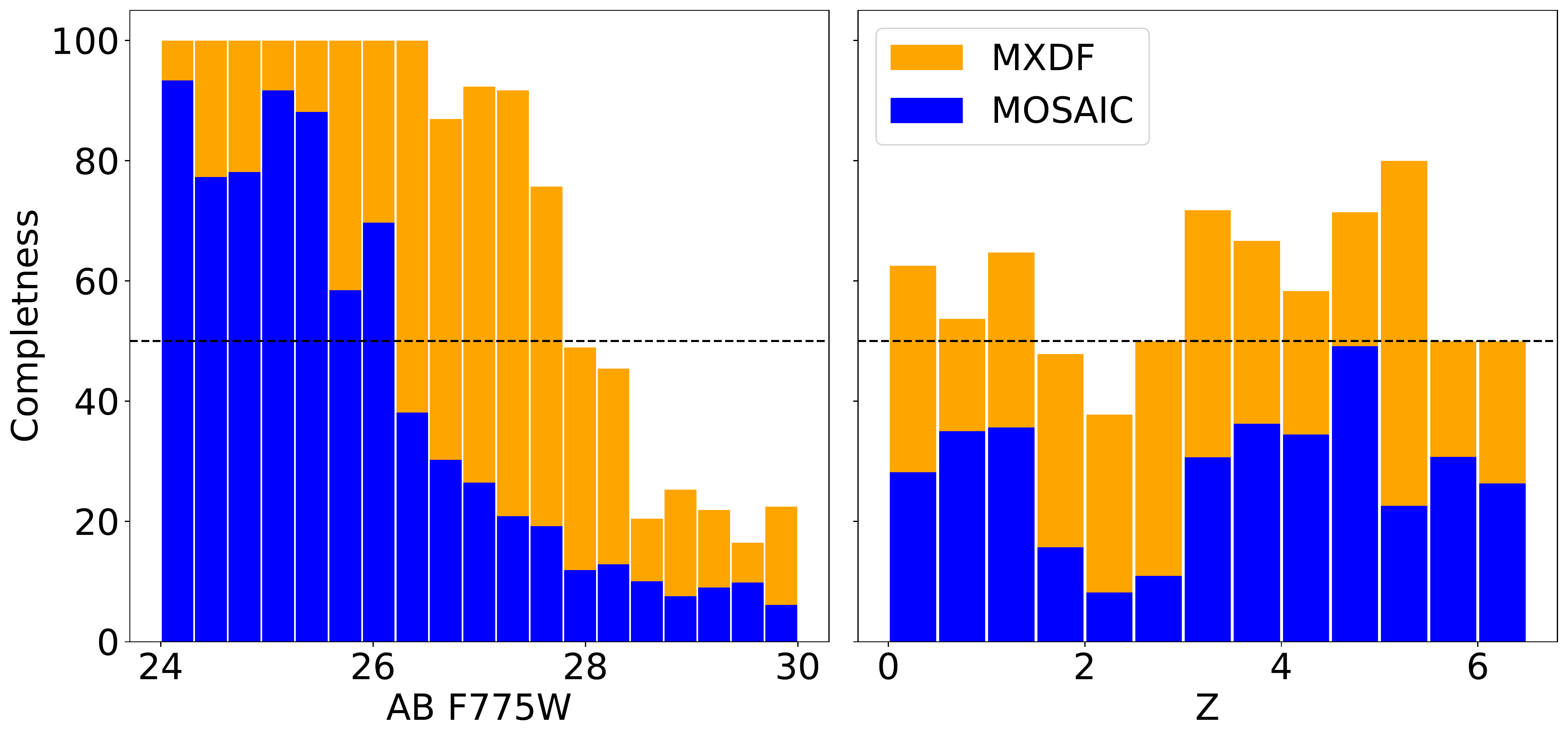}
\caption{
DR2 completeness with respect to HST deep imaging sample. Completeness is computed as the number of galaxies with a MUSE redshift in a 0.31 HST AB F775W magnitude (left panel) or 0.5 photometric redshift interval (right panel). The measurements have been performed for the deep \mxdf\ (+100-hour depth, orange bars) and \mosaic\ (+10-hour depth, blue bars) data sets and corresponding areas.
}
\label{fig:completeness}
\end{center}
\end{figure}

We performed the comparison between MUSE and the R15 spectro-photometric catalog in two areas: the \mxdf\ region limited to the 100+ hours of depth and the \mosaic\ area with 10+ hours of depth, after removing the \mxdf\ area. The former covers an area of 0.85 arcmin$^2$ and has 787 sources identified\footnote{We exclude all sources without measurable F775W magnitude.} in the R15 catalog, the latter covers 7.48 arcmin$^2$ and has 3619 sources identified in R15. The average depth achieved with MUSE in these two areas is 140.8 and 10.8 hours, respectively.
In each sample, we measure the fraction of HST sources that have a MUSE redshift in a given magnitude or redshift interval. The completeness evolution with respect to magnitude and photometric redshift is given in Fig.~\ref{fig:completeness}. 

For a given magnitude, the evolution of completeness with redshift is not flat: first, we see a clear drop in the $z = 2 - 3$ range, corresponding to the redshift range of the MUSE desert, then we measure a higher completeness at $z > 3$, indicating that the fraction of \laes\  is high in this redshift range. 
In the \mosaic\ sample at a 10-hour depth, the 50\% completeness is achieved at a magnitude of 26.6 (F775W) in the redshift range  $z = 0.8 - 1.6$ and 27.5 at $z \approx 4$. In the deepest \mxdf\ observations, the corresponding magnitude for the 50\% completeness are 27.6 and 28.7 for $z = 0.8 - 1.6$ and $z = 3.2 - 4.5$, respectively.

It is instructive to compare these figures with those obtained for the published spectroscopic redshifts sample (\citealt{Merlin2021} compilation). The 50\% completeness rate is achieved at a magnitude of 25.2 (F775W) in the $\rm z < 3$ redshift range. At higher redshifts ($\rm z > 3$) the completeness drops very rapidly with increasing magnitude: 20\% for AB 26, 5\% for AB 27.

\subsection{Physical properties}
\label{sec:sed}

In this section we derive the stellar masses and star-formation rates (SFRs) for our galaxy sample using spectral energy distribution (SED) fitting. Since the wavelength range of MUSE does not extend far enough into the infrared to provide sufficient constraints for the SED fit, we selected the subsample of 1664 galaxies (75\%) with $\rm ZCONF > 0$ that have been identified in the HST R15 catalog. The R15 catalog has 11 photometric bands\footnote{WFC3/UVIS F225W, F275W, and F336W; ACS/WFC F435W, F606W, F775W, and F850LP and WFC3/IR F105W, F125W, F140W, and F160W.} ranging from the NUV ($0.21 \mu m$)  to the WFC3/IR ($1.5 \mu m$). 

We used the high-z extension of the code \magphys\ \citep{daCunha2008, daCunha2015} with a minimum stellar mass of $\rm 10^6 M_\odot$ (see Sect.~3.2 of \citealt{Maseda2017}). The redshift of each object is fixed to the MUSE spectroscopic redshift.  In cases where the fitting results in un-physically small errors in stellar mass and/or SFR (i.e., when the models cannot accurately reproduce the photometric data points), we adopted a fixed error of $\pm 0.3$ dex (the typical uncertainty in determining stellar masses from fits to broadband photometry; e.g., \citealt{Conroy2013, Maseda2014}).

In addition to \magphys, we further utilized the \prospector\ SED fitting code \citep{Johnson2021} to derive stellar masses and SFRs for the same R15 photometric catalog. Through \prospector\ we are able to consider the contribution from emission lines to galaxy photometry. Given that a large fraction of our sources have strong emission lines, constraining the contribution of the emission lines in deriving galaxy properties is important.

In \prospector\, we used FSPS models \citep{Conroy2009} with the MILES spectral library and MIST isochrones to generate synthetic spectra to match with the observations. These models include contributions from nebular continuum and emission as described by \cite{Byler2017}. We used a nonparametric approach to fit the observed photometry using the ‘‘continuity\_sfh'’ template in \prospector. This computes the stellar mass produced in fixed time bins with a smoothness prior \citep{Leja2019}. We defined 7 time bins to fit for stellar mass. The first two bins were fixed at 0-30 and 30-100 Myr in lookback time. The furthest bin was fixed to be between  85\% - 100\% in lookback time defined by the MUSE spectroscopic redshift. The remaining four bins were evenly distributed between the second and final bin in logarithmic space. We used a \cite{Calzetti2000} dust attenuation law and allowed the V-Band dust optical depth to vary as a free parameter between 0 and 2.0. Gas-phase metallicity and stellar metallicity were tied together and were allowed to vary between 0.01 and $\approx$1.5 times the solar metallicity. We let the gas ionization parameter  vary between $\rm log(U)=-4$ and $-1$. In total we had 10 free parameters (6 SFH bins + stellar mass, dust, metallicity, and ionization parameter). Parameters were sampled using the Dynamic Nested Sampling method presented by \cite{Speagle2020} using 200 live points and batches and 50 nested walks. 

The stellar masses and the SFR averaged over the last 100 Myr are given in the main catalog (Sect.~\ref{sec:catalogs}). The error for the masses and SFR is based on the 1-sigma percentiles of the nested sampling chains. The stellar mass output by \prospector\ is the “total formed mass.” In order to convert this to the observed mass, we use the mass correction factor from the maximum a posteriori solution (the posterior sample with the highest posterior possibility) to derive an approximate correction factor for the marginalized stellar mass values.

\begin{figure}[htbp]
\begin{center}
\includegraphics[width=1\columnwidth]{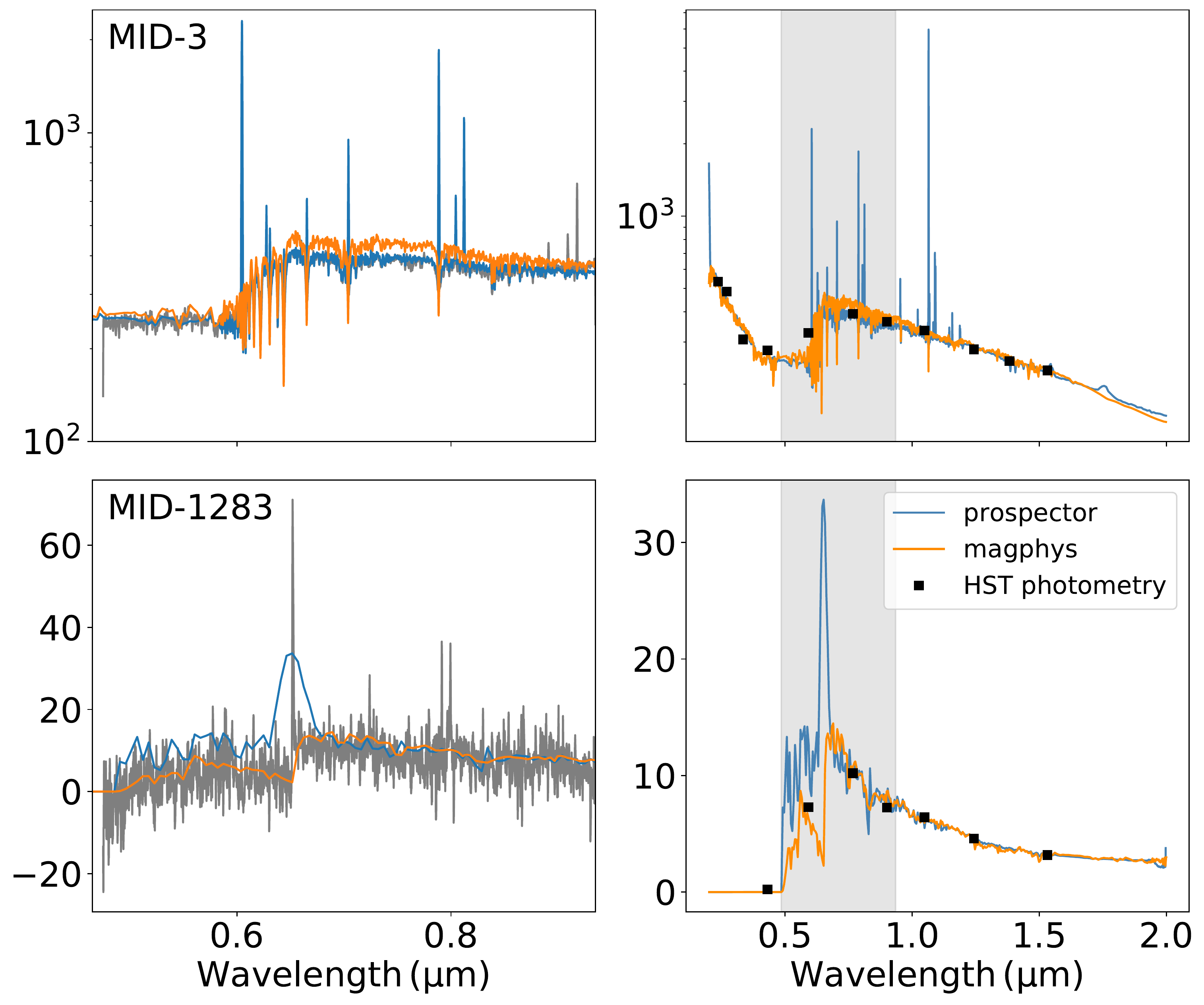}
\caption{Two example SED fits: the $z=0.62$ MID-3 \oii\ emitter (top panels) and the $z=4.36$ MID-1283 \lae\ (bottom panels). The left column shows the MUSE spectrum (in black) and the \prospector\ (blue line) and \magphys\ (orange line) SED fits. The right column displays the SED fits and the HST R15 photometry (black symbols). The shaded area in the right column indicates the MUSE wavelength range. Flux and wavelength units are \ergsa{-20} and $\rm \mu m$, respectively.
}
\label{fig:sed_ex}
\end{center}
\end{figure}

\begin{figure}[htbp]
\begin{center}
\includegraphics[width=1\columnwidth]{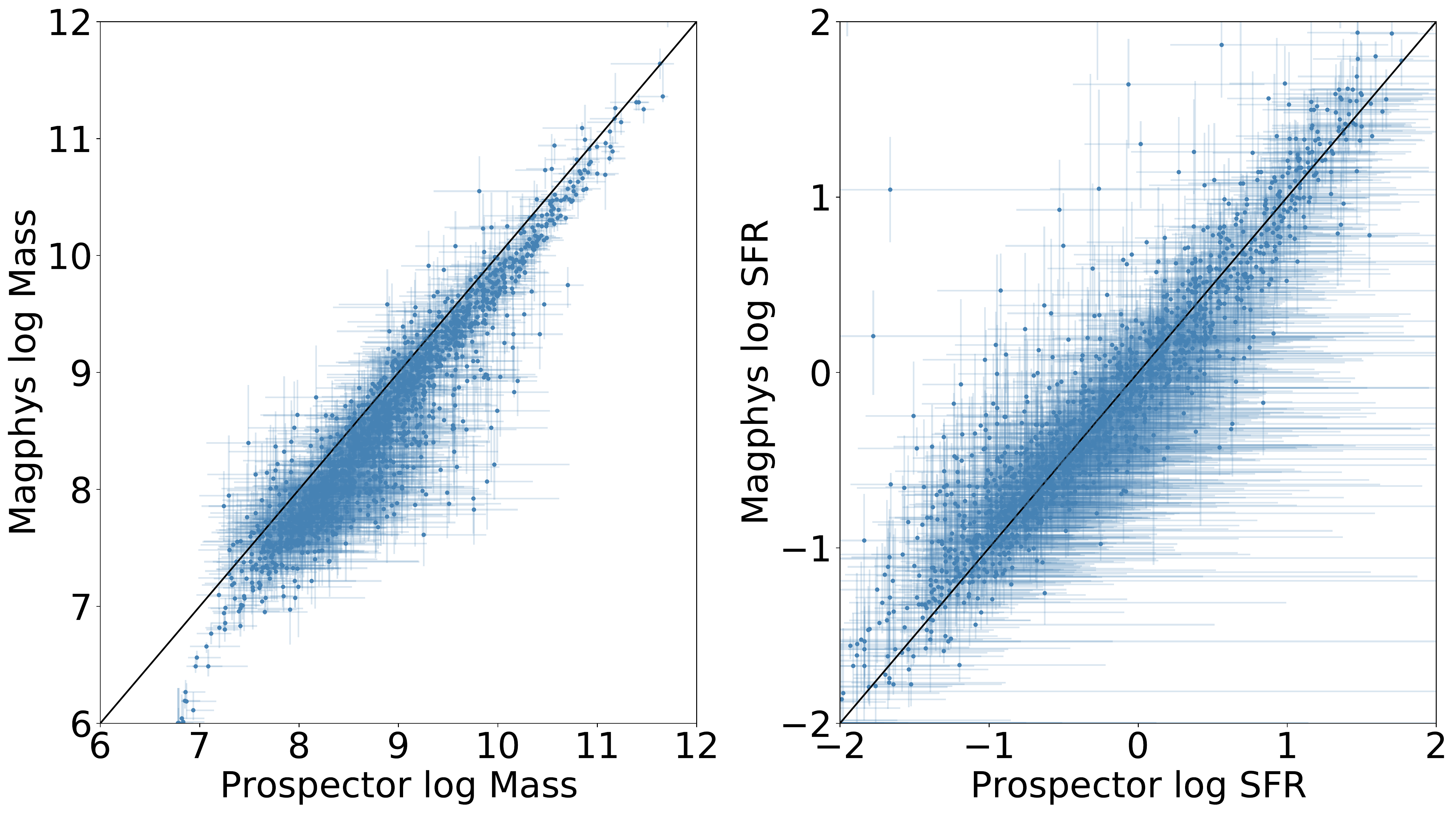}
\caption{Comparison of the \prospector\ and \magphys\ stellar mass and SFR results for the SED fits of all MUSE galaxies identified in the R15 catalog.
}
\label{fig:sed_comp}
\end{center}
\end{figure}

\begin{figure}[htbp]
\begin{center}
\includegraphics[width=1\columnwidth]{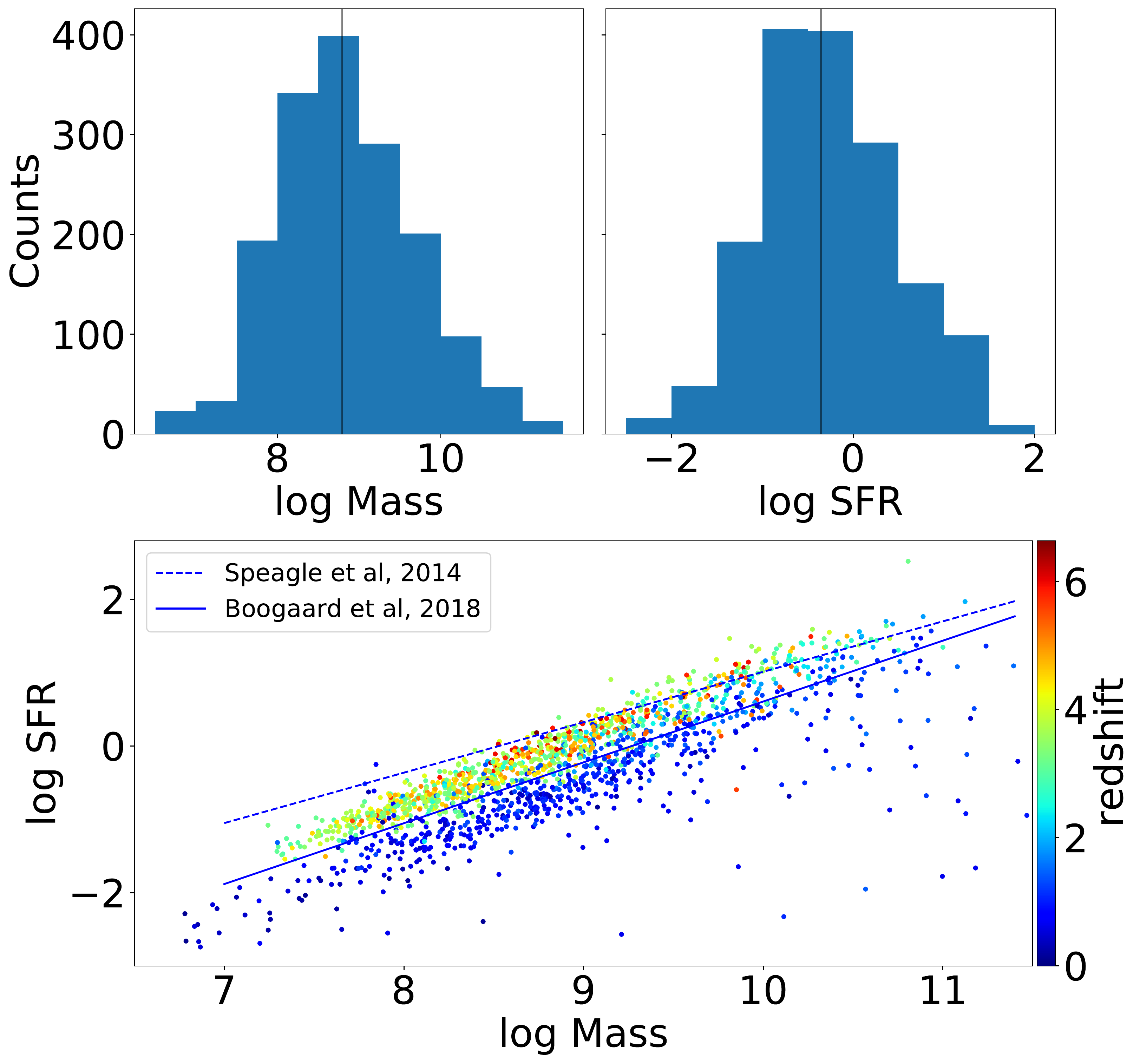}
\caption{Stellar mass (top left panel) and SFR (top right panel) distributions for the MUSE galaxies identified in the R15 catalog. Values are given for the \prospector\ SED fit in the $\approx$100 Myr lookback time. The stellar mass and the SFR are in $\rm log \, M_\odot$ and $\rm log \, M_\odot \, yr^{-1}$ units, respectively. The vertical black lines show the median values of the distributions. The bottom panel displays the star forming galaxies main sequence $\rm log\, SFR(log\, M)$. Symbols are colored according to the redshift. The z=1 main sequence derived by \cite{Boogaard2018} and \cite{Speagle2014} are shown as blue solid and dashed lines, respectively. See the main text for a discussion of the offset from our inferred values.
}
\label{fig:mass_sfr}
\end{center}
\end{figure}

Two examples of the SED fit are given in Fig.~\ref{fig:sed_ex}. In the case of MID-3 (upper left panel), it can be seen that the SED fit of \magphys, unlike \prospector, overestimates the continuum due to the presence of strong emission lines\footnote{
There are also some small differences in the photometry between MUSE and HST, as discussed in Sect.~\ref{sect:astro}, that may account for a small offset between the two sets of measurements.}. In Fig.~\ref{fig:sed_comp}, we compare the \magphys\ and \prospector\ stellar mass and SFR results for the full sample. We note that \prospector\ tends to derive higher masses than \magphys\ with a median offset of 0.25 dex. This is a known feature of \prospector\ that is documented in \cite{Leja2019,Leja2020}.
The stellar mass and SFR distributions are presented in the first two panels of Fig.~\ref{fig:mass_sfr}. As expected, the MUSE catalog probes low mass star forming galaxies with a median mass of $\rm 6.2 \times 10^8 \, M_\odot$ and a median SFR of $\rm 0.4 \, M_\odot yr^{-1}$. The last panel of Fig.~\ref{fig:mass_sfr} displays the traditional star forming main sequence. We note that our main sequence deviates significantly from the best-fit star forming main sequence at $z=1$ derived from a compilation of 25 studies by \cite{Speagle2014} (dashed blue line). Such an offset between \prospector\ SED results and previous studies has been identified and discussed by \cite{Leja2021}. As noted by the authors there are many reasons for the difference in inferred stellar mass and SFR between the \prospector's nonparametric star-formation histories  and the classical approaches (see their Sect. 6.2 for a detailed discussion of the origin of this difference). We also compared our results at $z=1$ with those of \cite{Boogaard2018} (Fig.~\ref{fig:mass_sfr} solid blue line). The \cite{Boogaard2018} study was based on DR1, still using HST photometry SED to estimate  stellar mass but with a different derivation of SFR based on the \ha\ and \hb\ recombination line flux. As shown in the plot, there is still a small offset between our estimate and the results from \cite{Boogaard2018}, but it is much smaller compared to the values of \cite{Speagle2014}. This residual offset is most likely due to the different estimate of the SFR.
Further investigations will be needed, for example into the  completeness and SED degeneracy, to interpret this main sequence, but such an analysis is beyond the scope of this paper.

Finally, we recall that this SED analysis does not take into account the population of 577 (25\%) galaxies without an HST counterpart in the R15 catalog. The majority of these sources are high equivalent-width \laes\ that are too faint in the HST broadband deep images to get meaningful SED measurements. 
However, as demonstrated by \cite{Maseda2018,Maseda2020}, this selection yields on average faint ($\mathrm{M_{UV} \approx -15}$), blue ($\mathrm{\beta \approx -2.5}$) star-forming galaxies.  Although we cannot directly determine the stellar masses from the UV continuum alone, an extrapolation to the \cite{Duncan2014} $\mathrm{M_{UV}-M_{\odot}}$ relation to $\mathrm{M_{UV}}$ = $-$15 implies that these galaxies should have stellar masses below $\mathrm{10^7 M_{\odot}}$ at these redshifts.

\subsection{Peculiar objects}
% objects.tex
\label{sec:objects}

\begin{figure*}[htbp]
\begin{center}
\includegraphics[width=1\textwidth]{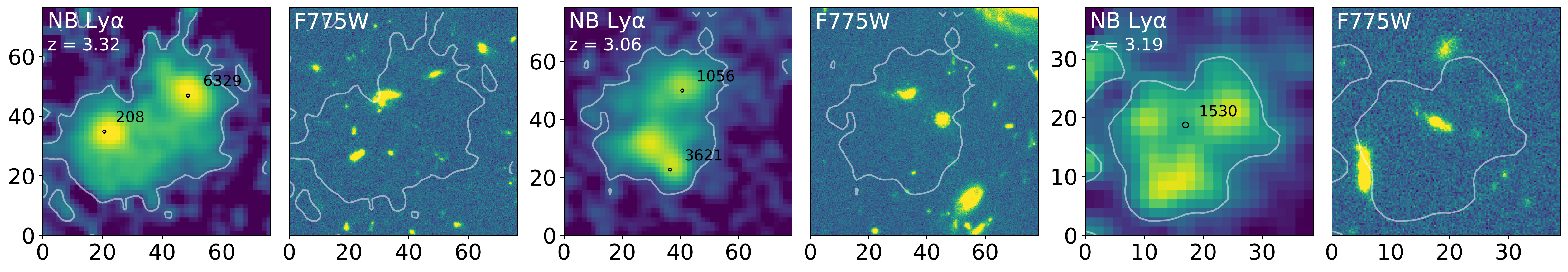}
\caption{Examples of sources with very extended \lya\ emission. For each source we show the \lya\ narrowband emission (NB \lya) and the HST F775W image. Corresponding catalog source locations and MIDs are indicated in black. The 2.0, 7.0 and \ergslineb{2.8}{-20} \lya\ flux contours are overlaid in each image from left to right. The units of the image axes are physical kpc.}
\label{fig:blobs}
\end{center}
\end{figure*}

Among this rich set of data, we give in what follows some examples of interesting objects. The first category is the \laes\ with very extended \lya\ emission. These objects are often called Lyman alpha blobs (LABs) in the literature. They are rare and mainly found in overdense regions like the two LABs discovered in the SSA22 supercluster \citep{Steidel2000}. 
While most \laes\ and rest-frame UV-selected galaxies exhibit diffuse \lya\ emission \citep{Wisotzki2016,Leclercq2017, Kusakabe2022}, LABs differ in terms of their high luminosity (typically brighter than \ergsluma{43}) and their large size (typically larger than 50 kpc; e.g., \citealt{Herenz2020}). A recent review is provided in \cite{Kimock2021}. 
We show in Fig.~\ref{fig:blobs}, three examples of \laes\ with such extended \lya\ emission. The first object (MID-208 and MID-6329) is a $z=3.32$ source detected in the \udft\ at a 29-hour depth. With a \lya\ luminosity of \ergslumb{8}{42} and a size of $\approx$60 kpc, it enters the LAB category. This object was already reported by \cite{Vanzella2017} using 20 hours of the same GTO observations. The second example (MID-1056 and MID-3621) was discovered in the \mosaic\ at an 11-hour depth. This $z = 3.06$ LAB is also bright (\ergslumb{1.4}{43} \lya\ luminosity) and extended ($\approx$60 kpc). This object coincides with a luminous type II active galactic nucleus identified in the Chandra 7Ms catalog \citep{luo2017}. The extended \lya\ emission was already reported by \cite{Brok2020} using the same data set. It is part of a 1.2 Mpc long \lya\ filament \citep{Bacon2021}. The third example, the $z=3.2$ MID-1530 \lae, is less extended (30 kpc) and luminous (\ergslumb{1.4}{42} \lya\ luminosity) than the other two and would not be formally categorized as a LAB. It is representative of the class of the large \lya\ halos found around \laes\ (e.g., \citealt{Leclercq2020}).

\begin{figure}[htbp]
\begin{center}
\includegraphics[width=1\columnwidth]{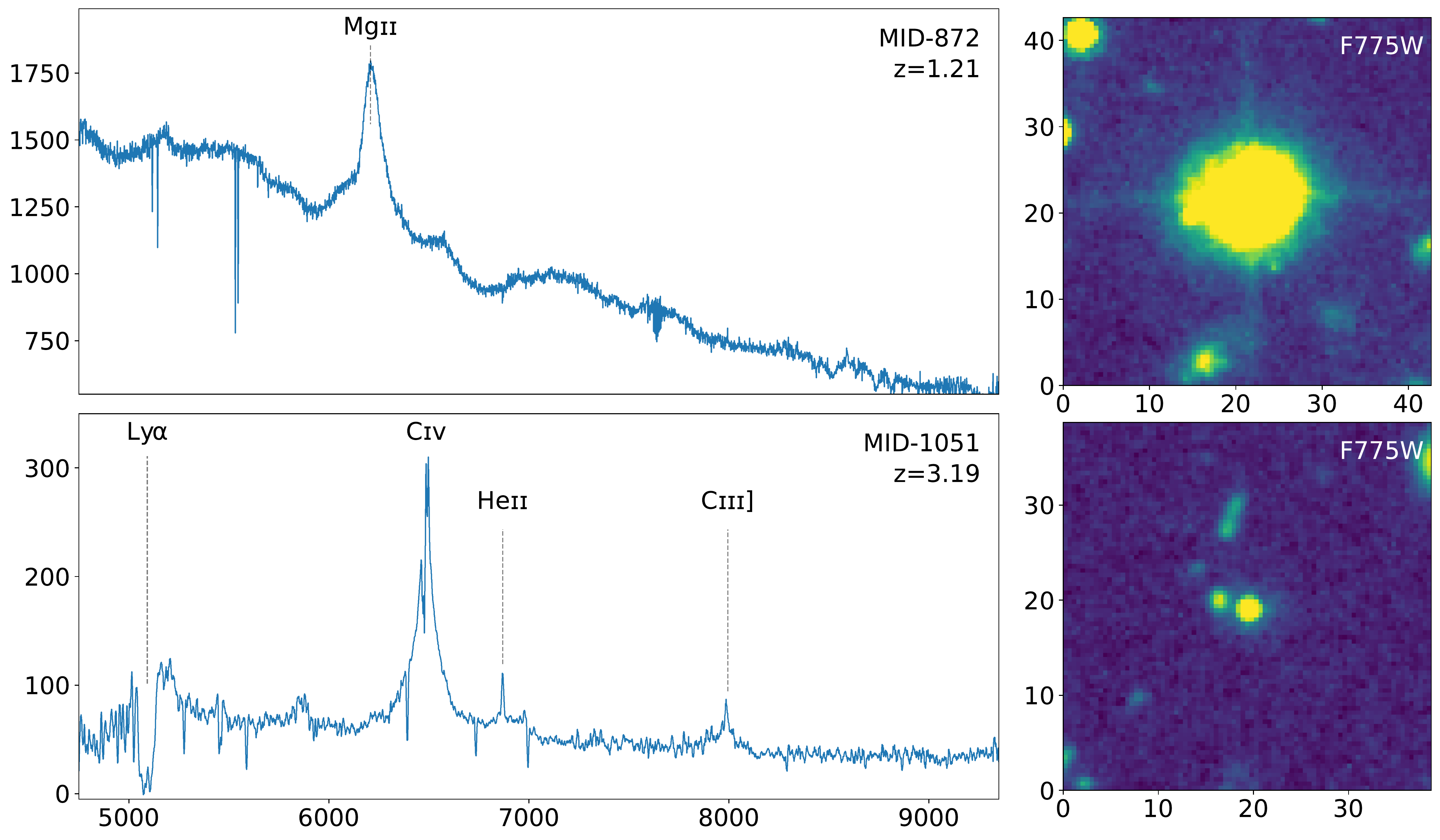}
\caption{Spectra of the two QSOs and their corresponding HST F775W images. The units of the image axes are physical kpc. Observed wavelength and flux units are respectively \AA\ and \ergsa{-20}.
}
\label{fig:qso}
\end{center}
\end{figure}

In addition to the MID-1056/3621 type II active galactic nucleus, we have two type I QSOs in the sample (Fig.~\ref{fig:qso}). The first is MID-977, a $z=1.2$ object with broad \mgii\ emission and some absorbers detected along the line of sight. The second is MID-1051, a $z=3.2$ QSO with damped \lya\ absorption and \civ, \heii,\ and \ciii\ emission.
 
\begin{figure}[htbp]
\begin{center}
\includegraphics[width=1\columnwidth]{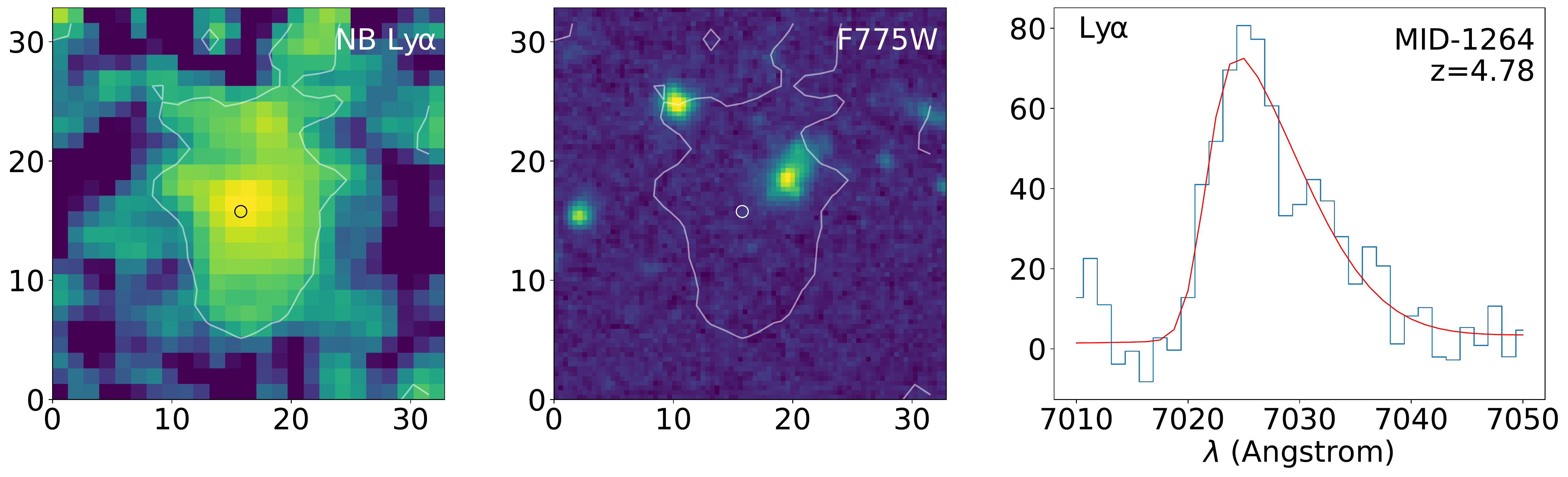}
\caption{Example of a \lae\ with extended \lya\ emission, offseted with respect to its potential HST counterpart. The two first panels display the \lya\ narrowband  and HST F775W images. The units of the image axes are physical kpc. The right panel shows the \lya\ profile and its fit with a skewed gaussian. Observed wavelength and flux units are respectively \AA\ and \ergsa{-20}.
}
\label{fig:offset_lya}
\end{center}
\end{figure}

As previously stated, \laes\ are typically surrounded by a \lya\ halo that is detected out to approximatively ten times the continuum size \citep[e.g.,][]{Leclercq2017}. Moreover, the peak of the \lya\ emission can be offset with respect to the UV continuum peak \citep[e.g.,][]{Ribeiro2020, Claeyssens2022}. We present an extreme example of this in Fig.~\ref{fig:offset_lya}, where the $z=4.8$ \lae\ (MID-1264) has a 4.1 kpc (0.6 arcsec) offset  between the UV and \lya\ narrowband centers. We note that such an offset is much larger than the $0.6 \pm 0.05$ kpc average offset observed by \cite{Ribeiro2020} in their deep VIMOS slit data. Nevertheless, these offset measurements rely on a correct match between the UV continuum source (in our case the deep HST  image) and the \lae. This is not always an easy task, and MID-1264 is a good example. The \lae\ was matched to RID-6956. This object is bright enough (25.6 AB F775W magnitude) to have a reliable photometric redshift. All HST catalogs give similar estimates, with for example $z_{\rm phot} = 4.77 \pm 0.07$ for the \astrod\ catalog. Our measurement of $z = 4.78$ fits well within the photo-z error bar, giving confidence that the \lae\ is matched with the HST source. In addition, all other nearby HST sources are much farther from the location of the \lya\ emission peak and their MUSE or photo-z redshifts indicate that they are foreground objects.
However, the high S/N of the \lya\ line profile shows evidence for a small dip at 7029 \AA\ that can be interpreted as the superposition of two \laes\ at $z=4.7806$ and $z=4.7885$ respectively. In this case, the fainter \lya\ line would be associated with RID-6956, while the brighter \lya\ line would be a \lae\ without an HST counterparts, satellite to the main galaxy. 

There are many examples where the HST sources have no reliable photo-z, are located in a crowded environment or have a complex morphology. In these cases,  matching \lya\ emission to HST imaging is a challenge. In general, we tend to favor HST-undetected \lae\ when the offset is large enough. This is in agreement with a recent study that used a large sample of lensed \laes\ observed with MUSE \citep{Claeyssens2022}; it shows that the average \lya\ offsets are small, $\rm 0.66 \pm 0.14$ kpc, and that the large offsets are likely due to satellite or merging galaxies.

\section{Summary and conclusions}
\label{sec:conclusions}

This second data release of the MUSE \textit{Hubble} Ultra Deep Field surveys is a major update of the DR1 published in \cite{Bacon2017} and \cite{Inami2017}. Like DR1, it is based on deep MUSE GTO observations of the HST UDF area. It covers a field of view of ${\rm 3 \times 3 \, arcmin^2}$ at a 10-hour depth (\mosaic) and ${\rm 1 \times 1 \, arcmin^2}$ at a 31-hour depth (\udft), with much improved data reduction and data analysis processes compared with DR1. A new data set, the MUSE extremely deep field (\mxdf), also located in the UDF XDF area and covering an approximatively circular area of 1 arcmin in diameter, has been pushed to a depth of 141 hours. This data set benefits from an improved spatial resolution of 0.55 arcsec at 7000\AA, thanks to the GLAO AOF system. 

The achieved $3\sigma$ emission line surface brightness sensitivity at 7000 \AA\ is 4.5, 2.3, and $1.0 \times \rm 10^{-19} \ergsurfb$ for the 10-hour depth \mosaic, 31-hour depth \udft,\ and 141-hour depth \mxdf\ datacubes, respectively. The corresponding $3\sigma$ point-source limiting flux for an unresolved emission line is  3.1, 1.5, and $0.6 \times \rm 10^{-19} \ergsline$.

Advanced source detection and extraction using the \origin\ blind detection \citep{Mary2020}  and the \odhin\ HST-based source de-blending (Sect.~\ref{sec:odhinmain}) software resulted in the secure identification of 2221 sources, a 41\% increase over DR1. Apart from eight stars, the collected sample is composed of 25 nearby galaxies ($z < 0.25$), 677 \oii\ emitters ($z=0.25-1.5$), 201 galaxies in the MUSE desert redshift range ($z=1.5-2.8$), and 1308 \laes\ ($z=2.8-6.7$).

We measure almost an order of magnitude more redshifts than the collections of all spectroscopic redshifts obtained before DR1 in the UDF (i.e., 2221 versus 292). At high redshift ($z > 3$), the difference is even more striking: 1308 versus 20.
Since 2004, extensive campaigns of spectroscopic observations have been carried out with most large ground-based telescopes: for example, VIMOS/VLT \citep{Lefevre2013, Garilli2021}, FORS2/VLT \citep{Mignoli2005, Vanzella2008}, and DEIMOS/KECK \citep{Doherty2005}.
It is surprising that, despite this significant effort in telescope time, so few spectroscopic redshifts have been obtained in the iconic HUDF field.
The reason is technical: ground-based spectroscopy has been obtained with multi-object spectrographs \citep[e.g., VIMOS;][]{Lefevre1998}, which are most efficient when performing spectroscopy of galaxies at the bright end of the LF. At the faint end, the density of galaxies increases to a level where it becomes very inefficient to build masks or to pack enough fibers to get spectra of sources in such crowded environments. The other reason is inherent to the multi-object spectroscopy (MOS) concept, where a preselection is needed prior to the observations. Objects not or barely detected in deep broadband imaging will not be selected for the spectroscopic observations. This effect is most pronounced at high redshifts and explains why the large population of high equivalent-width \laes\ uncovered by MUSE is almost completely absent from the MOS observations.

The other sources of published redshifts are the slitless spectroscopic observations performed with HST (e.g., \citealt{Brammer2012}). Like in integral field spectroscopy, slitless spectroscopy does not require imaging preselection. There are, however, some limitations in crowded areas due to overlapping spectra. This is mitigated by performing multiple observations at different rotations, but it remains difficult to use for faint sources in crowded environments. The main limitation is, however, the small HST 2.4 meter telescope aperture and the low spectral resolution ($\rm R \approx 50-100$), which limit its use to relatively bright compact sources.

Eighty percent of the galaxies in our final catalog have an HST counterpart identified in at least one of the published HST spectro-photometric catalogs. These galaxies are on average faint: a median AB F775W magnitude of 25.7 and 28.7 for the \oii\ and \lya\ emitters, respectively. We measured the completeness of our catalog with respect to HST and found that at a 10-hour depth, 50\% of galaxies of magnitude F775W AB 26.6 ($z=0.8-1.6$) and 27.5 ($z \approx 4$) are present in DR2. This number increases in the deepest 140-hour depth \mxdf\ area with 50\% completeness for F775W AB 27.6 and 28.7 at $z=0.8-1.6$ and $z=3.2-4.5$, respectively.

The comparison with three published photometric redshifts catalogs \citep{Rafelski2015, Momcheva2016, Merlin2021} confirms the results presented in \cite{Brinchmann2017}: the scatter and outlier fraction increase with magnitude and exceed the errors claimed in the corresponding studies. Among the three catalogs tested, the recent \astrod\ catalog \citep{Merlin2021} gives the smallest scatter and the smallest systematic, with $\Delta z = 0.02 - 0.04$ at F775W $\rm AB < 26$ and $\rm AB > 28.5$, respectively.  

We performed SED fitting based on the MUSE redshifts and the 11 HST broadband photometry measurements from the \cite{Rafelski2015} catalog. We used two different codes, \magphys\ \citep{daCunha2008, daCunha2015} and \prospector\ \citep{Johnson2021}, to derive the stellar mass and SFR. \prospector\ uses the MUSE emission line flux as an additional fitting constraint. Both codes give comparable results, although with some systematic differences and a large scatter for faint galaxies. The MUSE sample probes on average low-mass star-forming galaxies with a median mass of $\rm 6.2 \times 10^8 \, M_\odot$ and a median SFR of $\rm 0.4 \, M_\odot yr^{-1}$.

Twenty percent of our catalog, or 424 galaxies, have no counterpart in the HST source catalogs. While a few of these galaxies are clearly visible in the HST images and were just missed by the \sex\ source de-blending, the vast majority are too faint with respect to the HST limiting magnitude, which is between 29 and 30 AB magnitude depending on the filter. These new sources are high equivalent-width \laes\ that are detected by MUSE thanks to their bright and asymmetric broad \lya\ line. As expected, the fraction of HST-undetected \laes\ is a function of depth, with a sharp increase from 18\% in the 10-hour depth \mosaic\ to 49\% in the 140-hour depth \mxdf.
These figures confirm the existence of a large population of faint star-forming \laes\ revealed in MUSE deep fields \citep[e.g.,][]{Bacon2015, Bacon2017}. As shown by \cite{Maseda2018, Maseda2020}, these galaxies are on average very faint ($\rm M_{UV} \approx -15$) star-forming galaxies with a high ionizing photon production efficiency.

The \mxdf\ is the deepest spectroscopic survey ever performed. We believe it has great legacy value. The MUSE data, with their 3D content, amazing depth, wide spectral range, and excellent spatial and medium spectral resolution, are rich in information. Their location in the UDF area, which benefits from an exquisite collection of ancillary panchromatic information, is a major asset. We have made public the advanced data products, specific software programs, and a web interface to select and download data sets.

\begin{acknowledgements}
We warmly thank ESO Paranal staff for their great professional support during all GTO observing runs.
We thank the anonymous referee for a careful reading of the manuscript and many suggestion for improving its readability. 
We thank Marc Rafelski, Katherine Whitekaer, Emiliano Merlin and Adriano Fontana for providing information on their published spectrophotometric catalogs and associated segmented maps.   
RB, YG acknowledge support from the ANR L-INTENSE (ANR-20-CE92-0015).
NB acknowledges the support of the ANR 3DGasFlows (ANR-17-CE31-0017).
SLZ acknowledges support by The Netherlands Organisation for Scientific Research (NWO) through a TOP Grant Module 1 under project number 614.001.652.
\end{acknowledgements}

%-------------------------------------------------------- BIBLIO ------------------------------------------------------------
\bibliographystyle{aa}
\bibliography{dr2_biblio.bib}

% appendix.tex

\begin{appendix}

\section{Data products and software releases}
%products
\label{sec:products}

In this section we give practical information on the released data sets, namely images and datacubes, catalogs and source files. We also describe the features of the \amused\ web interface. Finally, we give the list of all specific software developed or updated in the context of this survey.

\subsection{Images and datacubes}
%datacubes
\label{sec:cubes}

For each data set (i.e., \mxdf, \udft,\ and \mosaic) we make the final datacubes and associated images public. These data are available for download through the \amused\ web interface and are described in Table~\ref{tab:cubes}.
Images and datacubes are in the multi-extension FITS format, with a data and optionally a variance extension. Masked data have NaN values. The primary header of each file contain a WCS extension and a number of keywords, including the spatial PSF model polynomial coefficients (Sect.~\ref{sec:fsf}). The file  format is described in detail in the corresponding \mpdaf\ documentation\footnote{ \href{https://mpdaf.readthedocs.io/en/latest/obj.html}{https://mpdaf.readthedocs.io/en/latest/obj.html}}. These files can be used directly with any FITS reading tool (e.g., \href{https://heasarc.gsfc.nasa.gov/fitsio/}{fitsio}, \href{https://sites.google.com/cfa.harvard.edu/saoimageds9}{ds9}), but we recommend using \mpdaf, which contains an easier and advanced manipulation of these files.

\begin{table*}[hbt!]
\caption{Released datacubes and images.}   
\label{tab:cubes}
\centering
\begin{tabular}{lllllc}
\hline 
File name & Description & data set & Type & Size & Version \\ 
\hline 
DATACUBE\_MXDF.fits & Main datacube & \mxdf & cube & 6.9 Gb & 3.0 \\ 
IMAGE\_MXDF.fits & White-light image & \mxdf & image & 2.0 Mb & 3.0 \\ 
EXPMAP\_MXDF.fits & Exposure map datacube & \mxdf & cube & 3.1 Gb & 3.0 \\ 
EXPMAP-IMAGE\_MXDF.fits & Exposure map image  & \mxdf & image & 0.9 Mb & 3.0 \\
\hline
DATACUBE\_UDF-10.fits & Main datacube & \udft & cube & 2.9 Gb & 1.1 \\ 
IMAGE\_UDF-10.fits & White-light image & \udft & image & 0.9 Mb & 1.1 \\ 
EXPMAP\_UDF-10.fit & Exposure map datacube & \udft & cube & 1.5 Gb & 1.1 \\ 
EXPMAP-IMAGE\_UDF-10.fits & Exposure map image  & \udft & image & 0.4 Mb & 1.1 \\ 
\hline  
DATACUBE\_MOSAIC.fits & Main datacube & \mosaic & cube & 25 Gb & 1.1 \\ 
IMAGE\_MOSAIC.fits & White-light image & \mosaic & image & 6.9 Mb & 1.1 \\ 
EXPMAP\_MOSAIC.fit & Exposure map datacube & \mosaic & cube & 13 Gb & 1.1 \\ 
EXPMAP-IMAGE\_MOSAIC.fits & Exposure map image  & \mosaic & image & 3.5 Mb & 1.1 \\ 
FIELDMAP\_UDF-MOSAIC.fits & Field map image & \mosaic & image & 6.9 Mb & 1.1 \\ 
\hline 
\end{tabular}
\tablefoot{The flux unit of datacubes and white light images is \ergsa{-20}. Variance is given in the square of the flux unit.
Units of exposure map images and datacubes are exposure counts, each of 25 mn duration. The exposure map image is obtained by taking the median over the wavelength axis. The \mosaic\ field map image codes the sub-fields identifier (1-9) used in the combination at each spaxel location (see \mpdaf\ fieldmap documentation\footnote{\href{https://mpdaf.readthedocs.io/en/latest/muse.html\#muse-mosaic-field-map}{https://mpdaf.readthedocs.io/en/latest/muse.html\#muse-mosaic-field-map}}).
The version refer to the version of the data reduction process. 
}
\end{table*} 

In addition we also provide a fits table DR2\_SENSITIVITY.fits with the median surface brightness and point-source limiting flux (Sect.~\ref{sec:noise}) as a function of wavelength. Column descriptions are given in Table~\ref{tab:sensitivity}.

\begin{table}[hbt!]
\caption{Description of the columns of the sensitivity table.}   
\label{tab:sensitivity}
\centering
\begin{tabular}{m{1.7cm}m{6.5cm}}
\hline 
Col. name & Description \\ 
\hline 
\texttt{WAVE} & Wavelength in \AA. \\ 
\texttt{SB\_MOSAIC} & Median surface brightness $1\sigma$ limiting emission line flux for the \mosaic\ data set (10-hour depth).\\ 
\texttt{SB\_UDF10} & Median surface brightness $1\sigma$ limiting emission line flux for the \udft\ data set (31-hour depth).\\ 
\texttt{SB\_MXDF} & Median surface brightness $1\sigma$ limiting emission line flux for the \mxdf\ data set (141-hour depth).\\ 
\texttt{PS\_MOSAIC} & Median point-source  $1\sigma$ limiting emission line flux for the \mosaic\ data set (10-hour depth).\\ 
\texttt{PS\_UDF10} & Median point-source  $1\sigma$ limiting emission line flux for the \udft\ data set (31-hour depth).\\ 
\texttt{PS\_MXDF} & Median point-source  $1\sigma$ limiting emission line flux for the \mxdf\ data set (141-hour depth).\\

\hline
\end{tabular}
\tablefoot{
Computations are performed for an unresolved emission line. Surface brightness and point-source $1\sigma$ limiting flux are given in \ergsurfb\ and \ergsline\ units, respectively.
}
\end{table}

%\FloatBarrier
\subsection{Catalogs}
% tables.tex
\label{sec:catalogs}

The main catalog gives a summary of source properties. It contains a subset of the main measured properties (i.e., the main emission and absorption line flux) and can be used standalone. For the full set of measurements three additional tables are given: the redshifts table, the line table and the narrow-band table. Catalogs are publicly available via the \amused\ web interface (Sect.~\ref{sec:amused}).

\begin{table*}[hbt!]
\caption{Released tables.}   
\label{tab:tables}
\begin{tabularx}{\textwidth}{lXcccc}
\hline 
File name & Description & Size & Rows & Version & Columns desc. \\ 
\hline 
dr2\_main\_09.fits & Main source catalog & 3.1 Mb & 2221 & 0.9 & Table.~\ref{tab:maintable}\\
dr2\_z\_09.fits & Redshift catalog & 0.9 Mb & 6722 & 0.9 & Table.~\ref{tab:redtable} \\
dr2\_lines\_09.fits & Emission and absorption lines catalog & 13 Mb & 56968 & 0.9 & Table.~\ref{tab:linetable} \\
dr2\_nb\_09.fits & Narrowband catalog & 4.2 Mb & 11938 & 0.9 & Table.~\ref{tab:nbtable} \\
dr2\_noise\_09.fits & Surface brightness and point source limiting emission flux & 0.2 Mb & 3705 & 0.9 & Table.~\ref{tab:sensitivity}\\ 
dr1\_dr2\_09.fits & DR1 sources that have been renamed & 12 Kb & 39 & 0.9 & Table.~\ref{tab:dr1_dr2}\\ 
\hline 
\end{tabularx}
\tablefoot{
These tables are only available in electronic form at the CDS via anonymous ftp to cdsarc.u-strasbg.fr (130.79.128.5) or via http://cdsweb.u-strasbg.fr/cgi-bin/qcat?J/A+A/ or via the AMUSED interface (Sect.~\ref{sec:amused}).
We also provide another format for the fits table, named "*\_astropy\_*." These tables are astropy fits tables with masked values\footnote{\href{https://docs.astropy.org/en/stable/api/astropy.table.MaskedColumn.html}{https://docs.astropy.org/en/stable/api/astropy.table.MaskedColumn.html}}.}
\end{table*}

%\subsubsection{Main catalog}
\label{subsect:maintable}

The main catalog is a table with 2221 entries, one line for each source. Sources have a unique MUSE identifier (MID). We note that sources already detected in DR1 have kept their identifier, except when the redshift or matching information was updated. In that case a new identifier was given. The catalog description is given in Table~\ref{tab:maintable}.

\begin{table}[hbt!]
\caption{Description of the columns of the main table.}   
\label{tab:maintable}
\centering
\begin{tabularx}{\columnwidth}{lX}
\hline 
Col. name & Description \\ 
\hline 
\texttt{ID} & MUSE source identifier (int)\\ 
\texttt{DATASET} & MUSE data set\tablefootmark{a} \\ 
\texttt{DEPTH} & exposure depth at source location in hours \\
\texttt{FROM} & spectrum extraction type\tablefootmark{b} (Sect.~\ref{sec:ext})\\ 
\texttt{ZCONF} & redshift confidence: 1 (low) -- 3 (high) (Sect.~\ref{sec:zconf})\\
\texttt{MCONF} & matching confidence: 0--3, (Sect.~\ref{sec:crossmatch}) \\
\texttt{IFLAG} & isolation flag: 1--3, (Sect.~\ref{sec:crossmatch}) \\
\texttt{ZSYS} & systemic redshift in vacuum \tablefootmark{c} (Sect.\ref{sec:zsyslae})\\
\texttt{ZSYS\_ERR} & error in systemic redshift (Sect.\ref{sec:zsyslae}) \\
\texttt{REFZ} & reference redshift line set \tablefootmark{d} (Sect.\ref{sec:platefit}\\  
\texttt{Z} & reference redshift value in vacuum \\
\texttt{Z\_ERR} & error in reference redshift \\
\texttt{DLYAFIT} & flag to indicate double \lya\ fit\\
\texttt{DV\_ttt} & velocity offset  with respect to reference redshift for redshift type \texttt{ttt} (\kms) \\
\texttt{DV\_ERR\_ttt} & velocity offset  with respect to reference redshift for redshift type \texttt{ttt} (\kms) \\
\texttt{RA} &  right ascension (J2000 degree), see astrometry Sect.~\ref{sect:astro}\\ 
\texttt{DEC} &  declination (J2000 degree) see astrometry Sect.~\ref{sect:astro}\\
\texttt{CENTER} & reference center \tablefootmark{e} (Sect.\ref{sec:narrowband})\\
\texttt{IN\_HST} & HST matching flag \tablefootmark{f}\\  
\texttt{IN\_ORI} & ORIGIN matching boolean flag \\
\texttt{IN\_DR1} & DR1 matching boolean flag \\
\texttt{IN\_MXDF} & source is located in \mxdf\ footprint (bool) \\
\texttt{IN\_UDF10} & source is located in \udft\ footprint (bool) \\
\texttt{RAF\_ID} & R15 catalog unique matched ID (int)\\
\texttt{RAF\_MIDS} & R15 catalog multiple matched IDs \tablefootmark{g}\\
\texttt{CANDELS\_ID} & \candels\ catalog unique matched ID (int)\\
\texttt{CANDELS\_MIDS} & \candels\ catalog multiple matched IDs \tablefootmark{g}\\
\texttt{C3DHST\_ID} & \tdhst\ catalog unique matched ID (int)\\
\texttt{C3DHST\_MIDS} & \tdhst\ catalog multiple matched IDs \tablefootmark{g}\\
\texttt{ASTRO\_ID} & \astrod\ catalog unique matched ID (int)\\
\texttt{ASTRO\_MIDS} & \astrod\ catalog multiple matched IDs \tablefootmark{g}\\
\texttt{MAG\_SRC} & source of magnitude \tablefootmark{h} \\
\texttt{MAG\_FLAG} & contamination flag (APER magnitude only) \tablefootmark{i} \\
\texttt{MAG\_xxx} & broadband AB magnitude in xxx HST filter \tablefootmark{j}\\
\texttt{MAGERR\_xxx} & AB magnitude error in xxx HST filter \tablefootmark{k}\\
\hline
\end{tabularx}
\tablefoot{
\\
\tablefoottext{a}{\texttt{MXDF,UDF10} or \texttt{MOSAIC}} \\
\tablefoottext{b}{\texttt{ORIGIN,ODHIN} or \texttt{NBEXT}} \\
\tablefoottext{c}{$\rm Z_{sys} \equiv Z$, except for for simple peak \laes} \\
\tablefoottext{d}{\texttt{BALMER,FORBIDDEN,LYALPHA,ABS,CIV1548} or \texttt{MGII2796}} \\
\tablefoottext{e}{\texttt{3DHST,CANDELS,CUSTOM,NB\_EMI,ORIGIN} or \texttt{RAFELSKI}} \\
\tablefoottext{f}{Ambiguous, Detected, Faint, Missed or Undetect}\\
\tablefoottext{h}{Coded as text with comma separator (e.g., 23,567)}\\
\tablefoottext{h}{\texttt{APER,3DHST,CANDELS} or \texttt{RAF}}. Aperture photometry (\texttt{APER}) is used for undetected HST source (Sect.~\ref{sec:nohst})\\
\tablefoottext{i}{If true indicate source contamination}\\
\tablefoottext{j}{Filters are \texttt{F606W, F775W} or \texttt{F850LP}} \\
\tablefoottext{k}{If \texttt{MAGERR\_xxx < 0}, then \texttt{MAG\_xxx} $\equiv$ noise stdev}\\
}
\end{table}

\begin{table}[hbt!]
\caption*{Continued.}   
\centering
\begin{tabularx}{\columnwidth}{lX}
\hline 
Col. name & Description \\ 
\hline 
\texttt{MASS\_ff} & $\rm log \, M / M_\odot $ where M is the stellar mass derived from the \texttt{ff}\tablefootmark{a} SED fit (Sect.\ref{sec:sed}).\\ 
\texttt{LERR\_MASS\_ff} & Lower $1\sigma$ percentile of $\rm log \, M / M_\odot $\\ 
\texttt{HERR\_MASS\_ff} & Upper $1\sigma$ percentile of $\rm log \, M / M_\odot $\\ 
\texttt{SFR\_ff} &  $\rm log \, SFR / M_\odot yr^{-1}$ where SFR is the star formation rate at 100 Myr lookback time as derived from the \texttt{ff}\tablefootmark{a} SED fit (Sect.\ref{sec:sed}).\\ 
\texttt{LERR\_SFR\_ff} &  Lower $1\sigma$ percentile of $\rm log \, SFR / M_\odot yr^{-1}$ \\ 
\texttt{HERR\_SFR\_ff} &  Upper $1\sigma$ percentile of $\rm log \, SFR / M_\odot yr^{-1}$ \\ 
\texttt{LINE\_SNR\_MAX} & name of emission or absorption line with max S/N \tablefootmark{b} \\
\texttt{SNR\_MAX} & max S/N \\
\texttt{FLUX\_MAX} & flux of the line with max S/N ($10^{-20} \ergsline$)\\
\texttt{lll\_EMI\_FLUX} & flux of the \texttt{lll}\tablefootmark{c} emission line ($10^{-20} \ergsline$) \\
\texttt{lll\_EMI\_SNR} & S/N of the \texttt{lll} emission line \\
\texttt{lll\_EMI\_EQW} & Rest frame equivalent width of the \texttt{lll} emission line (\AA) \\
\texttt{lll\_EMI\_VD} & Rest frame velocity dispersion\tablefootmark{d} of the \texttt{lll} emission line (\kms) \\
\texttt{lll\_ABS\_FLUX} & flux of the \texttt{lll}\tablefootmark{i} absorption line ($10^{-20} \ergsline$) \\
\texttt{lll\_ABS\_SNR} & S/N of the \texttt{lll} absorption line \\
\texttt{lll\_ABS\_EQW} & Rest frame equivalent width of the \texttt{lll} absorption line (\AA) \\
\texttt{lll\_ABS\_VD} & Rest frame velocity dispersion\tablefootmark{c} of the \texttt{lll} absorption line (\kms) \\
\hline
\end{tabularx}
\tablefoot{
\\
\tablefoottext{a}{\texttt{ff} is \texttt{PRO} for \prospector\ and  \texttt{MAG} for \magphys.}\\
\tablefoottext{b}{\texttt{b} at the end of line name indicate a blend (e.g., \texttt{OII3727b} is the sum of the \oiid\ doublet)}\\
\tablefoottext{c}{see table~\ref{tab:linename} for line names} \\
\tablefoottext{d}{corrected for instrumental velocity dispersion}
}
\end{table}

\label{subsect:ztable}

The redshift and line tables that result from the fitting performed on the reference spectrum by \pfit\ are described in Tables~\ref{tab:redtable} and \ref{tab:linetable}, respectively. The redshift table gives the common parameters (i.e., the redshift and the velocity dispersion) fitted for each line set. Some statistical  (e.g., maximum S/N) and fitting (e.g., $\chi^2$) information is also given. The line table contains detailed fitting information for each line.

\begin{table}[hbt!]
\caption{Description of the columns of the redshift table.}   
\label{tab:redtable}
\centering
\begin{tabular}{m{2.3cm}m{6.5cm}}
\hline 
Col. name & Description \\ 
\hline 
\texttt{ID} & MUSE source identifier \\ 
\texttt{DATASET} & MUSE data set\tablefootmark{a} \\ 
\texttt{LINESET} & line set\tablefootmark{b} \\
\texttt{Z} & redshift in vacuum \\
\texttt{Z\_ERR} & error in redshift \\
\texttt{VEL} & velocity offset with respect to reference redshift (\kms) \\
\texttt{VEL\_ERR} & velocity offset error (\kms) \\
\texttt{VDISP} & rest-frame velocity dispersion (\kms) \\
\texttt{VDISP\_ERR} & error in velocity dispersion (\kms) \\
\texttt{LINE} & name of line with the highest S/N \\
\texttt{SNRMAX} & maximum S/N for the fitted lines \\
\texttt{SNRSUM} & total S/N for all lines \\
\texttt{SNRSUM\_CLIPPED} & total S/N for lines with S/N $>$ 3 \\
\texttt{NL} & number of fitted lines \\
\texttt{NL\_CLIPPED} & number of fitted lines with S/N $>$ 3 \\
\texttt{RCHI2} & returned reduced $\chi^2$ by the minimization routine \\
\texttt{STATUS} & returned status of the fitting function\tablefootmark{c} \\
\hline
\end{tabular}
\tablefoot{
\\
\tablefoottext{a}{\texttt{MXDF,UDF10} or \texttt{MOSAIC}} \\
\tablefoottext{b}{\texttt{BALMER,FORBIDDEN,LYALPHA,ABS,CIV548} or \texttt{MGII2796}} \\
\tablefoottext{c}{LMFIT}
}
\end{table}

%'ID','data set','LINESET','Z','Z_ERR','VEL','VEL_ERR','VDISP','VDISP_ERR','LINE','SNRMAX','SNRSUM','SNRSUM_CLIPPED','NL','NL_CLIPPED','RCHI2','STATUS'

%\subsubsection{Lines table}
\label{subsect:linetable}

\begin{table}[hbt!]
\caption{Description of the columns of the line table.}   
\label{tab:linetable}
\centering
\begin{tabular}{m{1.7cm}m{6.5cm}}
\hline 
Col. name & Description \\ 
\hline 
\texttt{ID} & MUSE source identifier \\ 
\texttt{DATASET} & MUSE data set\tablefootmark{a} \\ 
\texttt{LINESET} & line set\tablefootmark{b} \\
\texttt{LINE} & line identifier (e.g., LYALPHA) \\
\texttt{LBDA\_REST} & rest wavelength (\AA) \\
\texttt{DNAME} & display name for the line (e.g., \lya), set to None for close doublets. \\
\texttt{FLUX} & total line flux ($10^{-20} \ergsline$) \\
\texttt{FLUX\_ERR} & line flux error ($10^{-20} \ergsline$) \\
\texttt{SNR}  & line S/N \\
%\texttt{VEL} & velocity offset with respect to initial redshift (\kms) \\
%\texttt{VEL\_ERR} & velocity offset error (\kms) \\
\texttt{Z} & redshift in vacuum \\
\texttt{Z\_ERR} & error in redshift \\
\texttt{VDISP} & rest-frame velocity dispersion (\kms) \\
\texttt{VDISP\_ERR} & error in velocity dispersion (\kms) \\
\texttt{SKEW}& skewness parameter \tablefootmark{c}  (Eq.~\ref{eq:asymline})\\ 
\texttt{SKEW\_ERR} & skewness error \tablefootmark{c} \\
\texttt{SEP} & separation between the two peaks \tablefootmark{d} (\kms) \\ 
\texttt{SEP\_ERR} & error in peak separation \tablefootmark{d} (\kms) \\
\texttt{VDINST} & instrumental velocity dispersion (\kms) \\
\texttt{LBDA\_OBS} & fitted position of the line peak in observed frame (\AA) \\
\texttt{PEAK\_OBS} & maximum flux of the line peak in observed frame ($10^{-20} \ergs$) \\
\texttt{LBDA\_LEFT} & observed wavelength at the left of the peak with half peak value (\AA) \\
\texttt{LBDA\_RIGHT} & observed wavelength at the right of the peak with half peak value (\AA) \\
\texttt{FWHM\_OBS} & full width at half maximum of the line in the observed frame (\AA) \\
\texttt{EQW} & restframe line equivalent width (\AA) \\
\texttt{EQW\_ERR} & restframe line equivalent width error (\AA) \\
\texttt{CONT\_OBS} & continuum mean value in observed frame ($10^{-20} \ergs$)  \\
\texttt{CONT} & continuum mean value in rest frame ($10^{-20} \ergs$)  \\
\texttt{CONT\_ERR} & error in continuum mean value in rest frame ($10^{-20} \ergs$)  \\
\texttt{NTSD} & $\rm log_{10}$ of the line fit relative error\\

\hline
\end{tabular}
\tablefoot{
\\
\tablefoottext{a}{\texttt{MXDF,UDF10} or \texttt{MOSAIC}} \\
\tablefoottext{b}{balmer, forbidden, lya, abs, civ1548 or mgii2796} \\
\tablefoottext{c}{restricted to \lya\ line} \\
\tablefoottext{d}{restricted to double peaked \lya\ line fit}
}
\end{table}

%'ID','data set','LINESET','LINE','LBDA_REST','DNAME','VEL','VEL_ERR','Z','Z_ERR',
%'VDISP','VDISP_ERR','SEP','SEP_ERR','VDINST','FLUX','FLUX_ERR','SNR','SKEW','SKEW_ERR',
%'LBDA_OBS','PEAK_OBS','LBDA_LEFT','LBDA_RIGHT','FWHM_OBS','EQW','EQW_ERR','CONT_OBS',
%'CONT','CONT_ERR','BLEND','ISBLEND','VEL_RTAU','VDISP_RTAU','FLUX_RTAU','SKEW_RTAU','SEP_RTAU'

%\subsubsection{Narrowbands table}
\label{subsect:nbtable}

The narrow-band table is described in Table~\ref{tab:nbtable} and contains information for all narrow bands derived from emission or absorption, single or combined lines.

\begin{table}[!htbp]
\caption{Description of the columns of the narrowband table.}   
\label{tab:nbtable}
\centering
\begin{tabular}{m{2.1cm}m{6.5cm}}
\hline 
Col. name & Description \\ 
\hline 
\texttt{ID} & MUSE source identifier \\ 
\texttt{DATASET} & MUSE data set\tablefootmark{a} \\ 
\texttt{LINE} & narrowband name\tablefootmark{b} (e.g., \texttt{NB\_EMI\_OII3727}) \\
\texttt{COMBINED} & boolean to indicate if the image is a combination of multiple narrow bands \\
\texttt{LINES} & list of emission lines combined together (e.g., OII3726,OII3729) \\
\texttt{DNAME} & display name for the line (e.g., \lya)\\
\texttt{SNR\_LINES}  & lines total S/N \\
\texttt{FLUX}\tablefootmark{c} &  narrowband flux measured over the segmented area ($10^{-20} \ergsline$) \\
\texttt{FLUX\_ERR} & narrowband flux error using datacube variance ($10^{-20} \ergsline$) \\
\texttt{FLUX\_ISO\_ERR} & narrowband flux error from SExtractor "isocontour" mode ($10^{-20} \ergsline$) \\
\texttt{SNR\_NB}\tablefootmark{c}  & narrowband S/N derived from \texttt{FLUX}/\texttt{FLUX\_ERR}\\
\texttt{FLUX\_AUTO} & narrowband  flux from  \sex\ "automatic" mode ($10^{-20} \ergsline$) \\
\texttt{FLUX\_AUTO\_ERR} & narrowband  flux error from  \sex\ "automatic" mode ($10^{-20} \ergsline$) \\
\texttt{LBDA\_INF} & lower limits\tablefootmark{d}  in wavelength used for narrowband computation (\AA) \\
\texttt{LBDA\_SUP} & upper limits\tablefootmark{d} in wavelength used for narrowband computation (\AA)  \\
\texttt{RA} &  right ascension of narrowband barycenter (J2000 degree) \\ 
\texttt{DEC} &  declination of narrowband barycenter (J2000 degree) \\
\texttt{OFFSET} & offset (arcsec) with respect to source reference center \\
\texttt{AREA} & segmented area ($\text{arcsec}^2$)\\
\texttt{PA} & principal axis ($\degree$)\\
\texttt{MAJAX} & major axis (arcsec) \\
\texttt{MINAX} & minor axis (arcsec) \\
\texttt{ELL} & ellipticity\\
\texttt{KRON} & KRON radius (arcsec)\\
\texttt{NSEG} & number of segments\\
\hline
\end{tabular}
\tablefoot{
\\
\tablefoottext{a}{\texttt{MXDF,UDF10} or \texttt{MOSAIC}} \\
\tablefoottext{b}{The names \texttt{NB\_EMI\_COMBINED} or \texttt{NB\_ABS\_COMBINED} are used for the optimal combination of emission or absorption lines (see Sect.~\ref{sec:narrowband})} \\
\tablefoottext{c}{These values are obtained over the narrowband-segmented area, they are generally different from the values reported in the lines table.} \\
\tablefoottext{d}{This is a list if more than one interval was used for the narrowband computation} 

}
\end{table}

%'ID','data set','LINE','LINES','DNAME','SNR_LINES','FLUX','FLUX_ERR','FLUX_ISO_ERR','SNR_NB',
%'FLUX_AUTO','FLUX_AUTO_ERR','LBDA_INF','LBDA_SUP','RA','DEC','OFFSET','AREA','PA',
%'MAJAX','MINAX','ELL','KRON','NSEG'

The list of DR1 sources that have been renamed in DR2 because of a new redshift assignment (Sect.~\ref{sec:dr1}) is part of the released data products. Its column description is given in Table~\ref{tab:dr1_dr2}.

\begin{table}[!htbp]
\caption{Description of the columns of the renamed DR1 sources table.}   
\label{tab:dr1_dr2}
\centering
\begin{tabular}{m{2.1cm}m{6.5cm}}
\hline 
Col. name & Description \\ 
\hline 
\texttt{DR1\_ID} & Previous DR1 source identifier \\ 
\texttt{DR1\_Z} & Previous DR1 redshift \\ 
\texttt{DR1\_ZCONF} & Previous DR1 redshift confidence \\ 
\texttt{DR2\_ID} & New DR2 source identifier \\ 
\texttt{DR2\_Z} & New DR2 redshift \\ 
\texttt{DR2\_ZCONF} & New DR2 redshift confidence \\
\texttt{DR2\_COM} & Comment\\ 
\hline
\end{tabular}
\end{table}

%\FloatBarrier
\subsection{Sources}
% sources.tex
\label{sec:sources}
Each source identified in the DR2 catalog is available as a specific \textit{Source} file, a multi-extension FITS file that gathers 
all source information in a single file \citep{Wells1981}. The \textit{Source} format is part of the MUSE \textsf{MPDAF} (\citealt{Bacon2016,Piqueras2019}). A DR2 \textit{Source} file contains generic information related to the source (e.g., its identifier and celestial coordinates), small images (e.g., white-light and narrow bands) and datacubes centered at the source location, spectra for various extraction scheme and tables (e.g., emission and absorption lines information).

The "Source" Python class is described in the \textsf{MPDAF} \href{https://mpdaf.readthedocs.io/en/latest/api/mpdaf.sdetect.Source.html}{on-line documentation}. A tutorial specific to the DR2 sources is also available online\footnote{\href{https://amused.univ-lyon1.fr/project/UDF/HUDF/help}{https://amused.univ-lyon1.fr/project/UDF/HUDF/help}}. We give in Tables~\ref{tab:source_hdr},\ref{tab:source_spectra},\ref{tab:source_images}, and \ref{tab:source_tables} the full description of the source content.

The DR2 source files, in addition of providing direct access to the datacube, spectrum, narrowband and white-light images contents for a given source, give access to a number of important ancillary information.
For example, the sources contains spectra derived from different extractions. We note that all parameters provided in the catalogs have been computed with a reference extraction method, given in the main catalog shown in Table~\ref{tab:maintable}, column \texttt{FROM}. However, this default extraction may not be optimal for a given science case as already discussed in Sect.~\ref{sec:ext}, and thus we provide all other available extractions in the source file. 
In the following we give our prescription for the extraction method to use in some typical conditions. 

\odhin\ extraction should be preferred for sources with detectable continuum, or when the continuum and absorption line information is important (e.g., passive galaxies). It is also the best method when the source blending is substantial and all neighboring sources are present in the HST segmentation map. 

\origin\ extraction is obviously recommended for galaxies undetected in HST. It is also best for faint line emitters as it provides generally the highest S/N. It also works better when the emission flux is spatially extended (e.g., \laes).

\nbext\ extraction is an alternative to \origin\ extraction. Although it provides generally lower S/N than the \origin\ extraction, it should be used in the few cases where the PCA continuum subtraction impacts the pseudo-narrow band. It should also be used in some case of close pairs of line emitters, when the \origin\ resulting pseudo-narrowband, which implies a PSF convolution \citep{Mary2020}, is not able to distinguish between the two sources. 

When available, we also give the spectrum derived from alternative data sets (i.e., \mosaic\ extraction for a source with \udft\ or \mxdf\ reference data sets). \pfit\ information is given in the form of tables  (PL\_LINES and PL\_Z) and spectra (named with the "PL\_" prefix). 

Some additional information for the \origin\ detected sources is given in the ORI\_LINES Table~\ref{tab:source_table_ori}. The SED fitting information (Sect.~\ref{sec:sed}) is optionally available in the SED Table~\ref{tab:source_table_sed}. 

The sources also contain catalog information on surrounding HST objects derived from the 3 HST catalogs (R15, \candels\ and \tdhst), \origin\ detected and DR2 neigbouring sources (named with "\_CAT" suffix). 
If available, spectra of neighboring sources derived from \odhin\ R15 extraction, are also given.

\begin{table}
\caption{Source content: header.}   
\label{tab:source_hdr}
\centering
\begin{tabular}{p{1.8cm}p{5.5cm}p{0.5cm}}
\hline 
Name & Description & Req. \\ 
\hline 
ID &  MUSE identifier & y \\ 
DATASET  & data set (MXDF, UDF10 or MOSAIC) & y \\ 
RA &  source right ascension (degree) & y \\ 
DEC &  source declination (degree) & y \\
FROM &  name of software used in source creation & y \\
\texttt{FROM\_V} & version of software used in source creation & y \\
CUBE &  MUSE datacube name & y \\
\texttt{CUBE\_V} & version of MUSE datacube  & y \\
SIZE &  source square size (arcsec) & y \\
CATALOG  & name of input catalog & y \\
ZCONF  & redshift confidence (1-3) & y \\
DEPTH  & average depth in hours at source location & y \\
EXPMEAN  & average number of exposures at source location & y \\
EXPMIN  & minimum number of exposures at source location & y \\
EXPMAX  & maximum number of exposures at source location & y \\
FSFMODE  & FSF (spatial PSF) model id (=2) & y \\
FSFLB1  & FSF blue normalization wavelength & y \\
FSFLB2  & FSF red normalization wavelength & y \\
FSF00FNC  & FSF FWHM number of polynomial coef (=4) & y \\
FSF00Fxx  & FSF FWHM polynomial coef value xx=00..03 & y \\
FSF00BNC  & FSF $\beta$ number of polynomial coef (=1) & y \\
FSF00Bxx  & FSF $\beta$ polynomial coef value xx=00 & y \\
REFSPEC & name of reference spectrum & y \\
REFZ & reference redshift & y \\
REFCENTER & identifier of reference center & y \\
\texttt{ccc\_ID} & matched source ID for catalog ccc & n \\
\texttt{ccc\_RA} & matched source RA for catalog ccc & n \\
\texttt{ccc\_DEC} & matched source DEC for catalog ccc & n \\
\texttt{ccc\_OFF} & matched source offset (arcsec) with respect to source center & n \\
\hline 
\end{tabular}
\tablefoot{ccc: catalog ORI (origin) RAF (Rafelski), CANDELS, C3DHST (3D-HST) \\
Req: required y (yes) or n (no)
} 
\end{table}

\begin{table}
\caption{Source content: spectra.}   
\label{tab:source_spectra}
\centering
\begin{tabular}{p{1.8cm}p{5.5cm}p{0.5cm}}
\hline 
Name & Description & Req. \\ 
\hline 
\texttt{ORI\_ds\_id} &  ORIGIN extraction for ds data set, id is the ORIGIN ID & n \\ 
\texttt{RAF\_ds\_id} &  ODHIN extraction based on Rafelski segmentation map for ds data set, id is the Rafelski ID & n \\ 
\texttt{CAN\_ds\_id} &  ODHIN extraction based on CANDELS segmentation map for ds data set, id is the CANDELS ID & n \\ 
\texttt{C3D\_ds\_id} &  ODHIN extraction based on 3D-HST segmentation map for ds data set, id is the 3D-HST ID & n \\ 
\texttt{EMI\_ds\_id} &  NBEXT extraction based on narrowband segmentation map for ds data set, id is the MUSE ID & n \\
\texttt{PL\_FIT} &  \pfit\ reference spectrum full fit (line emission + model continuum) & y \\
\texttt{PL\_CONT} &  \pfit\ reference spectrum model continuum fit & y \\
\texttt{PL\_LINE} &  \pfit\  continuum subtracted reference spectrum  & y \\
\texttt{PL\_LINEFIT} &  \pfit\ reference spectrum continuum subtracted fit & y \\
\texttt{PL\_FITP} &  \pfit\ reference spectrum full fit (line emission and absorption + polynomial continuum) & n \\
\texttt{PL\_ABSINIT} &  \pfit\ reference spectrum after subtraction of fitted emission lines & n \\
\texttt{PL\_ABSCONT} &  \pfit\ reference spectrum polynomial continuum fit & n \\
\texttt{PL\_ABSLINE} &  \pfit\ continuum and emission line subtracted reference spectrum  & n \\
\texttt{PL\_ABSFIT} &  \pfit\ reference spectrum absorption lines fit & n \\
\end{tabular}
\tablefoot{ds: MXDF, UDF10 or MOSAIC. id: catalog identifier. \\
Req: required y (yes) or n (no)
} 
\end{table}

\begin{table}
\caption{Source content: images.}   
\label{tab:source_images}
\centering
\begin{tabular}{p{3.0cm}p{4.5cm}p{0.5cm}}
\hline 
Name & Description & Req. \\ 
\hline 
\texttt{MUSE\_WHITE} & reconstructed white-light source image  & y \\ 
\texttt{MUSE\_EXPMAP} & exposure map source image (in number of exposure) & y \\ 
\texttt{ORI\_CORR\_REF} & reference correlation image for ORIGIN & n \\ 
\texttt{ORI\_CORR\_SEG} & reference segmentation image for ORIGIN & n \\ 
\texttt{ORI\_CORR\_lid} & correlation image for ORIGIN line lid & n \\ 
\texttt{ORI\_MAXMAP} & Maximum of correlation image for all ORIGIN sources & n \\ 
\texttt{HST\_ff} & HST image in filter ff centered on source & n \\ 
\texttt{HST\_SEGRAF} & HST segmentation image for Rafelski catalog and centered on source & n \\ 
\texttt{HST\_SEGCAN} & HST segmentation image for CANDELS catalog and centered on source & n \\ 
\texttt{HST\_SEG3D} & HST segmentation image for 3D-HST catalog and centered on source & n \\ 
\texttt{NB\_EMI\_lll} & narrowband source image for lll emission line & y \\ 
\texttt{NB\_ABS\_lll} & narrowband source image for lll absorption line & n \\ 
\texttt{SEG\_EMI\_lll} & narrowband source segmentation image for lll emission line & y \\ 
\texttt{SEG\_ABS\_lll} & narrowband source image for lll absorption line & n \\ 
\texttt{NB\_EMI\_COMBINED} & combined emission lines narrowband image & n \\ 
\texttt{NB\_ABS\_COMBINED} & combined absorption lines narrowband image & n \\ 
\texttt{SEG\_EMI\_COMBINED} & combined emission lines narrowband segmentation image & n \\ 
\texttt{SEG\_ABS\_COMBINED} & combined absorption lines narrowband segmentation image & n \\ 

\end{tabular}
\tablefoot{ff: F435W, F606W, F775W, F850LP or F160W \\
lll: narrowband line name (see table~\ref{tab:linename}).\\
Req: required y (yes) or n (no)
} 
\end{table}

\begin{table}
\caption{Source content: tables.}   
\label{tab:source_tables}
\centering
\begin{tabular}{p{1.8cm}p{5.5cm}p{0.5cm}}
\hline 
Name & Description & Req\tablefootmark{a} \\ 
\hline 
\texttt{PL\_Z} & table of fitted redshifts and related information for the source (columns description\tablefootmark{b} in Table~\ref{tab:ztable})& y \\ 
\texttt{PL\_LINES} & table of fitted lines parameters for the source (columns description\tablefootmark{b} in Table~\ref{tab:linetable})& y \\ 
\texttt{NB\_PAR} & table of narrow bands for the source (columns description\tablefootmark{b} in Table~\ref{tab:nbtable}) & y \\ 
\texttt{ORI\_LINES} & table of ORIGIN line detections for the source (columns description in Table~\ref{tab:source_table_orilines}) & n \\ 
\texttt{DR2\_CAT} & table of all DR2 sources in the source field of view (columns description in Table~\ref{tab:source_table_dr2}) & y \\ 
\texttt{ORIG\_CAT} & table of all ORIGIN sources in the source field of view (columns description in Table~\ref{tab:source_table_ori})& n \\
\texttt{HST\_CAT} & table of all HST RAFELSKI sources in the source field of view (columns description in Table~\ref{tab:source_table_hst})& n \\  
\texttt{CANDELS\_CAT} & table of all HST CANDELS sources in the source field of view (columns description in Table~\ref{tab:source_table_can}) & n \\  
\texttt{HST3D\_CAT} & table of all HST 3D-HST sources in the source field of view (columns description in Table~\ref{tab:source_table_3d})& n \\  
\texttt{SED} & table with \magphys\ and \prospector\ SED fit (columns description in Table~\ref{tab:source_table_sed}) & n \\  
\end{tabular}
\tablefoot{
\tablefoottext{a}{Required: y (yes) or n (no).}\\
\tablefoottext{b}{Note the omission of the two columns ID and data set in the source tables with respect to the main tables.} \\

} 
\end{table}

\begin{table}
\caption{Column description of source table \texttt{ORI\_LINES}}   
\label{tab:source_table_orilines}
\centering
\begin{tabular}{m{1.7cm}m{6.5cm}}
\hline 
Col. name & Description \\ 
\hline 
ID & ORIGIN source identifier \\ 
\texttt{num\_line} & ORIGIN line identifier \\ 
ra & right ascension ($\degree$) \\
dec & declination  ($\degree$) \\
lbda & wavelength (\AA) \\
comp & matched continuum segment \\
\texttt{T\_GLR} & correlation peaked value \\
STD & S/N peaked value \\
purity & purity estimate \\
\hline
\end{tabular}
\end{table}

\begin{table}
\caption{Column description of source table \texttt{DR2\_CAT}}   
\label{tab:source_table_dr2}
\centering
\begin{tabular}{m{1.7cm}m{6.5cm}}
\hline 
Col. name & Description \\ 
\hline 
ID & MUSE source identifier \\ 
DATASET & MUSE data set\tablefootmark{a} \\ 
Z & redshift in vacuum \\
ZCONF & redshift confidence (1-3) \\
RA & right ascension ($\degree$) \\
DEC & declination  ($\degree$) \\
\texttt{ORI\_ID} & matched ORIGIN identifier \\
\texttt{RAF\_ID} & matched RAFELSKI identifier \\
\texttt{CANDELS\_ID} & matched CANDELS identifier \\
\texttt{C3DHST\_ID} & matched HST-3D identifier \\
DIST & offset from source center (arcsec) \\
\hline
\end{tabular}
\tablefoot{
\tablefoottext{a}{MXDF, UDF10, or MOSAIC} \\
}
\end{table}

\begin{table}
\caption{Column description of source table \texttt{ORIG\_CAT}.}   
\label{tab:source_table_ori}
\centering
\begin{tabular}{m{1.7cm}m{6.5cm}}
\hline 
Col. name & Description \\ 
\hline 
ID & ORIGIN source identifier \\ 
ra & right ascension ($\degree$) \\
dec & declination  ($\degree$) \\
waves & list of detected wavelengths (\AA) \\
\texttt{T\_GLR} & correlation peaked value \\
STD & S/N peaked value \\
purity & purity estimate \\
DIST & offset from source center (arcsec) \\
\hline
\end{tabular}
\end{table}

\begin{table}
\caption{Column description of source table \texttt{HST\_CAT}.}   
\label{tab:source_table_hst}
\centering
\begin{tabular}{m{1.7cm}m{6.5cm}}
\hline 
Col. name & Description \\ 
\hline 
ID & RAFELSKI source identifier \\ 
RA & right ascension ($\degree$) \\
DEC & declination  ($\degree$) \\
\texttt{MAG\_F775W} & AB F775W magnitude \\
\texttt{Z\_BPZ} & Bayesian photometric redshift \\
\texttt{ZMIN\_BPZ} & lower limit for Bayesian photometric redshift \\
\texttt{ZMAX\_BPZ} & upper limit for Bayesian photometric redshift \\
DIST & offset from source center (arcsec) \\
\hline
\end{tabular}
\tablefoot{Columns description are described in \cite{Rafelski2015}}
\end{table}

\begin{table}
\caption{Column description of source table \texttt{CANDELS\_CAT}.}   
\label{tab:source_table_can}
\centering
\begin{tabular}{m{1.7cm}m{6.5cm}}
\hline 
Col. name & Description \\ 
\hline 
ID & CANDELS source identifier \\ 
RA & right ascension ($\degree$) \\
DEC & declination  ($\degree$) \\
\texttt{MAG\_F435W} & AB F435W magnitude \\
\texttt{MAG\_F606W} & AB F606W magnitude \\
\texttt{MAG\_F775W} & AB F775W magnitude \\
\texttt{MAG\_F850LP} & AB F850LP magnitude \\
\texttt{MAG\_F160W} & AB F160W magnitude \\
DIST & offset from source center (arcsec) \\
\hline
\end{tabular}
\tablefoot{Columns description are described in \cite{Whitaker2019}}
\end{table}

\begin{table}
\caption{Column description of source table \texttt{HST3D\_CAT}.}   
\label{tab:source_table_3d}
\centering
\begin{tabular}{m{1.7cm}m{6.5cm}}
\hline 
Col. name & Description \\ 
\hline 
ID & 3D-HST source identifier \\ 
RA & right ascension ($\degree$) \\
DEC & declination  ($\degree$) \\
\texttt{MAG\_F775W} & AB F775W magnitude \\
\texttt{z\_best\_s} & redshift type \\
\texttt{z\_best} & redshift \\
\texttt{z\_best\_l95} & lower limit for redshift \\
\texttt{z\_best\_u95} & upper limit for redshift \\
DIST & offset from source center (arcsec) \\
\hline
\end{tabular}
\tablefoot{Columns description are described in \cite{Skelton2014}}
\end{table}

\begin{table}
\caption{Column description of source table \texttt{SED}.}   
\label{tab:source_table_sed}
\centering
\begin{tabular}{m{1.9cm}m{6.1cm}}
\hline 
Col. name & Description \\ 
\hline 
wavelength & wavelength in \AA \\ 
flux\_prospector & \prospector\ SED flux\\
flux\_magphys & \magphys\ SED flux\\
hst\_wave & HST filter wavelength in \AA \\ 
flux\_hst\_obs & HST observed flux\\
\hline
\end{tabular}
\tablefoot{
Flux unit are \ergsa{-20}
}
\end{table}

%('wavelength','flux_prospector','flux_magphys','hst_wave','flux_hst_obs','flux_hst_prospector')

%\clearpage
%\newpage 
\subsection{The AMUSED web interface}
% amused.tex
\label{sec:amused}

\begin{figure}[hbt!]
\begin{center}
\includegraphics[width=1\columnwidth]{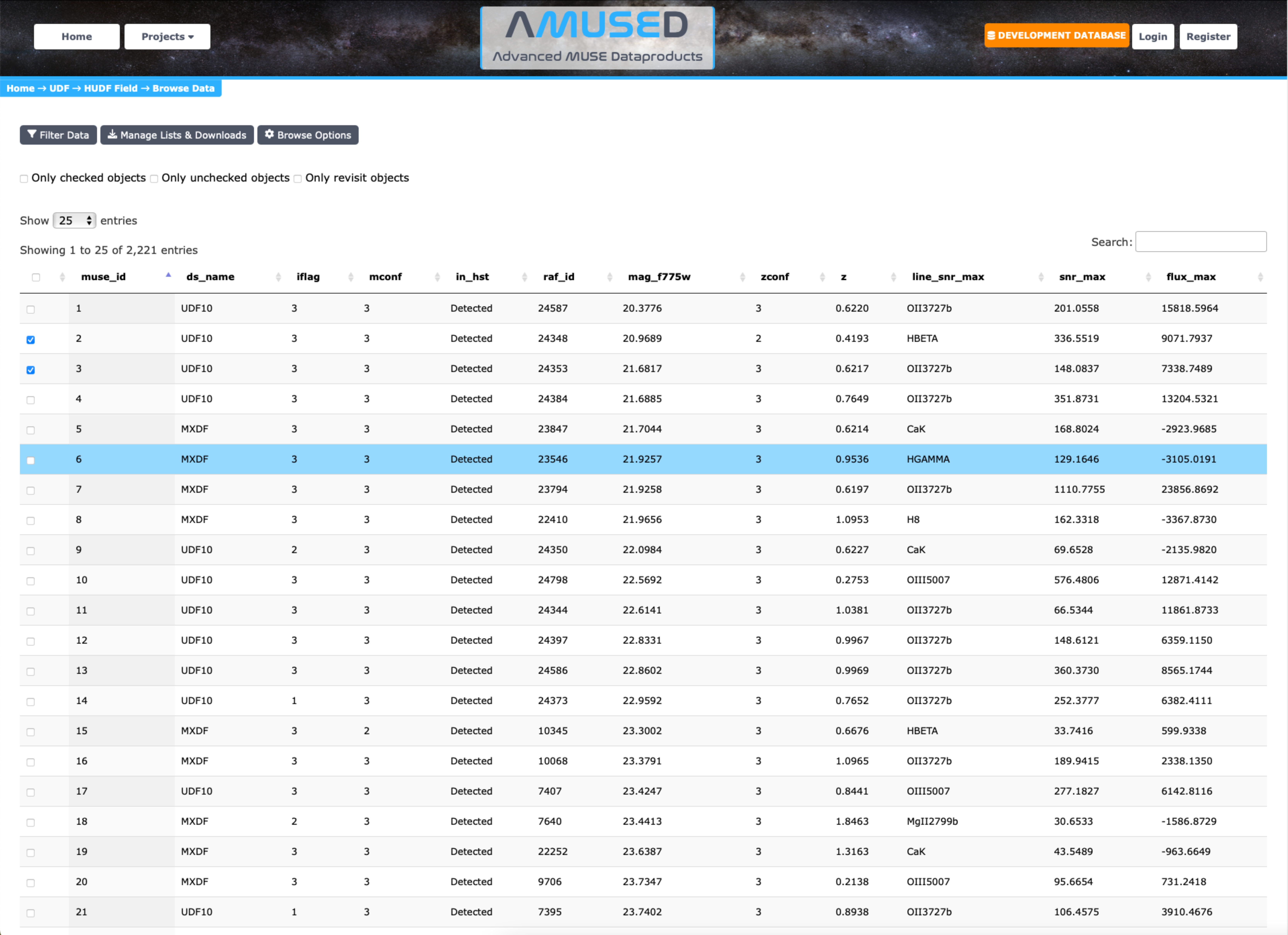}
\caption{AMUSED web interface source selection window.}
\label{fig:amused1}
\end{center}
\end{figure}

\begin{figure}[hbt!]
\begin{center}
\includegraphics[width=1\columnwidth]{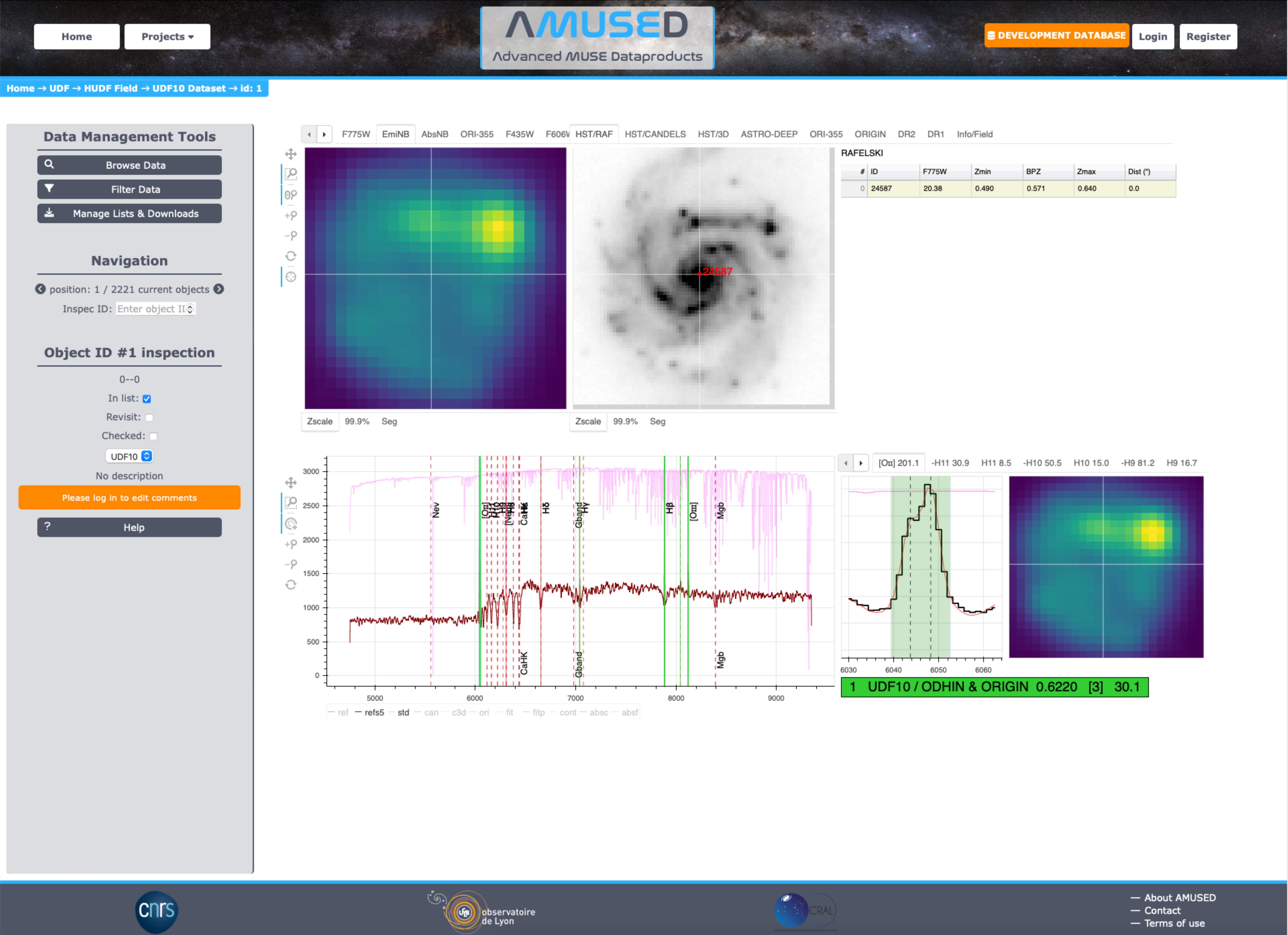}
\caption{AMUSED web interface source inspection window.}
\label{fig:amused2}
\end{center}
\end{figure}

AMUSED is a public web interface for inspection and retrieval of MUSE data products. The data are organized in a PostgresSQL relational database\footnote{\hyperlink{https://www.postgresql.org}{www.postgresql.org}}. User can perform advanced source selection: for example, selecting all MUSE sources with \lya\ flux higher than \ergslineb{8}{-19} and S/N $>$ 7, redshift in the $z = 3-4$ range, and without HST counterparts (Fig.~\ref{fig:amused1}). The resulting list can be visually inspected (Fig.~\ref{fig:amused2}) by looking to spectra, emission and absorption lines fit, reconstructed broadband and narrowband images and ancillary information (e.g., HST images). This process allows users to refine the selection, possibly taking notes for sources of interest, and then download the selected data. Exported data can be tables (Sect.~\ref{sec:catalogs}) in CSV or FITS format, interactive html visualization files and sources files in the MPDAF multi-fits format (Sect.~\ref{sec:sources}).

The user can also download the full data set without the need to inspect the sources. The final reduced datacubes for \mxdf, \udft,\ and \mosaic\ are also make available (Sect.~\ref{sec:cubes}). AMUSED is accessible at \hyperlink{https://amused.univ-lyon1.fr}{https://amused.univ-lyon1.fr}.

%\FloatBarrier

%\clearpage
%\newpage 
\subsection{Software}
\label{sec:soft}
In Table~\ref{tab:soft} we give the list of software developed in the context of this data release. We make those who are stable enough and at least partially documented available on the musevlt github directory\footnote{\href{https://github.com/musevlt}{https://github.com/musevlt}}. All software programs are python 3 modules. 

The released software should work for any other MUSE data sets. Some may require some minor adaptation for non-deep-field observations (e.g., \origin). Although some of these software programs are very specific to MUSE (e.g., \textsf{muse-psfr}), most of them should also work with non-MUSE IFU data, provided that the \mpdaf\ format is used.

\begin{table*}[htbt!]
\caption{List and status of specific software.}  

\begingroup 
\tiny
\label{tab:soft}
\centering
\begin{tabularx}{\textwidth}{lXllllX}
\hline 
Name & Description & Status & Availability & Version & Documentation & Reference\\ 
\hline 
\mpdaf & MUSE python data analysis framework & stable & pip, \href{https://github.com/musevlt/mpdaf}{github} & 3.6 & \href{https://mpdaf.readthedocs.io}{mpdaf.readthedocs.io} & \cite{Bacon2016, Piqueras2019} \\ 
\origin & Blind emission line source detection & stable & \href{https://github.com/musevlt/origin}{github} & 3.2 & \href{https://muse-origin.readthedocs.io}{muse-origin.readthedocs.io} & \cite{Mary2020} \\ 
\odhin & De-blending of MUSE source using HST images & stable & \href{https://github.com/musevlt/odhin}{github} & 1.0 & \href{https://odhin.readthedocs.io}{odhin.readthedocs.io}  & this paper and \cite{bacher2017thesis}\\
\pfit & Emission and absorption line fitting & stable & \href{https://github.com/musevlt/pyplatefit}{github} & 0.7 & \href{https://pyplatefit.readthedocs.io}{pyplatefit.readthedocs.io} & this paper\\
\pymarz & Redshift estimation & dev & & 0.2 \\
\se & Interactive source evaluation and inspection & dev & & 1.3 \\
\textsf{musered} & Data reduction management & stable & \href{https://github.com/musevlt/musered}{github} & 0.3 & \href{https://musered.readthedocs.io}{musered.readthedocs.io}  & this paper\\
\textsf{musex} & Database management & dev & & 0.3 \\
\textsf{imphot} & MUSE PSF estimation based on HST photometry & stable & \href{https://github.com/musevlt/imphot}{github} & 0.2 & \href{https://imphot.readthedocs.io}{imphot.readthedocs.io}  & \cite{Bacon2017}\\
\textsf{muse-psfr} & MUSE PSF estimation based on AO telemetry & stable & \href{https://github.com/musevlt/muse-psfr}{github} & 1.1 & \href{https://muse-psfr.readthedocs.io/en/latest}{muse-psfr.readthedocs.io} & \cite{Fusco2020}\\
\hline 
\end{tabularx}
\tablefoot{Software in development ("dev" in Status column) will be released later.}
\endgroup
\end{table*}
%\FloatBarrier

\clearpage
\newpage 
\section{Advanced data reduction}
\label{sec:datared}

\label{sec:DRS}
\subsection{Data reduction pipeline}

We first run the raw science data through the standard MUSE pipeline (\citealt{Weilbacher2020}), using the development version\footnote{Since v2.8.3 all development features used here are part of the public version.}. Individual exposures are processed by the \textit{scibasic} recipe with corresponding daily calibrations (flat fields, bias, arc lamps, twilight exposures) and geometry table (one per observing run) to produce the pixel tables (hereafter referred to as \textit{pixtable}).

For the MXDF, we applied two changes related to the dark subtraction and the overscan parameters. To improve the bias subtraction, the overscan parameter was set to \emph{vpoly:15,1.00001,1.00001} for all the recipes using that parameter (\citealt{Weilbacher2020} Sect.~3.1).
A sequence of long darks from June to August 2018 was used to produce a master dark, and then to derive a smooth model for each CCD as described in \cite{Weilbacher2020} Sect.~3.2.

The pipeline recipe \textit{scipost} is then used to perform astrometric and flux calibrations on each pixable. For the MXDF, we computed a median standard response for each GTO run, using the standard exposures observed during that run and excluding those taken under non-photometric conditions.

The \textit{scipost} recipe is run a first time in a "fast" mode (no sky subtraction, no Raman contamination correction ) to produce images that are used to compute the centering offsets. Those offsets are calculated with the PSF fitting algorithm described in \citet{Bacon2017}, Sect.~5.1, relative to the HST ACS images from the XDF data release \citep{Illingworth2013}.

The \textit{scipost} recipe is then run a second time with sky subtraction, Raman correction and self-calibration but without applying the astrometry (\emph{--astrometry=false}) %\todo{Peter: and no RV corr. I guess ?}\ans{Simon: no it was not desactivated}
 to produce pixtables that we can use for the {superflat} (see Sect. \ref{sec:superflat}). The same recipe is run again to produce datacubes from these pixtables, using the astrometry and the offsets computed previously. In this case \textit{scipost} will detect that the other steps have been done in the previous run so it will only need to produce the datacubes. The goal of this process is to minimize the computation time since we have to process hundreds of exposures.

For the different steps that require it (self-calibration, sky-subtraction and post-processing sky residual subtraction ZAP software) we use a source mask that we computed on the combined datacube from a previous version of the reduction.

\subsection{New version of the self-calibration}

We modified the recipe \textit{scipost} to include the self-calibration algorithm. This algorithm was previously developed in the Python package MPDAF \citep{Piqueras2019}. Since this algorithm uses the average sky level as a reference to calculate the correction, we had to interrupt the \textit{scipost} recipe before the sky subtraction, save the pixtables, reload them into Python and run the self-calibration, save again and resume processing with the MUSE pipeline. The overhead was huge, especially because the pixtables are very large files (8.6 GB) and saving and loading hundreds of them to remote storage takes time. Since the algorithm was already implemented as a C extension, moving it into the MUSE pipeline was an obvious choice to optimize processing, and it was also a great opportunity to make it available to the community.

Before porting the code to the MUSE pipeline, we developed a new version of the algorithm. When comparing the results of the first version used for DR1 with the flat field correction method \textit{CubeFix} developed by Cantalupo (in preparation; see \citealt{Cantalupo2019} for a description), it was clear that we could do better.
The new algorithm is inspired by both \textit{CubeFix} and the old version, and is described in more detail in \cite{Weilbacher2020}. 
While \textit{CubeFix} operates on datacubes, self-calibration operates directly on pixtables and could therefore be integrated into the \textit{scipost} recipe. 
%\todo{Should this paragraph discuss, *what* the change exactly is? I don't remember spelling that out in the pipeline paper. (And I actually don't know, since I never used the autocalib in the MPDAF version.)}
%\ans{Roland: I give a try, but then we need to explain the full detail of the algorithm and that would enter too much detail, I leave it like this for the moment}

\begin{figure}[htbp]
\begin{center}
\includegraphics[width=1\columnwidth]{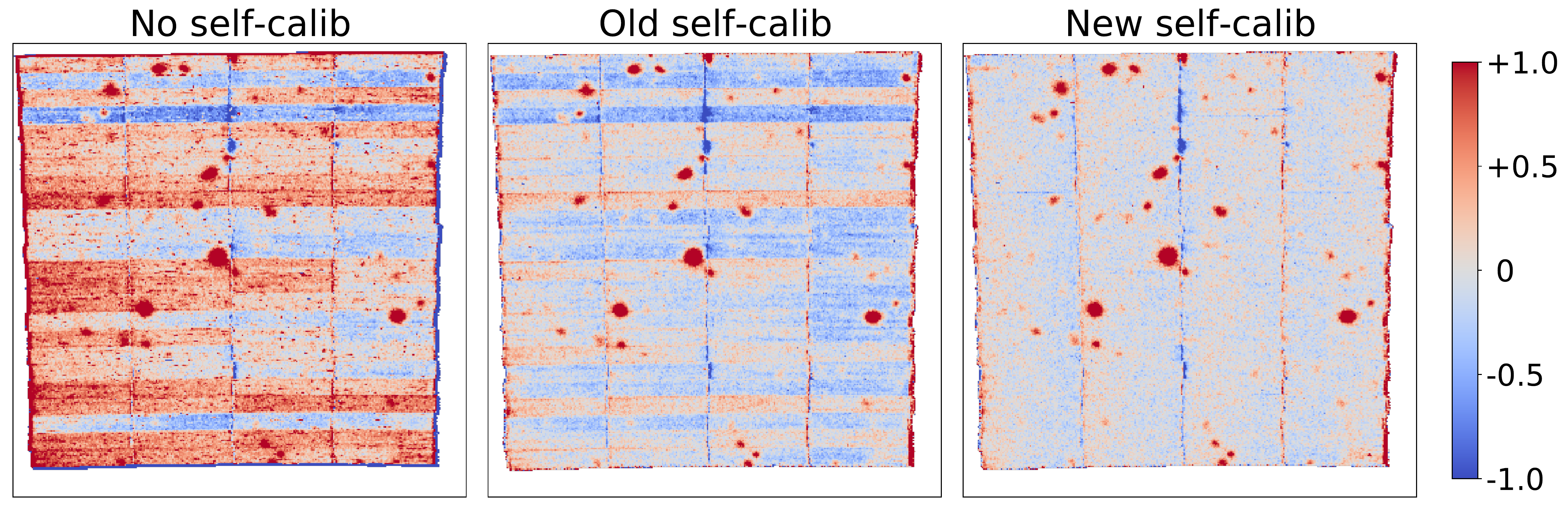}
\caption{Illustration of the new self-calibration. Left panel: background subtracted white-light image for one exposure after the \textit{scipost} recipe. Center panel: the same processed with the old version of the self-calibration. Right panel: with the new version. Flux units are \ergsa{-20}.}
\label{fig:autocalib-zap-comparison}
\end{center}
\end{figure}

\subsection{New version of ZAP (2.1)}
\label{sec:zap}

The ZAP software \citep{Soto2016} is a PCA-based tool developed to remove sky residuals in MUSE empty fields left by the imperfect sky subtraction pipeline process.

With the new version of self-calibration giving better results, it became clear that some of the artifacts we were observing in the reduced datacubes were due to ZAP. It introduced spatial variations for some pixels, which were visible in the white light images, and was also responsible for wiggles in the red part of the spectra. We corrected these problems with a new version of ZAP (version 2.1), where we made several changes: First, for the spatial variations we replaced the custom implementation of the PCA with the implementation from scikit-learn \citep{scikit-learn}, which is also much faster.
Second, for the wiggles in the red part, we used the median continuum filter instead of the weighted one (which is a median filter weighted by the median sky level) and we increased the width of the filter window to 300 pixels. The default values were changed as well in version 2.0 for the window size and 2.1 for the filter type.

Third, the new version uses only one sky segment by default, which means that the cube is no longer split along the wavelength axis. Originally ZAP used 11 segments, whose purpose was to have coherent groups of sky emission lines, with a smaller number of eigenvalues per segment. It also allowed the computation to be parallelized. But the segments were also responsible for continuum oscillations, and made the choice of the number of eigenvalues per segment very difficult and very sensitive. With only one segment the performance of the sky subtraction is much better, thanks to the higher correlation between sky lines on the whole wavelength range. It also easier to control the power of the sky subtraction since the number of eigenvalues used for the sky reconstruction must be chosen or determined automatically for each segment.

Fourth, another issue was that sometimes ZAP suppressed signal for very bright lines, such as bright \oii\ emitters. With the new version this is less likely to happen, thanks to the use of only one sky segment that provides a better estimation of the sky signal.

\begin{figure}[htbp]
\begin{center}
\includegraphics[width=1\columnwidth]{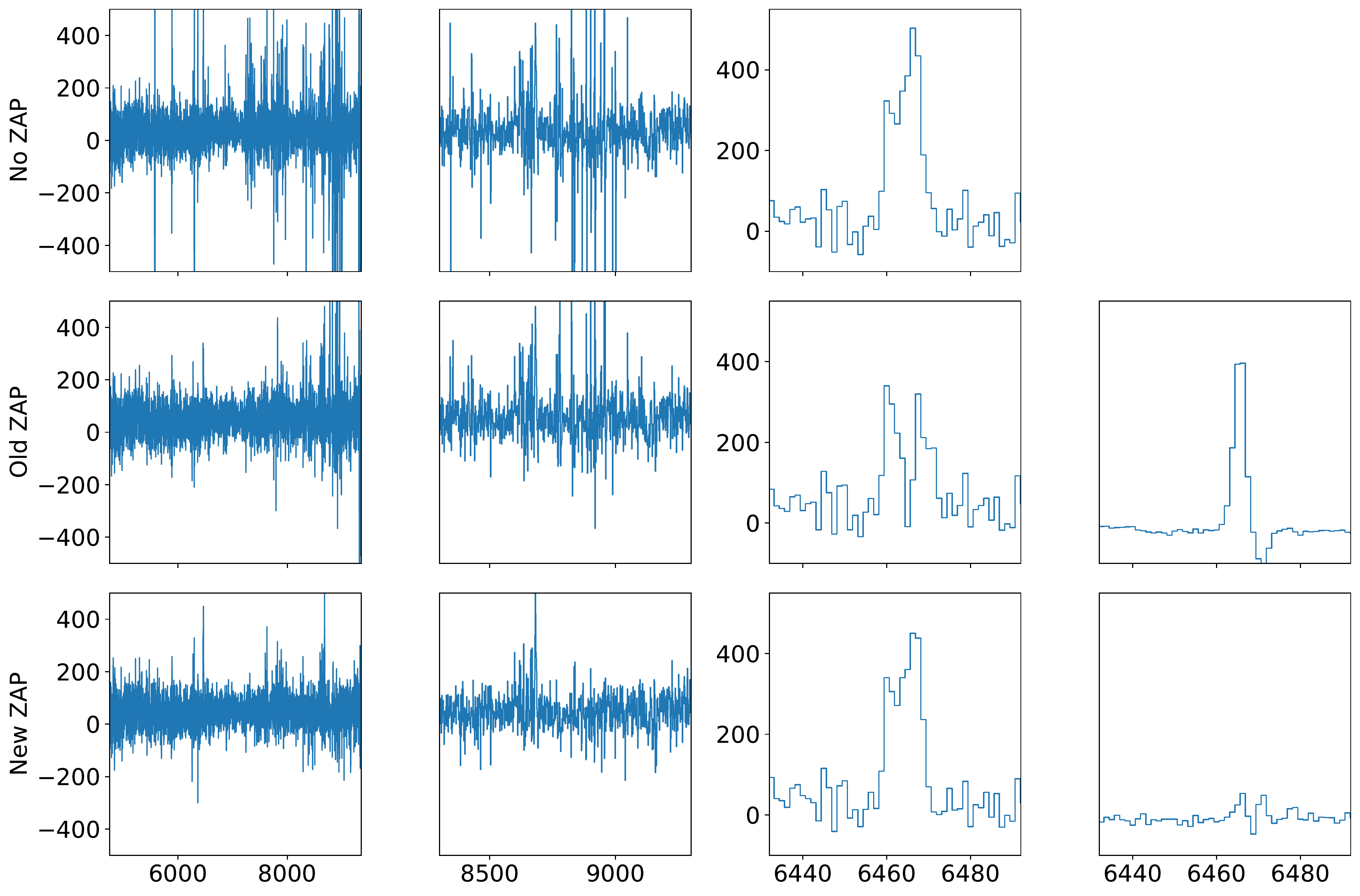}
\caption{Example of improved performance of the new ZAP version. The top, middle and bottom rows display respectively the non-zapped, old zapped, and new zapped spectra of an \oii\ emitter in the \mxdf. From left to right, the columns display: the full spectral range, the red part of the spectra, the \oiid\ emission lines and the difference with the no ZAP spectrum. Flux units are \ergsa{-20}.
}
\label{fig:zap-before-after}
\end{center}
\end{figure}

As shown in Fig.~\ref{fig:zap-before-after} this new version of ZAP has significantly improved the accuracy of sky subtraction and reduced the risk of capturing signal from bright emission lines. However, it revealed another issue: the flux at some spaxels are now sometimes rising or falling in the far red (above $8800\AA$; see the bottom left panel of Fig.~\ref{fig:superflat-autocalib}). This appears to be an instrumental effect, likely related to 2nd-order effects in the twilight-skyflat exposures used in the first step of the data reduction, and the old version of ZAP was removing this signal with its two last segment, instead of the sky emission in this wavelength range. Fortunately, this problem is now corrected by the {superflat} process described in the next subsection (see the bottom right panel of Fig.~\ref{fig:superflat-autocalib}).

Another change with respect to DR1 is that previously we were running ZAP on the individual exposures to get the best sky subtraction. With the {superflat} correction done on the individual datacubes (see the next section) this appeared not anymore necessary and we could run ZAP only once on the combined cube, which saves some computation time.

%\begin{figure}[htbp]
%\begin{center}
%\includegraphics[width=1\columnwidth]{zap-red-effect.pdf}
%\caption{\todo{Example of the red effect ? Also show the better sky sub \& how wiggles in the red are fixed}}
%\label{fig:zap-red-effect}
%\end{center}
%\end{figure}

\subsection{Superflat}
\label{sec:superflat}

%\begin{figure}[htbp]
%\begin{center}
%\includegraphics[width=1\columnwidth]{superflat.pdf}
%\caption{\todo{Superflat white-light + red}}
%\label{fig:superflat}
%\end{center}
%\end{figure}

In DR1, we masked the inter-stack holes in each individual exposure before combining them (\citealt{Bacon2017}, Sect.3.1.3). However, in addition to producing lower S/N region in the final datacube, the masking was never perfect and left some artifacts. To improve the process, we build a \textit{superflat} for each datacube, by combining many exposures where we cancel the rotation and dithering. This is done by adding a \texttt{MUSE\_SUPERFLAT\_POS} environment variable in the MUSE pipeline, which allows the RA, DEC, and DROT keywords to be overridden by the values of the exposure for which the superflat is built. This way the "instrumental grid" is the same for all the datacubes and the sources move because of the rotation and dithering of the field. As we have sparse fields, combining those exposures with sigma-clipping removes the astrophysical signal and produces the superflat, where only the instrumental residuals remain.
 
For each exposure a superflat is built with 36 exposures from the same observing run, possibly supplemented with additional exposures from the previous or next run when there were not enough exposures in a single run.
The superflat is subtracted from the science exposure\footnote{After a few experiments, we found that subtracting the superflat from the science exposure produced better results than conventional division as it is typically done in imaging.}.
 To minimize the computation time, we build the datacubes tailored to each exposure from the pixtables that have been computed previously and where the self-calibration, sky-subtraction and Raman correction have already been done. Therefore, it is sufficient to produce the datacubes for each exposure after applying the astrometry with the modified RA, DEC and DROT. Even with this shortcut, applying the superflat correction to hundreds of exposures is computationally intensive.

\begin{figure}[htbp]
\begin{center}
\includegraphics[width=1\columnwidth]{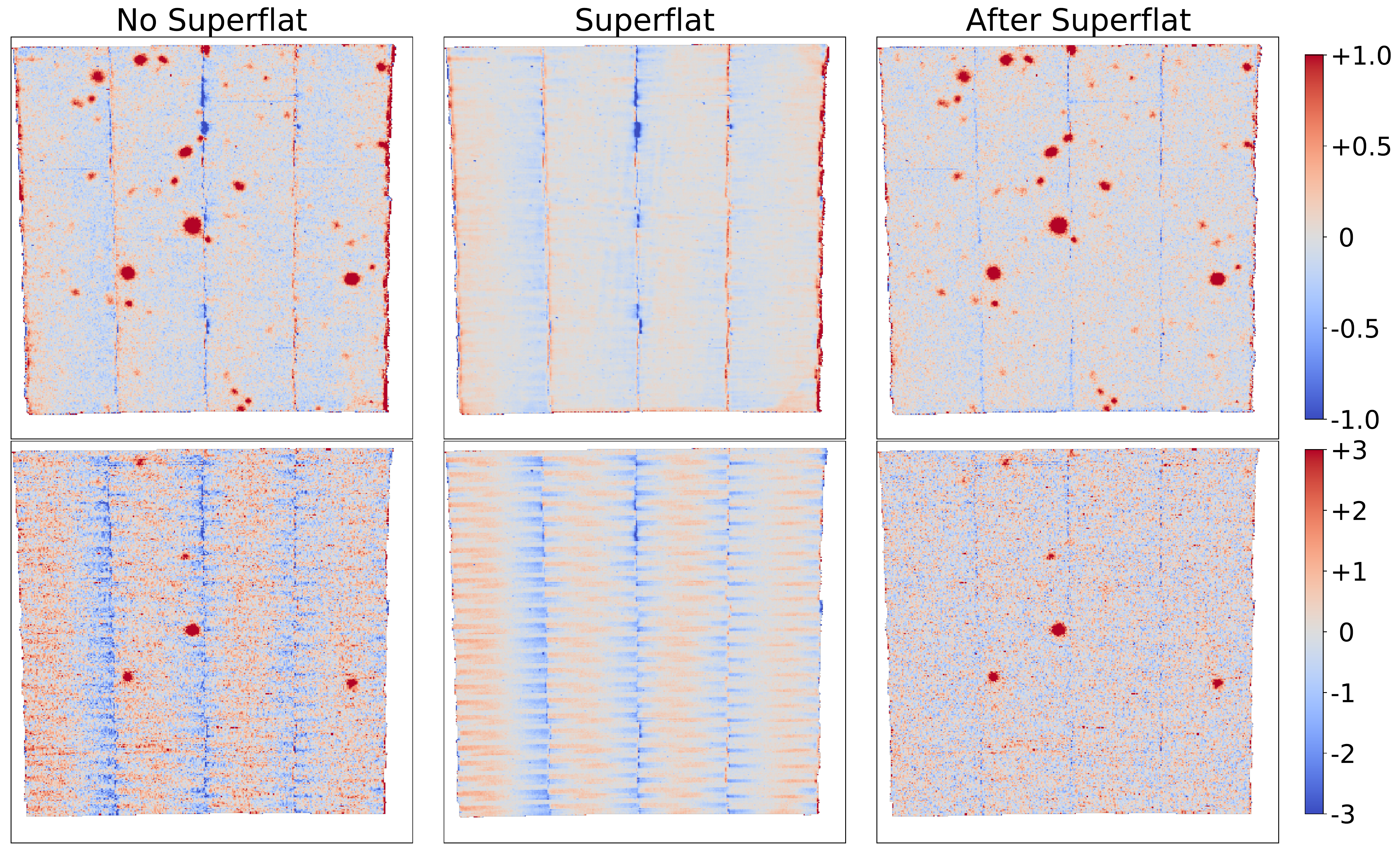}
\caption{Example of superflat applied to a single exposure. The top and bottom rows display respectively the white light and red part (9000-9350\AA) images. The left and right columns show the correction brought by the superflat (central column). Flux units are \ergsa{-20}. 
}
\label{fig:superflat-autocalib}
\end{center}
\end{figure}

As shown in the top panels of Fig.~\ref{fig:superflat-autocalib} the superflat gives excellent results on the white light image. Most of the residuals and holes in the inter-stacks are corrected, which means that we no longer need to mask these areas. We also note that red effect in the $9000-9350 \AA$ band (see the previous section) is also corrected very well (Fig.~\ref{fig:superflat-autocalib} bottom panels).

\subsection{\texttt{musered}}

To deal with the complexity of this data reduction, given the number of exposures and the number of reduction steps, we developed a Python package called \texttt{musered} on top of the MUSE pipeline. This package uses \texttt{python-cpl}\footnote{\href{https://github.com/olebole/python-cpl}{https://github.com/olebole/python-cpl}} to run the pipeline recipes, and it uses a database to gather information about the raw files and to keep track of everything that has been done, for each file, each recipe, and each version of the reduction. \texttt{musered} takes care of file associations when running a recipe, to find the correct calibration files, and provides a convenient command line interface to run the various recipes, either from the pipeline or from custom Python recipes. It also makes it possible to inspect what was done for each exposure, when it was processed, and with which parameters and which calibration files.

The \mxdf\ data set consists of 69 different nights and 373 exposures. It was then very time consuming to carefully inspect each exposure and each calibration frame. We therefore relied on several measurements and plots to identify problematic exposures, which sometimes led to the exclusion of a few problematic calibration sequences\footnote{1 BIAS, 1 FLAT, 1 WAVE and 1 TWILIGHT exposures were discarded.}. %\todo{Peter: Would it be interesting to mention, how many problematic calibrations were discarded?}\ans{Roland, footnote added}
In this case, the sequence is flagged and another sequence from the previous or next night is used.
The entire process costs 397 min of computing time per exposure on an 80-core workstation, of which 90\% was due to the superflat computation.
It took 103 days to completely reduce the \mxdf\ data set.

%\begin{table}
%\begin{center}
%\begin{tabular}{lr}
%Recipe & Median time (min) \\
%muse\_scibasic & 6.24 \\
%muse\_scipost\_rec & 2.72 \\
%imphot & 91.66 \\
%muse\_scipost & 16.01 \\
%muse\_scipost\_make\_cube & 3.71 \\
%superflat & 202.67 \\
%imphot\_scale\_superflat & 47.62 \\
%fsf & 0.73 \\
%Total & 397.00 \\
%\end{tabular}
%\caption{Median computation time of the 377 exposures of the MXDF for the different recipes.}
%\label{tab:computation-times}
%\end{center}
%\end{table}

\subsection{Post-processing}
\label{sec:postdrs}

Before combining the datacubes of all individual exposures, we identify those that should be discarded due to poor quality. The spatial PSF of each exposure is estimated with the \psfrec\ tool, as described in Sect.~\ref{sec:fsf}. Twenty-two exposures with a PSF at 7000 \AA\ greater than 0\farcs8 were rejected. The remaining 338 datacubes were combined using a 5$\sigma$ sigma clipping scheme. The residuals left by sky subtraction were then removed using the new version of ZAP, as described in Sect.~\ref{sec:zap}. An additional check was performed on the resulting datacube to confirm that no offset was left in the background level at each wavelength plane. 

As shown in \citealt{Weilbacher2020}, Sect.~4.6, the pipeline propagated variance of the datacube does not take into account the additional correlated noise due to the interpolation process. In DR1, we replaced the propagated variance of each individual datacube with a noise estimate derived from the datacube after masking the light sources, which is then rescaled to account for the impact of the correlated noise (\citealt{Bacon2017}, Sect.~3.5.1). The limitation is that this noise estimate is not valid for bright sources that are not sky-dominated. Another inaccuracy is that this noise estimate is by construction constant across the field of view and thus does not account for possible flat field variation. Thus, to address these limitations, we implemented a different scheme for \mxdf: we computed the median ratio between the propagated variance and the variance estimated on the datacube itself after adjusting for correlated noise. This ratio has a slight evolution with the wavelength from 2.45 in the blue to 2.55 in the red. The propagated variance is then simply multiplied by this factor.

\clearpage
\newpage 
\section{\odhin}
% odhin.tex annex
\label{sec:odhin}

%%%%% Macro
\def\b0{\boldsymbol{0}}

\def\ba{\boldsymbol{a}}
\def\bb{\boldsymbol{b}}
\def\bd{\boldsymbol{d}}
\def\bg{\boldsymbol{g}}
\def\bs{\boldsymbol{s}}
\def\bt{\boldsymbol{t}}
\def\bu{\boldsymbol{u}}
\def\bv{\boldsymbol{v}}
\def\bx{\boldsymbol{x}}
\def\by{\boldsymbol{y}}
\def\bz{\boldsymbol{z}}

\def\bA{\boldsymbol{A}}
\def\bB{\boldsymbol{B}}
\def\bC{\boldsymbol{C}}
\def\bD{\boldsymbol{D}}
\def\bE{\boldsymbol{E}}
\def\bF{\boldsymbol{F}}
\def\bG{\boldsymbol{G}}
\def\bH{\boldsymbol{H}}
\def\bI{\boldsymbol{I}}
\def\bL{\boldsymbol{L}}
\def\bM{\boldsymbol{M}}

\def\bP{\boldsymbol{P}}
\def\bQ{\boldsymbol{Q}}
\def\bR{\boldsymbol{R}}
\def\bS{\boldsymbol{S}}
\def\bT{\boldsymbol{T}}
\def\bU{\boldsymbol{U}}
\def\wbU{\widetilde{\boldsymbol{U}}}
\def\bV{\boldsymbol{V}}
\def\bX{\boldsymbol{X}}
\def\bY{\boldsymbol{Y}}
\def\bZ{\boldsymbol{Z}}

\def\beps{\boldsymbol{\epsilon}}
\def\bSigma{\boldsymbol{\Sigma}}
\def\bbeta{\boldsymbol{\beta}}
\def\br{\boldsymbol{r}}
\def\bw{\boldsymbol{w}}
\def\bW{\boldsymbol{W}}
\def\Fm{F^{-\textrm{min}}}
\def\FM{F^{\textrm{max}}}
\def\mean#1{\left< #1 \right>}

%%%%%

For more details of this whole procedure please refer to chapter 3 of \cite{bacher2017thesis}.
\subsection{Model}
\label{method_unmix}

\subsubsection{Notations}
\label{ssec:notations}
Let $\bM$ be a matrix composed of the elements $m_{i,j}$ and $\bv$ be a vector. Let $\diag(s_j)$ be a diagonal matrix with diagonal elements $s_j$.
$||\bM||_2$ is the Frobenius norm of matrix $\bM$.
$\Tr(\bM)$ is the trace of matrix $\bM$. $\bM^+$ is the pseudo-inverse of matrix $\bM$.  Hadamard (element-wise) product between a matrix $\bA$ and a matrix $\bB$ is noted $\bA \circ \bB$.

Let $\bY$ be the data matrix from MUSE of size $n \times \lambda$ where $n=pq$ is the number of spaxels (p,q are the two spatial axis) and $\lambda$ is the number of spectral bands.
Let $\bZ$ be the data matrix from HST of size  $N \times \Lambda$ with $N$ the number of image pixels, $\Lambda$ the number of filters ($\approx 4$). $\Lambda \ll \lambda$ and $n \ll N $.

Let $\bX$ be the unknown field with a high spatial and spectral resolution of size $N \times \lambda$.

\subsubsection{Model}
Data observed by MUSE is assumed to be the addition of spatially degraded $\bX$ data field and measurement noise $\mathcal{N} $. Spatial degradation is the multiplication of a convolution operation (by the FSF) and a subsampling operation. The spatial PSF (FSF) varies with the wavelength. We then have the following model:

\begin{equation}\begin{aligned}
\textrm{For } 1\leq l \leq \lambda, \quad \underset{n \times 1}{\bY_l} = \underset{n \times N}{\bB_l} \times \underset{N \times 1}{\bX_l}  + \underset{n \times 1}{\mathcal{N}}
\label{eq:MUSE}
\end{aligned},\end{equation}
with $\bB_l$ the spatial degradation matrix (subsampling and convolution), which can be written as $\bB_l=\bS\times \bC_l$ where $\bC_l$ is the convolution matrix by FSF at wavelength $l$ and $\bS$ is the subsampling matrix.

The HST observation, composed of images from different filters, is considered the result of a spectral degradation of $\bX$:
\begin{equation}\begin{aligned}
\underset{N \times \Lambda}{\bZ} = \underset{N \times \lambda }{\bX} \times \underset{\lambda \times \Lambda}{\bA}
\label{eq:HST}
\end{aligned}
,\end{equation}
with $\bA$ the matrix of the HST filter responses.

Let $\bH \in \mathbb{R}^{N\times k }$ be the segmentation matrix at HST resolution. This segmentation is obtained using \citep{Rafelski2015}.
Elements of $\bH$ are defined as $$h_{ij} = \left\{
\begin{array}{ll}
1 \textrm{ if the source }j \textrm{ is present on pixel }i,\\ 
0 \textrm{ otherwise}.
\end{array}
\right.
$$

It should be noted that we assume that MUSE and HST data have already been well aligned spatially.
We also note that we do not take the LSF into account  as it is negligible here.

\subsubsection{Assumptions}
We take the following assumptions and approximation: \begin{hypothesis}[Spatial separability]
All sources are spatially separated on HST images.
\end{hypothesis}

\begin{hypothesis}[Spatial invariability]
The spatial shape of a source does not vary with the wavelength
\end{hypothesis}

\begin{hypothesis}[Spectral invariability]
Sources are modeled with a unique spectrum.
\end{hypothesis}

\begin{hypothesis}[FSF invariability]
The transfer function HST-MUSE is known, spatially invariant, and can be approximated as constant over a given spectral band.
\end{hypothesis}

Assumption A1 is valid for almost all sources in the considered fields. When it is not valid, the proposed method cannot do further de-blending.
Assumptions A2 and A3 are also approximations that allow the sources to be separated. Note however that some sources, like close-by galaxies that are spatially resolved on MUSE, have velocity fields that will deform the observed spectra, breaking the A3 hypothesis. Other sources have emissions at certain wavelengths that differ greatly from the continuum, or even have no continuum, making the HST information irrelevant in these cases.
Assumption A4 is in practice quite accurate. The FSF of HST is now well known and MUSE FSF has also been well studied \citep{villeneuve2011psf,carfantan2014hdr}. The latter is modeled by a 2D circular Moffat 2D function. The Moffat scale parameter varies slowly with the wavelength such that FSF can be considered constant on tens or even hundreds of MUSE spectral bands.

Assumption A2 allows the model to be placed within the framework of a linear mixing model:
$$\bX\approx \underset{N \times k}{\bU} \times \underset{k \times \lambda}{\bD},$$
where $k$ is the number of sources in the data ($k  \ll N $), $\bD  \in \mathbb{R}^{k\times \lambda }$ is the sources spectra matrix , and  $\bU  \in \mathbb{R}^{n \times k }$ the intensity matrix.
We note that here we are talking of intensities, and not abundances like in other hyperspectral contexts (remote sensing), so we cannot use the classical sum-to-one constraint on the rows of this matrix.

All of these assumptions allow us to deduce an intensity matrix $\bU$ using HST images. From the MUSE data and this matrix $\bU$, we therefore try to directly estimate the matrix of the $\bD$ spectra, without explicitly reconstructing $\bX$.

\subsection{Method}

The strategy is as follows:

\begin{algorithm}[htb]
\caption{}
\label{alg:unmix_whole}
\begin{algorithmic}[1]

\For{ each (vectorized) image $\bZ_i$ of the HST}
\State Get intensity matrix $\bU_i$ from $\bZ_i$.
\For{ each spectral block  $l$ }
\State Degrade $\bU_i$ to MUSE spatial resolution to get $\widetilde{\bU}_{i,l}$ 
\State Reverse the system $\bY_bl=\widetilde{\bU}_{i,l}\bD_{i,l}$ to find $\widehat{\bD}_{i,l}$
\EndFor
\EndFor
\State Combine results: $\widehat{\bD}_l=\sum\limits_i a_{i,l}\widehat{\bD}_{i,l}, \quad 1\leq l \leq \lambda$
 where $a_{i,l}$ is the spectral response of the filter $i$ at the wavelength $l$.
\Statex
\end{algorithmic}
\end{algorithm}

%\begin{enumerate}
%\item For each (vectorized) image $\underbrace{\bZ_i}_{N\times 1}$ of the HST:
%\begin{enumerate}
%\item We obtain the intensity matrix $\bU_i$ from $\bZ_i$.
%\item For each spectral block $bl$, we spatially degrade $\bU_i$ to the spatial resolution of MUSE and get $\widetilde{\bU}_{i,l}$.
%\item For each spectral block  $bl$, we reverse the system $\bY_l=\widetilde{\bU}_{i,l}\bD_{i,l}$ to find $\widehat{\bD}_{i,l}$ 
%\end{enumerate}
%\item We then combine the different estimates made on each HST band by making a weighted average by the filter responses: 
%\begin{equation}\begin{aligned}
%\widehat{\bD}_l=\sum\limits_i a_{i,l}\widehat{\bD}_{i,l}, \quad 1\leq l \leq \lambda,
%\label{eq:comb_spe}
%\end{aligned}\end{equation} 
% where $a_{i,l}$ corresponds to the spectral response of the filter $i$ at the wavelength $l$.
%\end{enumerate}

Steps of lines (4) and (5) are costly to compute on all 3600 wavelength MUSE. As FSF varies slowly along wavelengths, we work on blocks (200 or 300 spectral bands) where we estimate a mean FSF over the block. This considerably reduces the computational cost since it is sufficient to pseudo-invert a single matrix $\widetilde{\bU}_{i,l}$ for each block $l$.
We note that we nevertheless obtain a full spectrum estimation, only the intensity matrix is assumed constant on the spectral bloc (A2).
Along this process of linear operations we propagates the variance given with the MUSE data. Thus, we get at the end a variance for each estimated spectrum.

\subsubsection{Intensity matrix estimation}
First step (line 2) is done using the segmentation map of HST sources. From this map we build a $N\times k$ matrix $\bH$, each column $j$ of $\bH$ corresponds to the presence/absence binary mask of the source $j$ in the vectorized HST field.
We then apply element-wise product between $\bH$ and HST image $\bZ_i$ to get the intensity matrix $\bU_i$:
\begin{equation}\begin{aligned}
\bU_i =  \bH\circ\bZ_i
\label{eq:segU}
\end{aligned}.\end{equation}

In step (4), we first perform a multiplication by the convolution transfer matrix between HST and MUSE. The convolution kernel from HST to MUSE $K_{HM}$ is obtained from the convolution kernels of HST $K_H$ and MUSE $K_M$, assumed to be known. 
Then the convolution result is resampled to the resolution of MUSE, using a linear interpolation.

We then have the intensity matrix at MUSE resolution $\widetilde{\bU}_{i,l}=\bB_l\bU_{i}$.

Then for each object defined by the segmentation map, we get its own intensity map (even if the objects are not separable on the MUSE data).

\subsubsection{Spectra estimation}

Step (c) could be done by simply solving a linear mixing model in the least squares sense. But when the intensity matrix is poorly conditioned, that is to say, when objects are spatially very close, solving the problem in the least squares sense may over-fit the noise and create spectral artifacts (see Sect. \ref{subs:simu}).

In order to cope with this phenomenon, a classical approach is to add constraints to regularize the problem.
There are a large number of possible regularizations in the literature, among which we can notably mention the penalization regularizations of the ridge type \citep {hoerl1970ridge} or LASSO \citep{tibshirani1996regression} and the regularizations by informational criterion like the Bayesian information criterion \citep[BIC:][]{schwarz1978estimating}. 
To answer to MUSE data specificities, we chose to exploit both ridge and sparse regularizations, as spectra are composed of a smooth spectral continuum and a small set of sharp emission/absorption lines. 

We thus separately process lines and continuum. For this, the procedure is as follows:
(i) obtaining the line cube $\bY^r$ by subtracting the continuum cube estimated by a robust filtering (e.g., a median filter); (ii) reconstruction of the $\widehat{\bD}^r$ lines associated with the objects from the line cube obtained previously: we will see that a parsimonious regularization approach with model selection has been chosen;
(iii) subtraction of the estimated contribution of the lines to obtain the remaining data $\bY^c = \bY-\widetilde\bU\widehat{\bD}^r$; (iv) estimation of the spectral continua $\widehat{\bD}^c$ using a ridge regularization on the data $\bY^c$; and (v) combination of the two estimates $\widehat{\bD} =\widehat{\bD}^r+\widehat{\bD}^c$.

\subsubsection*{Spectral line estimations}
Sparse regularization of spectral lines avoids over-fitting the noise and favors line attribution to a minimal number of spectra, which is the most physically probable.
Spectral lines are assumed to extend over several spectral slices. We can thus exploit this assumption to ensure a good lines reconstruction and avoid spectral discontinuities (where part of a line is associated with one object and the other part to another). 
Thus, we seek to do the model selection jointly over all the spectral support of the spectral line. To do so, it is thus necessary to estimate this spectral support.
Strategy is thus as follows: (i) fast detection of all potential spectral supports of lines and (ii) for each segment, selection of objects with nonzero spectrum, then estimation of spectra using only selected objects.

The lines support detection is based on local extrema detection and is detailed in \citep{bacher2017thesis}.
It allows all the spectral supports of the potential lines in the studied area to be obtained quickly. We can then estimate the spectra of objects with regularization by selecting models on each of these supports.

The sparse regularization is done using BIC selection.
BIC is written as $\BIC = K\log(n) -2\log(\hat{L})$ where $\hat{L}$ is the model likelihood maximum, $K$ is the number of free parameters, $n$ is the number of samples (here pixels in the area). In our case we assume that the noise is Gaussian. On a spectral band, for a model $\mathcal{M}$ with $k$ objects with nonzero spectrum, we have
\begin{equation}\begin{aligned}
\BIC(\mathcal{M}) = K\log(n) + \log(\widehat{\sigma_{\mathcal{M}}}^2)
\label{eq:bic_exp}
\end{aligned}
,\end{equation}

\noindent where $\widehat{\sigma_{\mathcal{M}}}^2$ is the empirical variance of the residuals obtained using the $ \mathcal{M} $ model chosen. We have a nonzero spectrum free parameter plus one for the noise variance, and hence $ K = k + 1 $. To minimize the criterion BIC we see that there is a compromise to be found between a model that best explains the data ($ \widehat {\sigma} ^ 2 $ minimal) and a simplest possible model ($ k $ minimal) .
The selection of the best model is then reduced to a combinatorial problem where all the possible combinations of nonzero spectra are tested and the one that minimizes the BIC is retained.
For computational reasons, instead of exploring all combinations we use a greedy variant for the BIC.

For each line $l$, we select the model $\mathcal{M}_l$ over the spectral support of $l$ using
\begin{equation}\begin{aligned}
\mathcal{M}_l = \argmin\limits_{\mathcal{M}} \BIC(\mathcal{M}),
\label{eq:sel_BIC}
\end{aligned}\end{equation} 
where $\BIC(\mathcal{M})$ comes from Eq. \eqref{eq:bic_exp}.

If we then note $\wbU_\mathcal{M}$ the intensity matrix composed only of the sources selected in $\mathcal{M}$, we can compute the associated spectra $\widehat{\bD}^r_l$ for line $l$ with  
\begin{equation}\begin{aligned}
\widehat{\bD}^r_l = \argmin\limits_{\bD} ||\bY^r_l-\wbU_{\mathcal{M}_l}\bD||^2_2
\label{eq:calc_spe_raie}
\end{aligned}
.\end{equation}

\subsubsection*{Continuum regularization}
After subtracting the estimated contribution of the lines, we now estimate the spectral continua of the objects using ridge regularization.
For a given block of sheets $l$ and an HST image $i$ we note the column vector $\by = \underbrace{\bY_{l}}_{n\times 1}$, $\widetilde{\bU}= \widetilde{\bU}_{i,l}$ and $\bd = \underbrace{\bD_{i,l}}_{k\times 1}$.  
Ridge regularized least squares admits an analytical solution of the form
\begin{equation}\begin{aligned}
\widehat{\bd} = \left(\widetilde{\bU}^T\widetilde{\bU}+\alpha\bI_k\right)^{-1}\widetilde{\bU}^T\by,
\label{eq:ls_exp_regul}
\end{aligned}\end{equation}
with $\bI_k$ the identity matrix of $\mathbb{R}^{k\times k}$ and where $\alpha\geq 0$ is the regularization parameter to be fixed.

The optimal value (in the Mean Square Error sense) of regularization parameter $\alpha$ is estimated using generalized cross validation \citep[GCV;][]{golub1979generalized}.
In order to increase the robustness of the cross-validation, only one regularization parameter is searched per block of a few tens of spectral sheets. This amounts to considering that the signal-to-noise ratio of the data (deprived of the lines) evolves weakly according to the wavelength. This hypothesis is valid if we consider that the sky's lines have been sufficiently well subtracted.

\subsubsection*{Flux correction}
A natural drawback of ridge regularization is that the estimated solution is biased toward zero, reflected in our case by a loss of flux on the estimated spectra, which is of course not desired.
To overcome this drawback, we take advantage of the redundancy of penalized estimators along the spectrum: by looking for a multiplicative factor per spectral block of sufficiently large size (on the order of a hundred sheets), we can make sure that the average flow (on the block) is preserved, while preserving the benefit of regularization. This correction factor is obtained in the following way: one looks for a diagonal matrix $ \bF = \diag (\{f_j \} _ {1 \leq j \leq k}) $ such as one has for continuum $ \bY ^ c = \wbU \bF \widehat {\bD} ^ c $. The matrix $ \bF $ being diagonal this amounts to defining 
\begin{equation}\begin{aligned}
f_j = \frac{\left\langle\left(\bY^c\left(\widehat{\bD}^{c}\right)^+\right)_j,\wbU_j\right\rangle}{||\wbU_j||_2^2}, \textrm{ for } 1\leq j \leq k.
\label{eq:corr_fact}
\end{aligned}\end{equation}
We note that if this least-squares correction was done wavelength by wavelength it would in fact cancel the ridge regularization.

The regularized approach of spectra estimation for an HST image $i$ within a FSF-constant spectral block $j$ is summarized in the algorithm \ref{alg:unmix_regul}.
We note that the whole method stays unsupervised as regularizations are without external parameters (for lines estimation) or with auto-estimated parameter (for continuum estimation).

\begin{algorithm}[htb]
\caption{Regularized procedure of spectra estimation}
\label{alg:unmix_regul}
\begin{algorithmic}[1]
\State \emph{Input:} data $\bY\leftarrow\bY_j$, intensity matrix $\widetilde{\bU}\leftarrow\widetilde{\bU}_{i,j}$  
\Statex
\State Estimation of $\bY^c$
\Comment Continuum estimation (robust filtering)
\State Lines estimation $\bY^r = \bY - \bY^c$
\Comment Continuum subtraction
\State Line detections
\For{ line $l$ }
\State Computation of $\mathcal{M}_l$ using eq. \eqref{eq:sel_BIC}
\Comment 
\State Computation of $\widehat{\bD}^r_l$ using eq. \eqref{eq:calc_spe_raie}
\EndFor
\State Construction of $\widehat{\bD}^r$
\Comment Spectral concatenation
\State Computation of $\bY^c = \bY - \bY^r$
\For{ bloc spectral $b$ }
\State Computation of $\alpha_{m+1s}$ using GCV
\State Computation of $\widehat{\bD}^r$
\Comment  Continuum estimation using ridge
\EndFor
\State Construction of $\widehat{\bD}^c$
\Comment Spectral concatenation
\State Computation of $\bF$ using \eqref{eq:corr_fact}
\Comment Flux correction factors
\State Computation of $\widehat{\bD} = \widehat{\bD}^r + \bF\widehat{\bD}^c$
\Comment Lines + Continuum
\Statex
\State \emph{Output:}
$\widehat{\bD}$
\Comment Sources spectra (for image HST $i$ and FSF-constant spectral band $j$)
\end{algorithmic}
\end{algorithm}

Finally, in order to take into account the residuals of the atmosphere spectrum subtraction, we add throughout the unmixing procedure, the estimation of a spectrum spatially constant in the considered area, corresponding to the sky background spectrum. This spectrum can be seen as the intercept of the studied regression problem.

\subsection{Validation on simulated data}
\label{subs:simu}

\subsubsection{Simulation settings}
In order to evaluate the validity of this method, we built a series of data with sources that are increasingly close to each other spatially. To do this, we built a datacube ($160 \times 160$) in flux units with high resolution (called HST data by abuse of language), which was then degraded following the MUSE model (same subsampling and core of convolution) to obtain data at low spatial resolution (which will be called MUSE data by abuse of language). An additive Gaussian noise is then applied, slightly correlated spatially by the application of a 3 by 3 spatial core.

\begin{figure}[hbtp]
\centering
\includegraphics[width=0.9\linewidth]{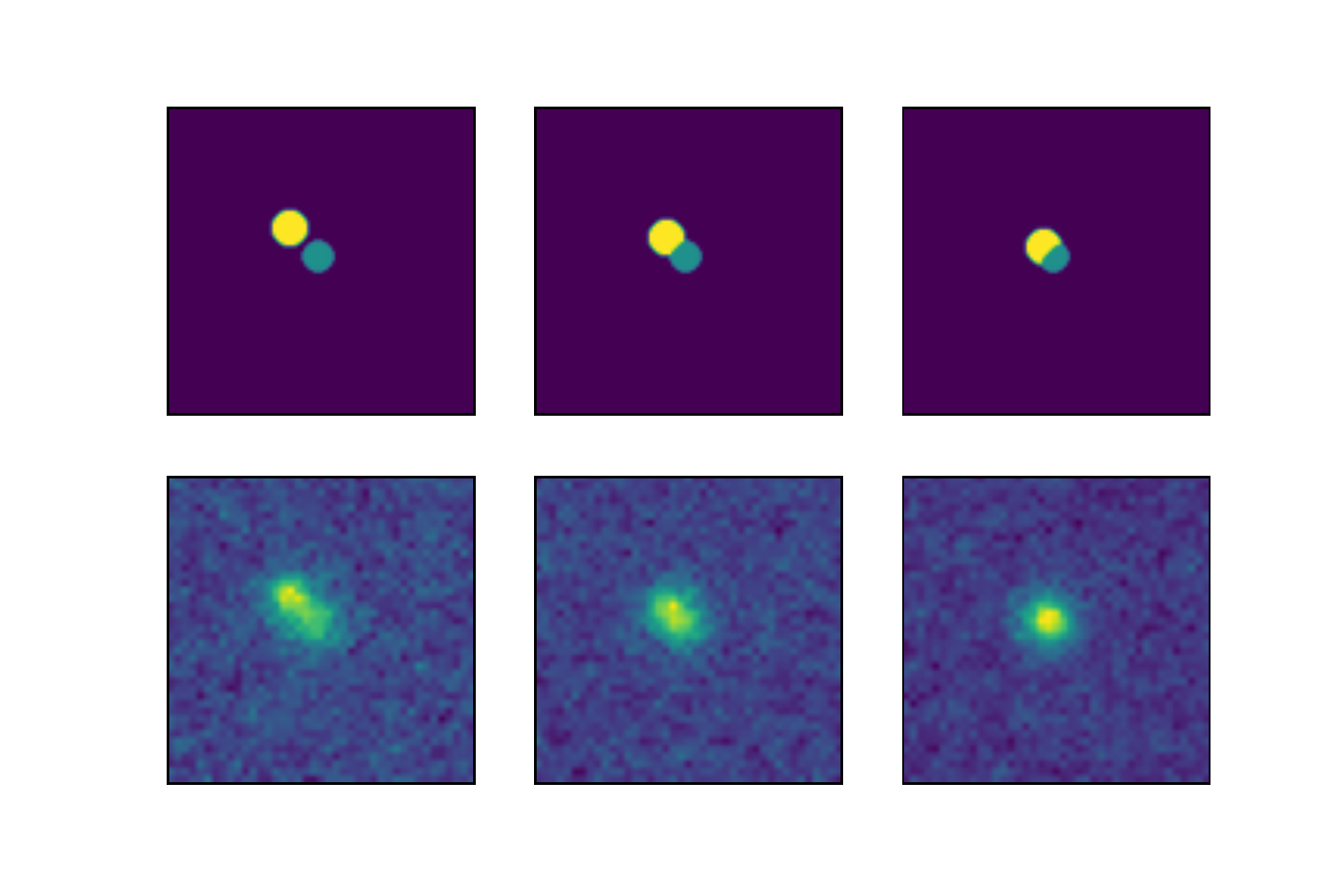}
\caption{Simulation example. First row: "HST" segmentation map. Second row: "MUSE" white image. Distance of sources is characterized by the unit-less conditioning number $c$, increasing here from 1.2 (left column) to 3.7 (right column).}
\label{fig:simu_segmap_white}
\end{figure}

\begin{figure}
\centering
\includegraphics[width=0.8\linewidth]{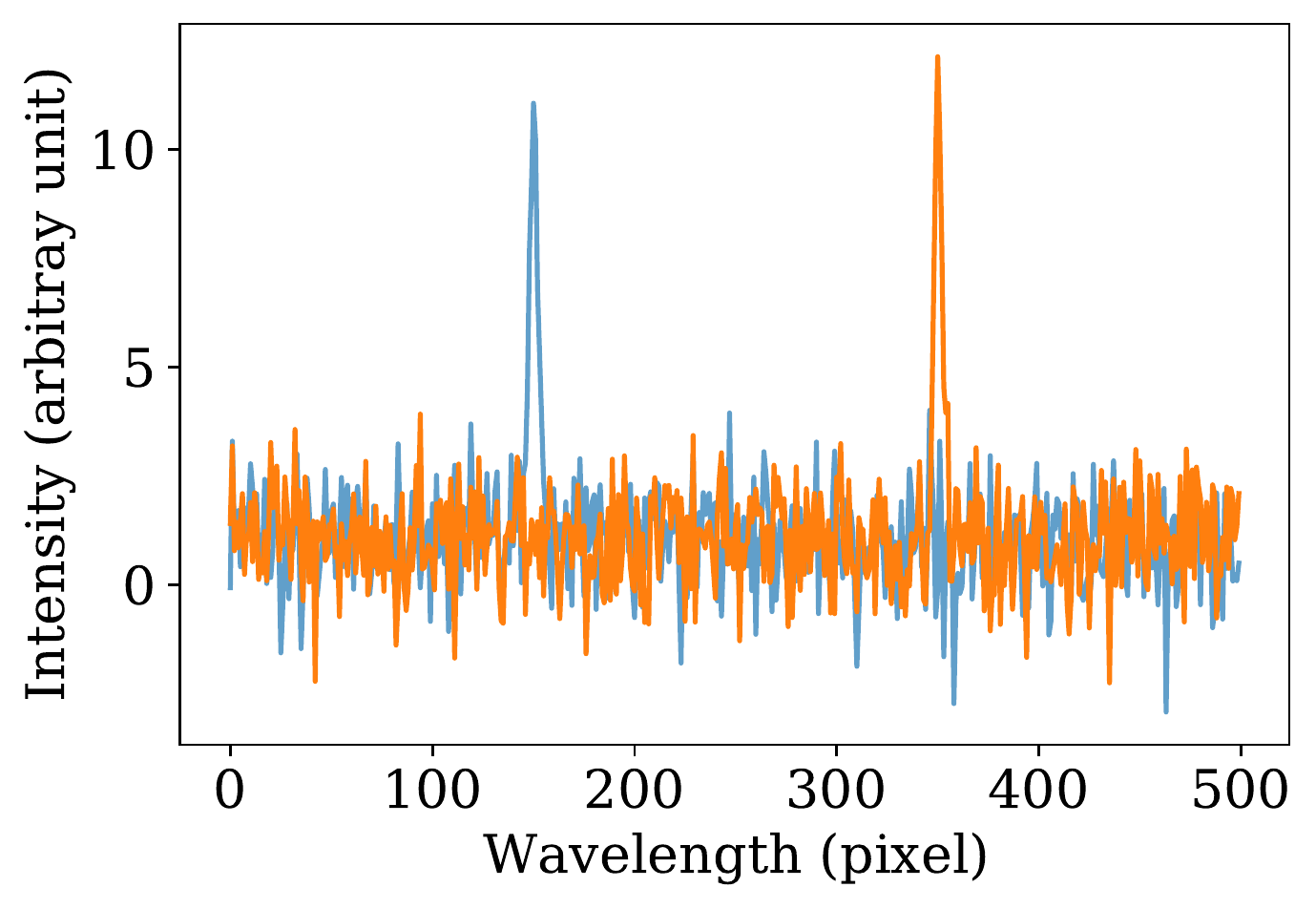}\caption{Ground truth spectra of the two simulated sources.}
        \label{fig:spectra}
\end{figure}

We can see in Fig. \ref{fig:simu_segmap_white} the simulated cube white image at MUSE and HST resolution.
Simulated spectra are composed of an emission line for each object and a spectral continuum (see fig. \ref{fig:spectra}).

De-blending difficulty is measured using the conditioning number $c$ of intensity matrix  $\widehat{\bU}$ (at MUSE spatial resolution): 
\begin{equation}
c=\frac{s_{max}(\widehat{\bU})}{s_{min}(\widehat{\bU})},
\label{eq:cond_number}
\end{equation}
 where $s_{max}(\widehat{\bU})$ and $s_{min}(\widehat{\bU})$ are the minimum and maximum singular values of the intensity matrix, and $c$ is a bound of estimation error of $\widehat{\bD}$ relative to a perturbation on data $\bY$.
The closer $c$ is to 1, the easier the inversion problem is. On the contrary, high values of $c$ will imply instability and over-fitting of the noise.

We then measured the fidelity to the ground truth  (the mean of the intercorrelations between each estimated spectrum,  $\widehat{\bd_j}$, and its ground truth, $\bd_j$),
 $$
 f = \frac{1}{k}\sum\limits_{j=1}^k \frac{\langle\widehat{\bd_j},\bd_j\rangle}{||\widehat{\bd_j}||.||\bd_j||},
$$
and the inter-correlation, 
  $$
  ic =\frac{\langle\widehat{\bd_0},\widehat{\bd_1}\rangle}{||\widehat{\bd_0}||.||\widehat{\bd_1}||},
  $$
  between the two estimated spectra.

\subsubsection{Results}
\begin{figure}[hbtp]
\centering
\includegraphics[width=0.8\linewidth]{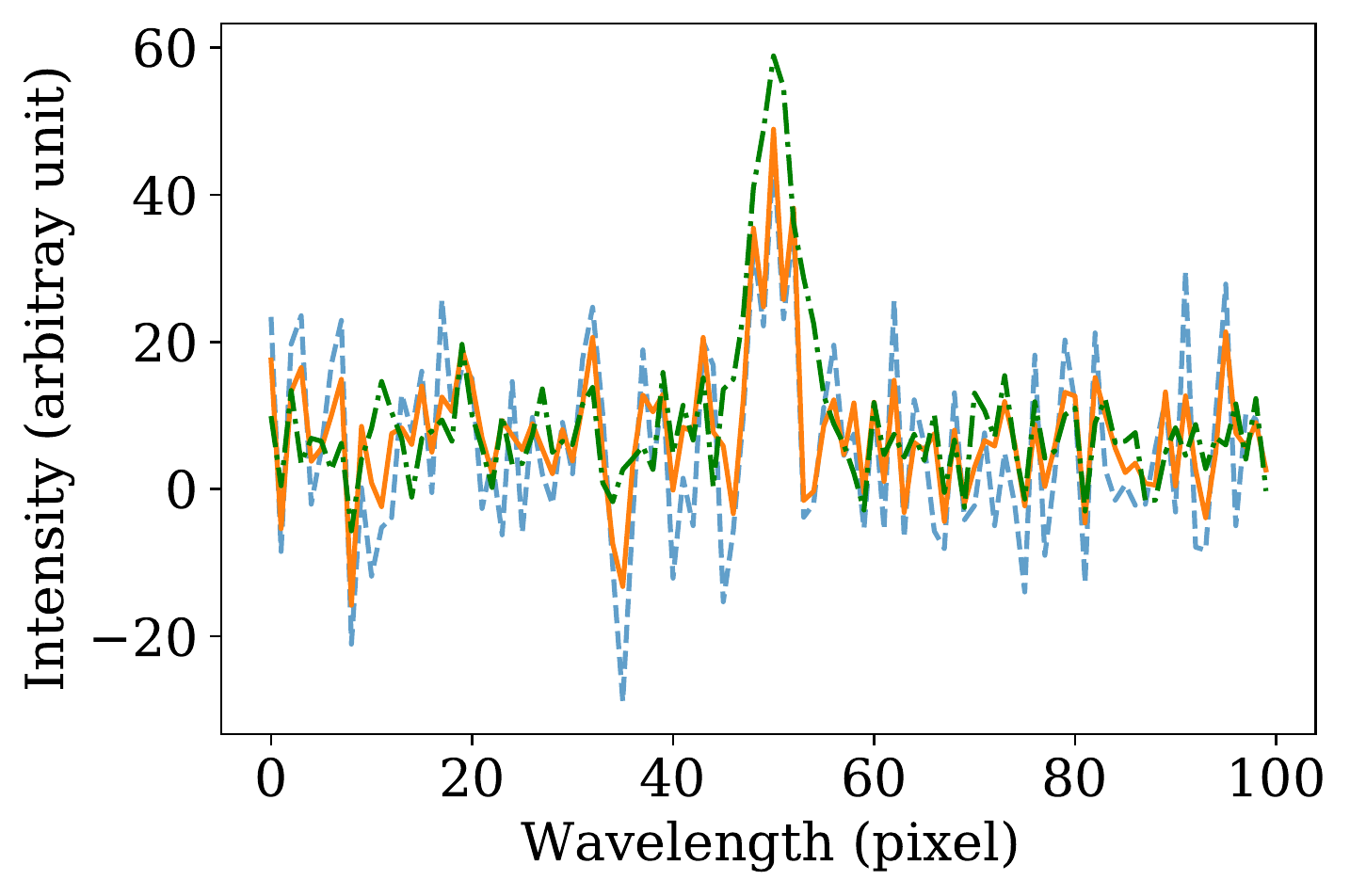}
\caption{Ground truth (green dashed-dot) versus estimated spectra (proposed method in orange, least squares in dashed blue) for a difficult case (condition number equals 6)}
        \label{fig:deblending}
\end{figure}

We compared the proposed regularization method to a simpler least squares approach that has a closed-form solution given by
\begin{equation}\begin{aligned}
\widehat{\bD}_{i,l} =(\widetilde{\bU}_{i,l}^T\widetilde{\bU}_{i,l})^{-1} \widetilde{\bU}_{i,l}^T\bY_l.
\end{aligned}\end{equation}Figure \ref{fig:deblending} shows one of the two estimated spectra, around its spectral emission line, and when the sources are very close spatially ($c\approx6$).
We can see that both the proposed method and a non-regularized least squares approach estimate accurately the emission line but the proposed method is better at avoiding increased variance around the peak.

\begin{figure}[hbtp]
\centering
\includegraphics[width=0.75\linewidth]{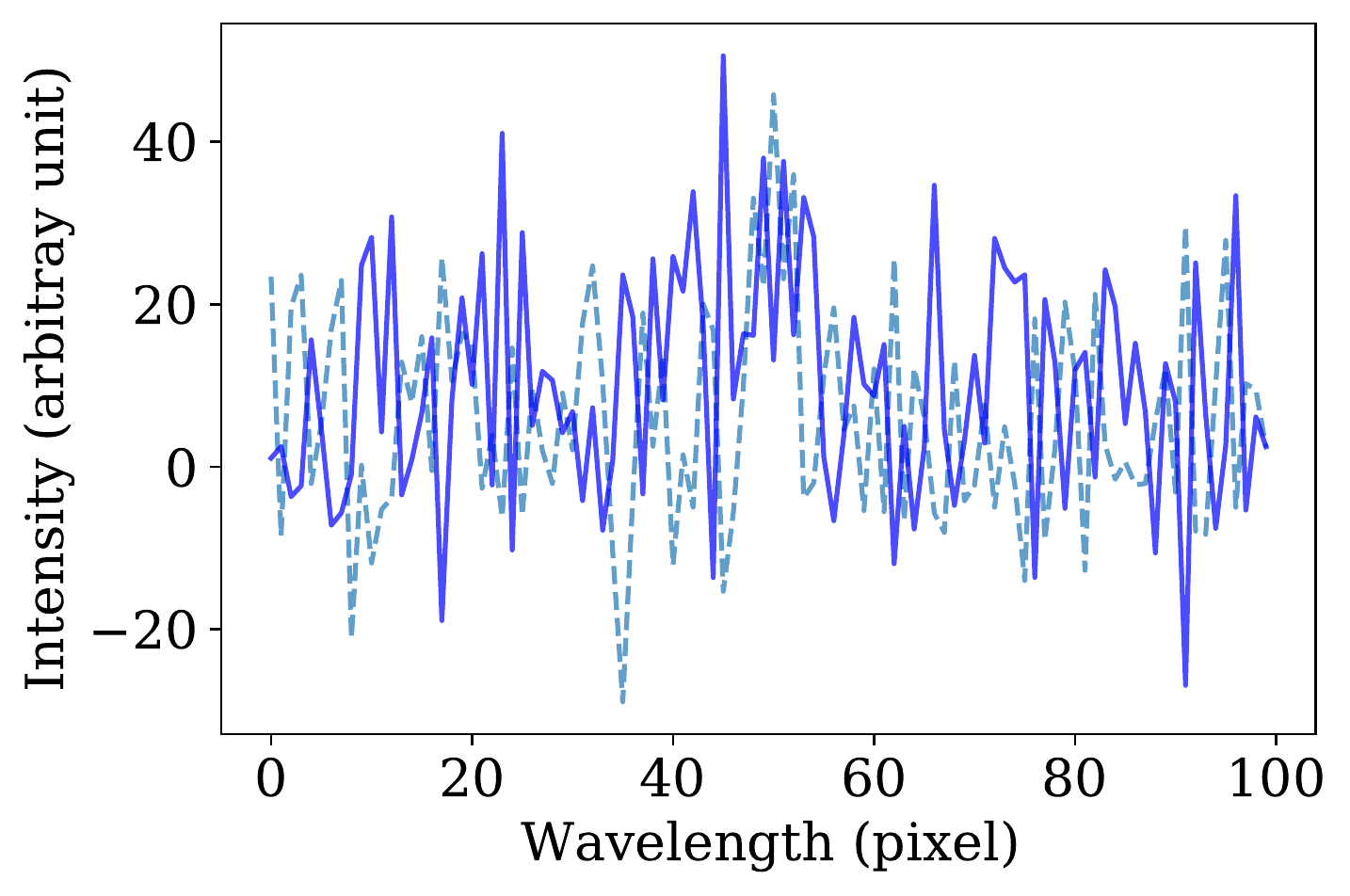}
\caption{Zoom over a part of the non-regularized least squares estimation of both sources spectra, for a difficult case (condition number equals 6). Some anticorrelated artifacts are visible.}
        \label{fig:anticorr}
\end{figure}

\begin{figure}[htbp]
        \centering
        \subfloat[Variance of spectra residuals]{\includegraphics[width=0.75\linewidth]{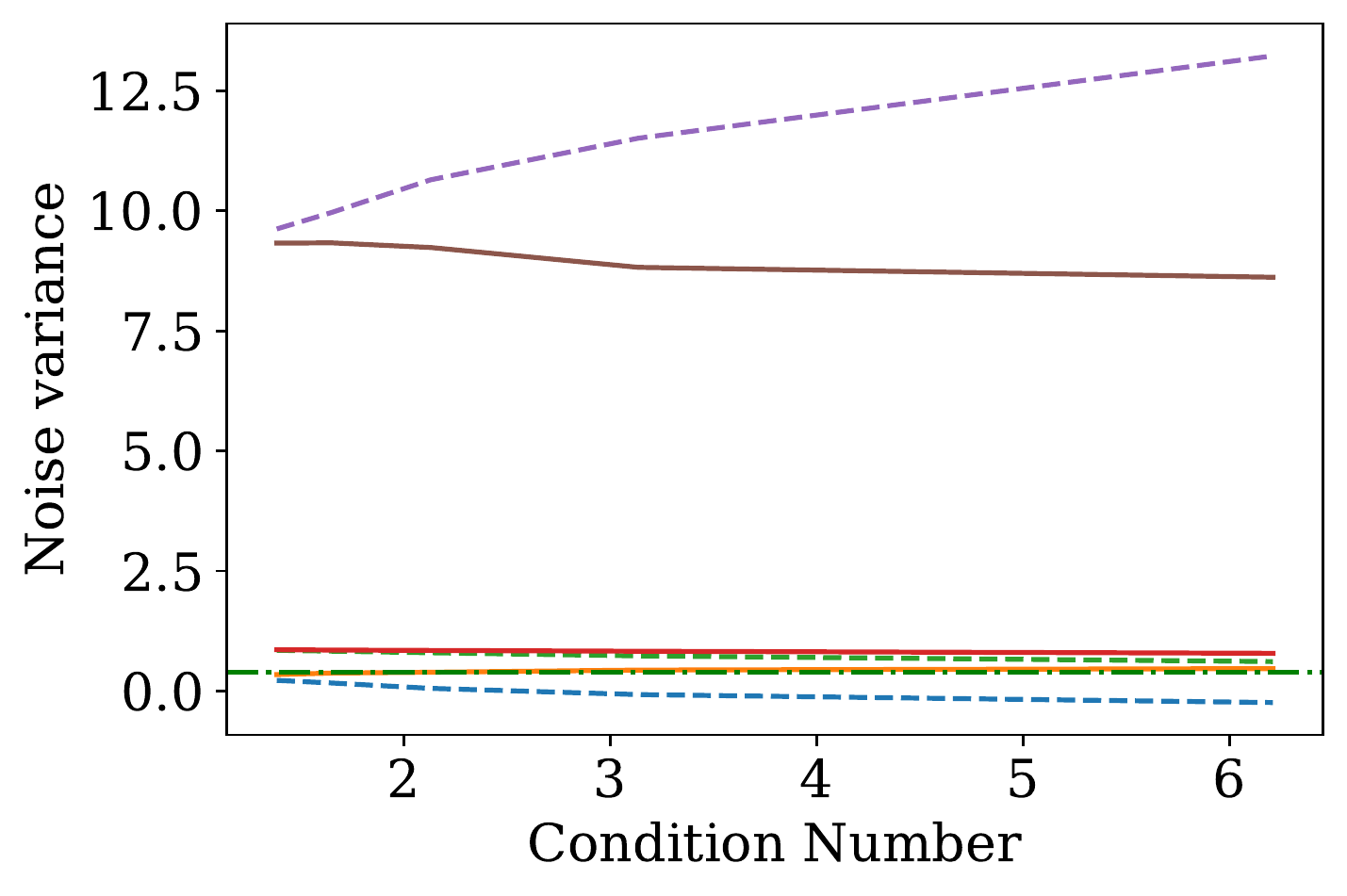}}\\
                \subfloat[Intercorrelation between objects spectra (dashed-dot green line is the ground truth)]{\includegraphics[width=0.75\linewidth]{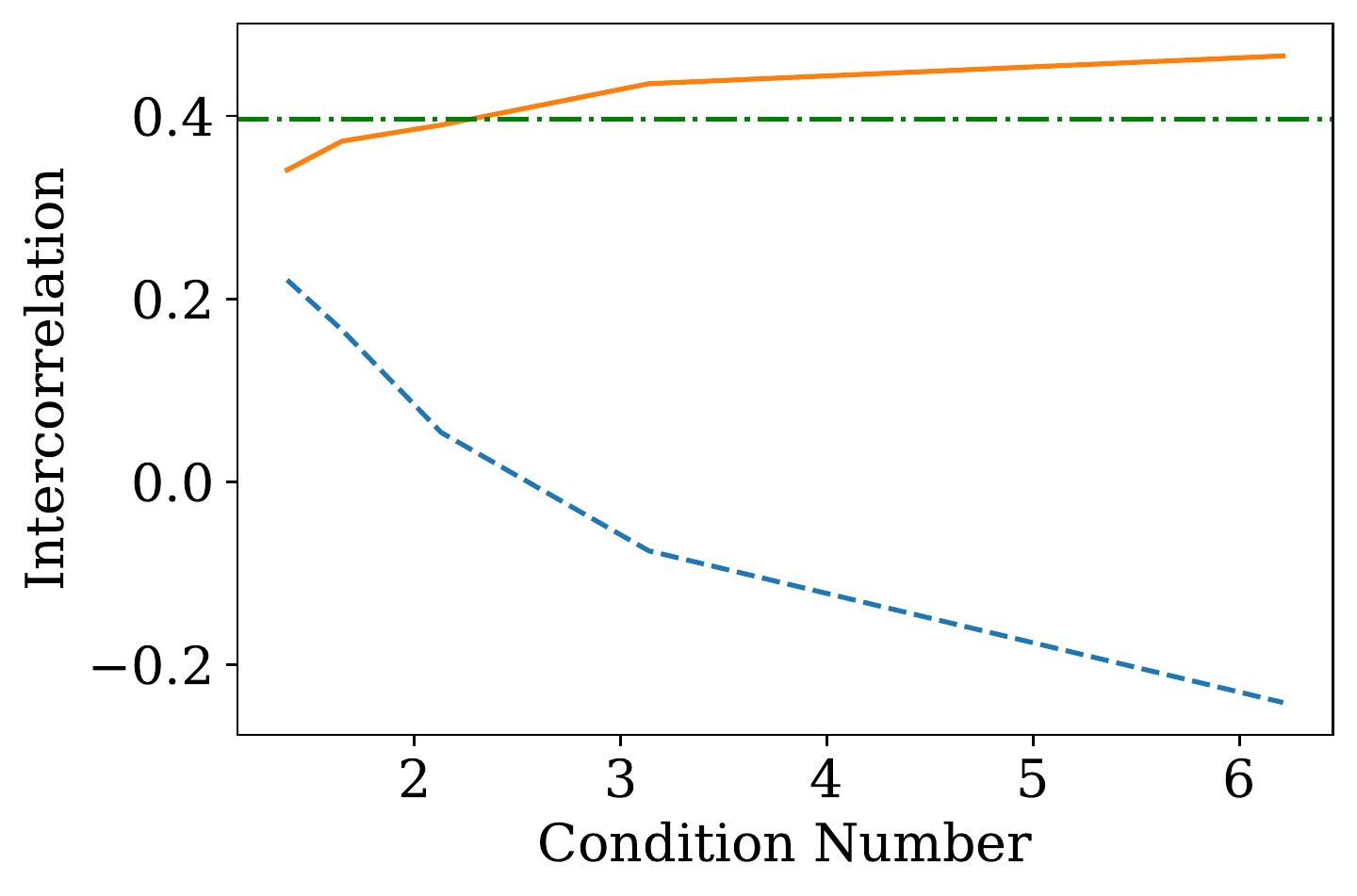}}\\
                                \subfloat[Fidelity to ground truth]{\includegraphics[width=0.75\linewidth]{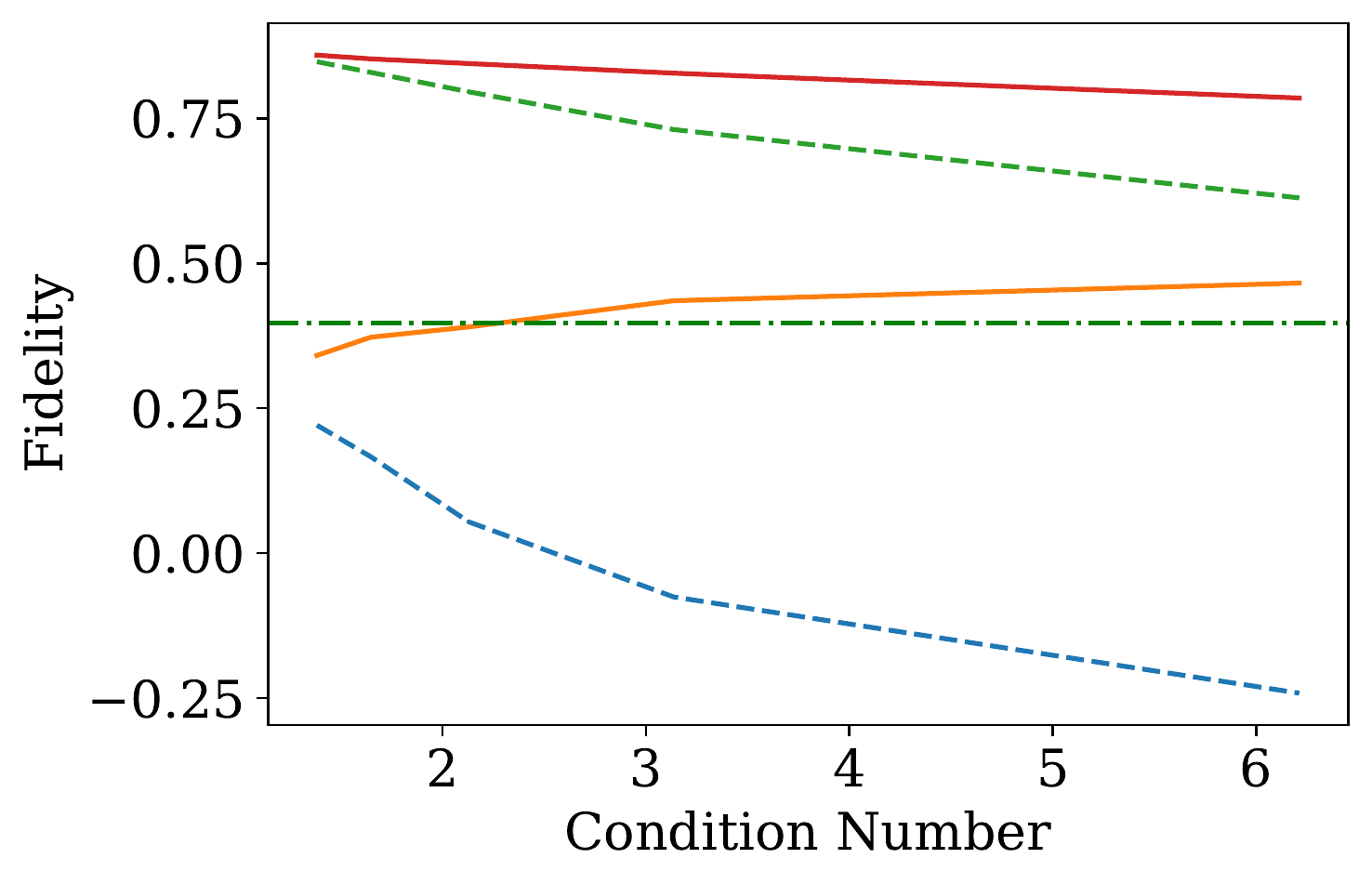}}\\
        \caption{Performance evolution function of conditioning. The dashed blue line indicates no regularization and the orange line with regularization. Results are averaged over ten Monte Carlo runs.}
        \label{fig:compare_regul}
\end{figure}

The results shown in Fig. \ref{fig:compare_regul} describe the evolution of performance as the conditioning degrades. Performances of the proposed method are clearly more robust to ill-conditioning than a simple least squares approach. In particular, variance of the estimated spectra residuals increases and the estimated spectra become strongly negatively correlated. This anticorrelation is explained by the positivity of the $\widetilde\bU$ intensity matrix. This phenomenon is also illustrated by Figs. \ref{fig:deblending} and \ref{fig:anticorr}, where the unmixed spectra when the sources are very close spatially ($c\approx6$) are shown. The robustness to anticorrelated patterns in the proposed method allows it to keep a good fidelity to the sources spectra (fig \ref{fig:compare_regul}c).
. 

\section{Emission and absorption lines}
\label{sec:linename}
The lines input table used in \pfit\ is given in Tab.~\ref{tab:linename}. We note that the default line table can also be retrieved directly from the \pfit\ python interface\footnote{\href{https://pyplatefit.readthedocs.io/en/latest/tutorial.html\#Selecting-and-updating-emission/absorption-lines}{pyplatefit.readthedocs.io/en/latest/tutorial.html\#Selecting-and-updating-emission/absorption-lines}}.

The lines are grouped into 3 different families (column {\em Set}): the Balmer series ({\em bal}), non-balmer emission lines ({\em for}) and ISM absorption lines ({\em ism}). The vacuum rest wavelength in \AA\ is given in the column {\em Wave}. The multiplet ({\em Mult}) column is used to group lines together when performing the fit. {\em Mult} is an integer value following the adopted notation in the literature (e.g., [OII]3727 for the \oiid\ doublet).
The {\em Main} boolean column can be used to filter main and secondary lines. The two boolean columns, {\em Emi} and {\em Abs}, indicate if the line is in emission and absorption, respectively. The {\em Res} column is used to perform an independent fit for some major resonant lines. We note that some emission lines like OVI and NV, although resonant, are not fitted separately because they were too faint in our galaxy sample.

\begin{table}[!htbp]
\tiny
\caption{\pfit\ lines definition.}   
\label{tab:linename}
\begin{tabular*}{\columnwidth}{lccccccc}
\hline
Name & Set & Wave & Mult & Main & Emi & Abs & Res \\
\hline
OVI1032 & for & 1031.91 & 1033 & F & T & F & F \\
OVI1038 & for & 1037.61 & 1033 & F & T & F & F \\
LYALPHA & bal & 1215.67 & 0 & T & T & F & T \\
NV1238 & for & 1238.82 & 1240 & F & T & F & F \\
NV1242 & for & 1242.80 & 1240 & F & T & F & F \\
SiII1260 & ism & 1260.42 & 0 & F & F & T & F \\
OI1302 & ism & 1302.17 & 1303 & F & F & T & F \\
SiII1304 & ism & 1304.37 & 1303 & F & F & T & F \\
CII1334 & ism & 1334.53 & 0 & F & F & T & F \\
SiIV1394 & for & 1393.76 & 1403 & F & T & T & F \\
OIV1397 & for & 1397.23 & 1403 & F & T & F & F \\
OIV1400 & for & 1399.78 & 1403 & F & T & F & F \\
SiIV1403 & for & 1402.77 & 1403 & F & F & T & F \\
NIV1486 & for & 1486.50 & 0 & F & T & F & F \\
SiII1527 & ism & 1526.71 & 0 & F & F & T & F \\
CIV1548 & for & 1548.20 & 1549 & T & T & T & T \\
CIV1550 & for & 1550.77 & 1549 & T & T & T & T \\
FeII1608 & ism & 1608.45 & 1610 & F & F & T & F \\
FeII1611 & ism & 1611.20 & 1610 & F & F & T & F \\
HeII1640 & for & 1640.42 & 0 & F & T & F & F \\
OIII1660 & for & 1660.81 & 1663 & F & T & F & F \\
OIII1666 & for & 1666.15 & 1663 & F & T & F & F \\
AlII1671 & ism & 1670.79 & 0 & F & F & T & F \\
NIII1750 & for & 1749.67 & 0 & F & T & F & F \\
AlIII1854 & ism & 1854.10 & 0 & F & F & T & F \\
AlIII1862 & ism & 1862.17 & 0 & F & F & T & F \\
SiIII1883 & for & 1882.71 & 1886 & F & T & F & F \\
SiIII1892 & for & 1892.03 & 1886 & F & T & F & F \\
CIII1907 & for & 1906.68 & 1909 & T & T & F & F \\
CIII1909 & for & 1908.73 & 1909 & T & T & F & F \\
CII2324 & for & 2325.40 & 2326 & F & T & F & F \\
CII2325 & for & 2326.11 & 2326 & F & T & F & F \\
CII2326 & for & 2327.64 & 2326 & F & T & F & F \\
CII2328 & for & 2328.84 & 2326 & F & T & F & F \\
FeII2344 & ism & 2344.21 & 0 & F & F & T & F \\
FeII2374 & ism & 2374.46 & 0 & F & F & T & F \\
FeII2382 & ism & 2382.76 & 0 & F & F & T & F \\
NeIV2422 & for & 2421.83 & 2424 & F & T & F & F \\
NeIV2424 & for & 2424.42 & 2424 & F & T & F & F \\
OII2470 & for & 2471.02 & 0 & F & T & F & F \\
FeII2586 & ism & 2586.65 & 0 & F & F & T & F \\
FeII2600 & ism & 2600.17 & 0 & F & F & T & F \\
MgII2796 & for & 2796.35 & 2799 & F & T & T & T \\
MgII2803 & for & 2803.53 & 2799 & F & T & T & T \\
MgI2853 & ism & 2852.97 & 0 & F & F & T & F \\
NeV3426 & for & 3426.85 & 0 & F & T & F & F \\
OII3726 & for & 3727.09 & 3727 & T & T & F & F \\
OII3729 & for & 3729.88 & 3727 & T & T & F & F \\
H11 & bal & 3771.70 & 0 & F & T & T & F \\
H10 & bal & 3798.98 & 0 & F & T & T & F \\
H9 & bal & 3836.47 & 0 & F & T & T & F \\
NeIII3869 & for & 3870.16 & 0 & T & T & F & F \\
HeI3889 & for & 3889.73 & 0 & F & T & F & F \\
H8 & bal & 3890.15 & 0 & F & T & T & F \\
CaK & ism & 3933.66 & 0 & F & F & T & F \\
CaH & ism & 3968.45 & 0 & F & F & T & F \\
NeIII3967 & for & 3968.91 & 0 & F & T & F & F \\
HEPSILON & bal & 3971.20 & 0 & F & T & T & F \\
HDELTA & bal & 4102.89 & 0 & T & T & T & F \\
CaG & ism & 4304.57 & 0 & F & F & T & F \\
HGAMMA & bal & 4341.68 & 0 & T & T & T & F \\
OIII4363 & for & 4364.44 & 0 & F & T & F & F \\
HBETA & bal & 4862.68 & 0 & T & T & T & F \\
OIII4959 & for & 4960.30 & 0 & T & T & F & F \\
OIII5007 & for & 5008.24 & 0 & T & T & F & F \\
MgB & ism & 5175.44 & 0 & F & F & T & F \\
HeI5876 & for & 5877.25 & 0 & F & T & F & F \\
NaD & ism & 5891.94 & 0 & F & F & T & F \\
OI6300 & for & 6302.05 & 0 & F & T & F & F \\
NII6548 & for & 6549.85 & 0 & F & T & F & F \\
HALPHA & bal & 6564.61 & 0 & T & T & T & F \\
NII6584 & for & 6585.28 & 0 & T & T & F & F \\
SII6717 & for & 6718.29 & 0 & T & T & F & F \\
SII6731 & for & 6732.67 & 0 & T & T & F & F \\
ARIII7135 & for & 7137.80 & 0 & F & T & F & F \\
\end{tabular*}
\end{table}

\end{appendix}

\end{document}